\documentclass[referee,a4paper,12pt,traditabstract]{jswsc} 
\usepackage{amsmath}
\usepackage{graphicx}
\usepackage{txfonts}
\usepackage{subfigure}
\usepackage{epstopdf}
\usepackage[mathlines]{lineno}
\usepackage[authoryear,round]{natbib}
\usepackage[backref]{hyperref}
\usepackage{url}
	\usepackage{comment}
	\usepackage{outlines}
	\usepackage[english]{babel}
	\usepackage{csquotes}
	\usepackage{floatrow}
	\usepackage{multirow}
	\usepackage[table]{xcolor}
	\usepackage{setspace}
	\usepackage{enumitem}
		
	\usepackage{longtable}
	\usepackage{tabu}
	\newcommand{\equationname}{equation}
	\renewcommand{\figurename}{Fig.}
	\renewcommand{\tablename}{Table}
	\newcommand{\sectionname}{Sect.}

	\newcommand{\ie}{{i.e.,}}
	\newcommand{\cf}{{cf.}}
	\newcommand{\eg}{{\em e.g.,}}
	\newcommand{\alf}{Alfv\'en }
	
	\newcommand{\grad}{ \overline{\mathbf{\nabla}} }
	\newcommand{\divB}{ \overline{\mathbf{\nabla}} \cdot \overline{\bB} }
	\newcommand{\divg}{ \overline{\mathbf{\nabla}} \cdot }
	\newcommand{\curl}{ \overline{\mathbf{\nabla}} \times }
	\newcommand{\unitI}{ \bar{\bar{\bI}} }

	\newcommand{\bB}{ \mathbf{B} }
	\newcommand{\bD}{ \mathbf{D} }
	\newcommand{\bU}{ \mathbf{U} }
	\newcommand{\bE}{ \mathbf{E} }
	\newcommand{\bF}{ \mathbf{F} }

	\newcommand{\bI}{ \mathbf{I} }
	\newcommand{\bJ}{ \mathbf{J} }

	\newcommand{\bP}{ \mathbf{P} }

	\newcommand{\bS}{ \mathbf{S} }
	
	\newcommand{\bV}{ \mathbf{V} }

	\newcommand{\ba}{ \mathbf{a} }
	\newcommand{\bb}{ \mathbf{b} }
	\newcommand{\bc}{ \mathbf{c} }
	\newcommand{\be}{ \mathbf{e} }

	\newcommand{\bj}{ \mathbf{j} }

	\newcommand{\Bn}{ \mathbf{n} }

	\newcommand{\br}{ \mathbf{r} }
	
	\newcommand{\bu}{ \mathbf{u} }
	\newcommand{\Bv}{ \mathbf{v} }
	
	\newcommand{\bx}{ \mathbf{x} }

	\newcommand{\sA}{ {\scriptscriptstyle{\rm A}} }
	\newcommand{\sB}{ {\scriptscriptstyle{\rm B}} }
	\newcommand{\sC}{ {\scriptscriptstyle{\rm C}} }
	\newcommand{\sD}{ {\scriptscriptstyle{\rm D}} }
	\newcommand{\sE}{ {\scriptscriptstyle{\rm E}} }
	
	\newcommand{\sG}{ {\scriptscriptstyle{\rm G}} }
	\newcommand{\sH}{ {\scriptscriptstyle{\rm H}} }
	\newcommand{\sI}{ {\scriptscriptstyle{\rm I}} }

	\newcommand{\sM}{ {\scriptscriptstyle{\rm M}} }
	
	\newcommand{\sO}{ {\scriptscriptstyle{\rm O}} }
	
	\newcommand{\sR}{ {\scriptscriptstyle{\rm R}} }

	\newcommand{\sSun}{ {\scriptscriptstyle{\rm \odot}} }
	\newcommand{\sperp}{ {\scriptscriptstyle{\rm \perp}} }
	\newcommand{\spar}{ {\scriptscriptstyle{\rm \parallel}} }
	\newcommand{\snull}{ {\scriptscriptstyle{\rm 0}} }

	\newcommand{\muo}{ \mu_\snull }
	\newcommand{\epso}{ \epsilon_\snull }

    \DeclareMathOperator\sgn{sgn}
    
	\hypersetup{colorlinks=true,citecolor=cyan,urlcolor=cyan,linkcolor=blue}
    
\begin{document}
\title{What Sustained Multi-Disciplinary Research Can Achieve: The Space Weather Modeling Framework} 
    \titlerunning{Physics-Based Space Weather Modeling with SWMF}
    \authorrunning{Gombosi et al.}
    \author{Tamas I. Gombosi\inst{1},
          Yuxi Chen\inst{2},
          Alex Glocer\inst{3},
          Zhenguang Huang\inst{2},
          Xianzhe Jia\inst{2},
          Michael W. Liemohn\inst{2},
          Ward B. Manchester\inst{2},
          Tuija Pulkkinen\inst{4},
          Nishtha Sachdeva\inst{2},
          Qusai {Al Shidi}\inst{2},
          Igor V. Sokolov\inst{2},
          Judit Szente\inst{2},
          Valeriy Tenishev\inst{2},
          Gabor Toth\inst{2},
          Bart van der Holst\inst{2},
          Daniel T. Welling\inst{5},
          Lulu Zhao\inst{2}
          \and
          Shasha Zou\inst{2}}
	\institute{Department of Climate and Space Sciences and Engineering (CLaSP), 
			University of Michigan, Ann Arbor, MI 48109, USA;
			\email{\href{mailto:tamas@umich.edu}{tamas@umich.edu}}
		\and
			CLaSP, University of Michigan Ann Arbor
		\and
			NASA Goddard Space Flight Center
		\and
			CLaSP, University of Michigan Ann Arbor. Also at: Aalto University, Espoo, Finland
		\and
			Department of Physics, University of Texas at Arlington}
\abstract
{
    MHD-based global space weather models have mostly been developed and maintained at academic institutions. While the ``free spirit'' approach of academia enables the rapid emergence and testing of new ideas and methods, the lack of long-term stability and support makes this arrangement very challenging. This paper describes a successful example of a university-based group, the Center of Space Environment Modeling (CSEM) at the University of Michigan, that developed and maintained the Space Weather Modeling Framework (SWMF) and its core element, the BATS-R-US extended MHD code. It took a quarter of a century to develop this capability and reach its present level of maturity that makes it suitable for research use by the space physics community through the Community Coordinated Modeling Center (CCMC) as well as operational use by the NOAA Space Weather Prediction Center (SWPC).
}
	\maketitle
    \pagenumbering{arabic}
    \setcounter{page}{1}
    \setcounter{tocdepth}{4}
    \tableofcontents

\section{Introduction}
Over the past few decades there has been an increasing awareness of the potentially devastating impact that the dynamic space environment can have on human assets. Extreme ``space weather" events, driven by eruptive solar events such as Coronal Mass Ejections (CMEs), are widely recognized as critical hazards whose consequences cannot be ignored.

Because of society's reliance on the electrical grid, the internet, high-frequency communication, GPS (Global Positioning System) navigation signals and an increasing array of digital electronic devices, space weather events -- such as severe solar storms -- can wreak havoc on technological systems and trigger losses from business interruption and damaged physical assets \cite[cf.,][]{Baker:2009a}. While power outages from space weather are low-frequency events, they have the potential to cause crippling long-term damage. In fact, the risk of high impact damages due to space weather fits the profile of a market-changing catastrophe such as hurricane Katrina, the 9/11 attack, or the Japanese earthquake and tsunami \cite[\cf][]{FEMA:2019a}. All were unprecedented and believed to be highly unlikely -- and yet they occurred.

There is an additional, less publicized reason that policymakers care about space weather: its association to electromagnetic pulses (EMPs) \cite[cf.,][]{Gombosi:2017a}. An EMP is a natural or anthropogenic burst of electromagnetic energy that can damage all kinds of electronic and even physical objects. Understanding and mitigating space weather effects also have national defense implications.

Space weather involves a vast domain extending from the Sun to beyond Earth's orbit, with regions governed by very different physics at different spatial and temporal scales. Simulating and predicting space weather with first-principles models requires space physics expertise for the various sub-domains and advanced numerical algorithms. Since the sub-domain models keep changing and evolving, they need to be coupled in a flexible manner using proper software engineering. Finally, the simulation needs to run faster than real time, which means that a deep understanding of high-performance computing is required. Clearly, developing a first-principles space weather model requires sustained multi-disciplinary collaboration of space physicists, applied mathematicians, computer scientists and software engineers. 

Presently there are only a couple of physics-based space weather models that are capable of spanning the entire region from the low solar corona to the edge of the heliosphere. One is the European Space Agency's Virtual Space Weather Modelling Centre \cite[VSWMC,][]{Poedts:2020a} and the other one is the Space Weather Modeling Framework \cite[SWMF,][]{Toth:2005swmf, Toth:2012a}.
In this paper we describe the evolution and current capabilities of the SWMF and its unique capabilities to address the myriad of processes involved in studying and predicting space weather.  
In the main text we focus on the the broad range of space weather simulations made possible by the advanced capabilities of BATS-R-US (Block Adaptive-Tree Solar-wind Roe-type Upwind Scheme) and SWMF. The fundamentals of the BATS-R-US and SWMF codes are described in detail in Appendix~\ref{sec:creation}. The extended physics and algorithmic advances incorporated in these codes are important and we present a concise summary of these advances in Appendices~\ref{sec:physics} (physics) and \ref{sec:algorithms} (algorithms). Finally, Appendix~\ref{sec:epic} describes our most advanced simulation capability that embeds fully kinetic domains inside extended MHD models.

\section{Evolution of Space Weather Models}
\label{sec:evolution}

Models capable of predicting space weather can be loosely divided into three broad categories: Empirical models, black box (mainly machine learned) models, and physics-based models. 

\subsection{Empirical Models}
\label{subsec:empirical}
Empirical models aggregate data in different ways to make specific predictions of the current and future state of the system based on how the system has responded historically. Such models are mostly data driven and typically make limited or no assumptions of the underlying physics. The quality of the models is heavily dependent on the data coverage in space and over different geomagnetic conditions. Widely used examples are the MSIS (Mass Spectrometer and Incoherent Scatter) model \cite[]{Hedin:1987a, Hedin:1991a} of the upper atmosphere and the \cite{Tsyganenko:1989a, Tsyganenko:1995a, Tsyganenko:2002a, Tsyganenko:2002b} models of the terrestrial magnetic environment.

The MSIS model \cite[]{Hedin:1987a, Hedin:1991a} brings together mass spectrometer and incoherent scatter data to build an empirical model of the thermosphere. The model provides estimates of temperature and the densities of atmospheric constituents such as N$_2$, O, O$_2$, He, Ar, and H. Low-order spherical harmonics expansion is used to describe spatial (latitude, local time), and temporal (annual, semiannual) variations. The model is often used for data comparisons and theoretical calculations requiring a background atmosphere, for example in calculations of satellite orbital decay caused by atmospheric drag.

The extension of the geodipole field to the magnetosphere is sustained by currents flowing in the geospace. The magnetic field variations from these currents can be deduced from space-borne magnetic field measurements, and have been collected into a large database. The Tsyganenko models \cite[]{Tsyganenko:1989a, Tsyganenko:1995a, Tsyganenko:1996a, Tsyganenko:2002a, Tsyganenko:2002b, Tsyganenko:2005a} describe the large-scale current systems with parametrized empirical functions, and the parameter values are found through least-squares fitting to the large observational database. The models have been extensively used \eg to connect magnetospheric substorm and storm dynamic processes to their ionospheric signatures \cite[]{Pulkkinen:1992a, Baker:1996a, Pulkkinen:2006a}.

\subsection{Black-Box Models}
\label{subsec:blackbox}

Linear prediction filters have been used to build models for a variety of space weather parameters, including the auroral electrojet (AE) indices and the ring current Dst (Disturbance storm-time) index. 
Predictions of magnetospheric storm conditions have been done using neural networks to construct nonlinear models to forecast the AL and/or Dst index using various solar wind driver parameters \cite[]{Lundstedt:1994a, Weigel:2003a}. 

Recent machine learning models have been quite successful in predicting geomagnetic indices \cite[see][]{Camporeale:2019a, Leka:2018b}. To support their use in space weather research requires open-access, robust, and effective software tools. Typically, the models are custom-made and making use of a stack of standard computational frameworks for learning. Machine learning methods have also been employed for prediction of the ionospheric total electron content (TEC) \cite[\cf][]{Liu:2000a} and solar flares \cite[\cf][]{Chen:2019a_ML, Jiao:2020a, Wang:2020a}. However, as most machine learning models are not interpretable, they typically do not help us to understand the underlying physics.

\subsection{Physics-Based Models}
\label{subsec:physics-based}
Physics based models directly solve equations representing the underlying physical processes in the system, often with observations based inputs, in order to study the evolution and dynamics of the space environment. 
Physics-based space weather models have been found to be particularly valuable for predicting both the rare extreme events as well as more commonly observed space weather. 

Extreme space weather events with the most severe implications for human assets and activities are low-frequency events that create challenges for forecasting and prediction. Since the dawn of the space age, there have been a handful of events with major space weather impacts, as well as other events with more modest effects. For example, the 13 March 1989 event was a particularly strong case with a minimium Dst of -589 nT that induced currents in the power grid leading to the ultimate collapse of the Hydro-Quebec power system \cite[]{Bolduc:2002a}. There is a great deal of interest in both being able to predict such events in advance, as well as quantifying how strong events could result in wide-spread disruptions. The low frequency of such events is particularly challenging for empirical or machine learning models, which struggle with out of sample predictions. 

Global magnetohydrodynamics (MHD) models for space science applications were first published in the early 1980s \cite[]{LeBoeuf:1981a, Wu:1981a, Brecht:1981a, Brecht:1982a}. Later models applied more advanced algorithms to solve the MHD equations. These models include the Lyon-Fedder-Mobarry (LFM) \cite[]{Lyon:1986a, Lyon:2004a}, the OpenGGCM  \cite[Open Geospace General Circulation Model,][]{Raeder:1996a, Raeder:1995a}, the Watanabe-Sato \cite[]{Watanabe:1990a, Usadi:1993a}, the GUMICS \cite[Grand Unified Magnetosphere Ionosphere Coupling Simulation model,][]{Janhunen:1996a}, and the Integrated Space Weather Prediction Model (ISM) \cite[]{White:1998a, Siscoe:2000a} models of the Earth's magnetosphere. 
The solar codes include models for the solar corona \cite[Magnetohydrodynamics Around a Sphere (MAS),][]{Linker:1994a, Linker:1999a},  \cite[]{Hayashi:2013a}, the heliosphere \cite[]{Usmanov:1993a, Usmanov:2000a}, the inner heliosphere ENLIL \cite[]{Odstrcil:2003a, Odstrcil:2009a}, as well as combined models of the corona and inner heliosphere \cite[Solar–interplanetary adaptive mesh refinement space–time conservation element and solution element MHD model (SIP-AMR-CESE MHD Model),][]{Feng:2014a,Feng:2014b}.  
More general-use models include Ogino's planetary magnetosphere code \cite[]{Ogino:1986a}, Tanaka's 3D global MHD model \cite[]{Washimi:1996a}, Winglee's multifluid Hall MHD code \cite[]{Winglee:1998a, Winglee:2005a}, Toth's general MHD Versatile Advection Code (VAC) \cite[]{Toth:1996b} and its modern version, MPI-AMRVAC \cite[]{Keppens:2021a}, KU Leuven's European heliospheric forecasting information asset \cite[EUHFORIA,][]{Pomoell:2018a} and the University of Michigan's BATS-R-US \cite[]{Powell:1999a, Toth:2012a} model.

\section{The Origins of BATS-R-US \& SWMF}
\label{sec:origin}
Advanced space plasma simulation codes became possible when leading applied mathematicians and computer scientists became integral parts of the teams developing models to solve physical systems. In the early 1990s, two pioneers of high-order \cite{Godunov:1959a} schemes that revolutionized computational fluid dynamics (CFD), Bram van Leer \cite[\cf][]{vanLeer:1973a, vanLeer:1974a, vanLeer:1977a, vanLeer:1977b, vanLeer:1979a} and Philip Roe \cite[\cf][]{Roe:1981a}, became interested in space physics problems.This interest resulted in the extension of modern CFD methods to rarefied magnetized plasma flows and the development of the first modern, high performance MHD code, BATS-R-US \cite[]{Powell:1999a}. \figurename~\ref{fig:batsrus} summarizes the present capabilities of BATS-R-US; the algorithms are discussed in detail in Appendix~\ref{sec:algorithms}.

\begin{figure}[bht]
\centering
\includegraphics[width=1\textwidth]{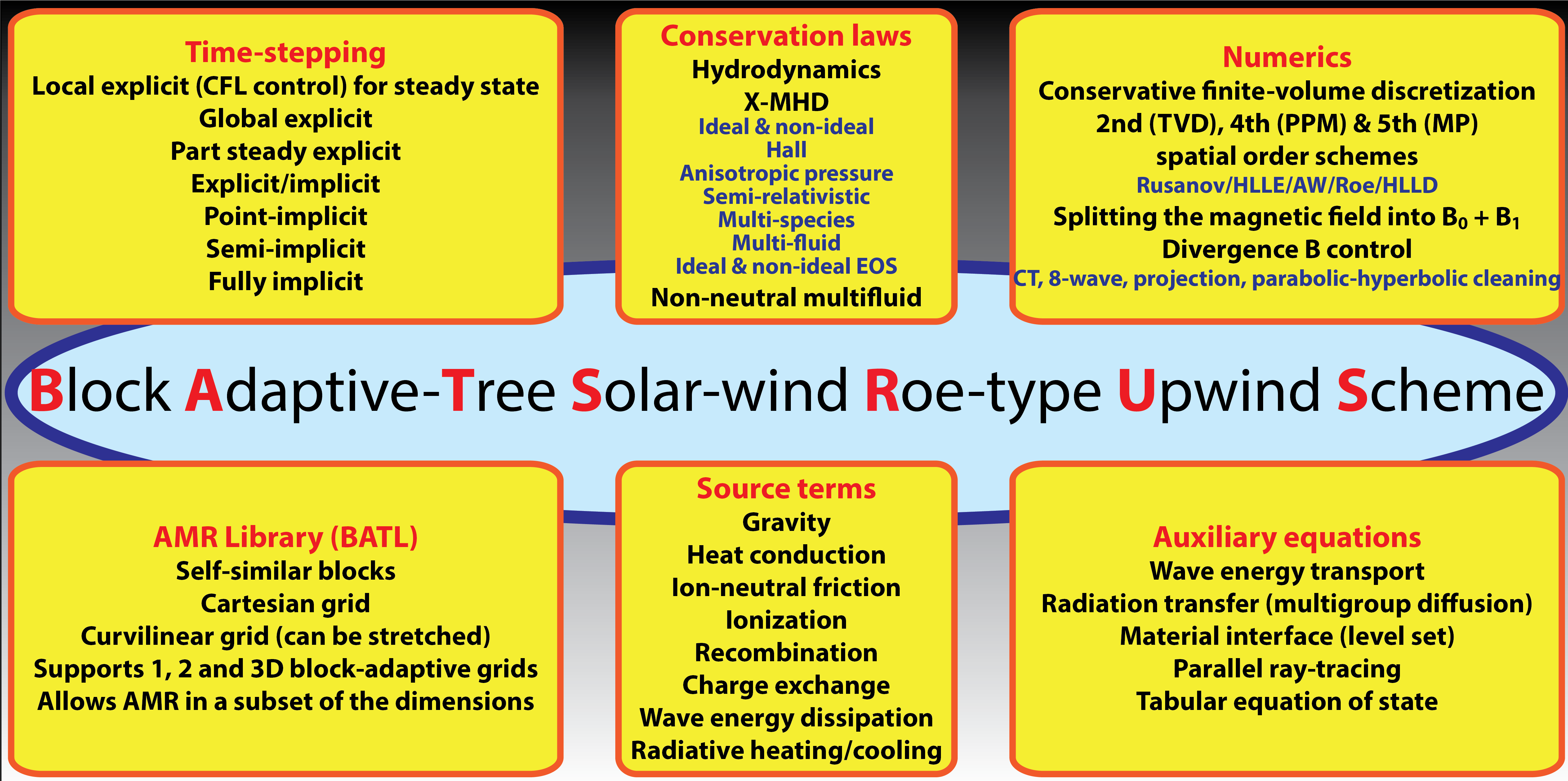}
\caption{Overview of the BATS-R-US multiphysics code.}
\label{fig:batsrus}
\end{figure}

\begin{figure}[tbh]
\centering
\includegraphics[width=1\textwidth]{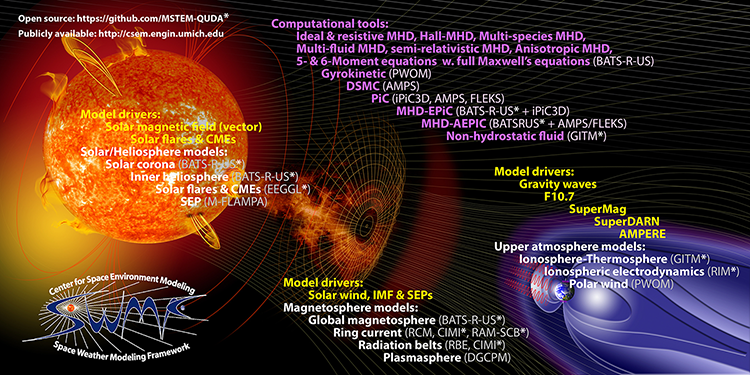}
\caption{Schematic diagram of the Space Weather Modeling Framework. The SWMF and its core models are open source (\url{https://github.com/MSTEM-QUDA}), while the full SWMF is available via registration under a user license (\url{http://csem.engin.umich.edu/tools/smmf}).}
\label{fig:swmf}
\end{figure}

The BATS-R-US \cite[]{Powell:1999a, Toth:2012a} is a versatile, high-performance, generalized magnetohydrodynamic code with adaptive mesh refinement (AMR) that can be configured to solve the governing equations of ideal and resistive MHD \cite[]{Powell:1999a}, semi-relativistic \cite[]{Gombosi:2002a}, anisotropic \cite[]{Meng:2012a}, Hall \cite[]{Toth:2008a}, multispecies \cite[]{Ma:2002a} and multi-fluid \cite[]{Glocer:2009a} extended magnetofluid equations (XMHD) and, most recently, non-neutral multifluid plasmas \cite[]{Huang:2019a}. BATS-R-US is used to model several physics domains (see \figurename~~\ref{fig:swmf}). The efficiency of BATS-R-US is crucial to reach faster than real-time performance with the SWMF while maintaining high resolution in the domains of interest.

In a number of fields in which computer-based modeling of complex, multi-scale,
multi-physics problems plays an important role, \textit{software frameworks} have been developed. In the area of computational space physics there are only two operational software frameworks, the Space Weather Modeling Framework \cite[SWMF,][]{Toth:2005swmf, Toth:2012a} and the Virtual Space Weather Modelling Centre \cite[VSWMC,][]{Poedts:2020a}.
Other frameworks are either under development \cite[]{Zhang:2019b}, abandoned \cite[]{, Luhmann:2004a}, or are rarely used for space weather applications \cite[]{Hill:2004a}.

The SWMF \citep{Toth:2005swmf, Toth:2012a} is a fully functional, documented software that provides a high-performance computational capability to simulate the space-weather environment from the upper solar chromosphere to the Earth's upper atmosphere and/or the outer heliosphere. 
The SWMF tackles the wide range of temporal and spatial scales as well as the different physical processes governing the different heliophysics domains through a modular approach. Each physics domain is covered by a numerical model developed particularly for that purpose. The framework couples several of these components together to execute the simulation in a setup best suited for the problem at hand.

\section{The SWMF Today}
\label{sec:swmftoday}
In 2021, the Space Weather Modeling Framework (SWMF) \cite[]{Toth:2012a} consists of a dozen physics domains and a dozen different models that provide a flexible high-performance computational capability to simulate the space-weather environment from the upper solar chromosphere to the Earth's upper atmosphere and/or the outer heliosphere. It contains over 1 million lines of Fortran 2008 and C++ code, dozens of Perl, Python and Julia scripts, IDL visualization tools and XML descriptions of the input parameters. \figurename~\ref{fig:swmf} summarizes the main features and capabilities of the current SWMF.

The full SWMF suite, developed and maintained at the University of Michigan, has been openly available for a long time via registration under a user license (\url{http://csem.engin.umich.edu/tools/swmf}). Recently, a major part of the SWMF has been released on Github under a noncommercial open-source license (\url{https://github.com/MSTEM-QUDA}). \figurename~\ref{fig:swmf} shows the open source and registration controlled components of the SWMF. 

In addition, SWMF runs can be requested via the Community Coordinated Modeling Center (CCMC) at the NASA Goddard Space Flight Center (\url{https://ccmc.gsfc.nasa.gov/index.php}), where people even with little experience in advanced computer simulations can request specific runs through a user-friendly web interface. The user specifies the domains and the driving input parameters, and the CCMC runs-on-request system carries out the simulation. Once the CCMC completes the run, the output files and standard visualization images are made available through the web interface (\url{https://ccmc.gsfc.nasa.gov/index.php}). 



For space weather related simulations, the SWMF is typically used in two basic configurations: The \alf Wave Solar-atmosphere Model (AWSoM/AWSoM-R) and the SWMF/Geospace Model.

AWSoM/AWSoM-R \cite[]{vanderHolst:2010a, vanderHolst:2014a, Sokolov:2013a, Gombosi:2018a, Sokolov:2021a} describes the solar corona (SC) from the low transition region where the plasma temperature is about $5\times10^4$K and goes out to about $20R_\sSun$. This is the region where the hot, supersonic solar wind is generated. It also simulates the 3D inner heliosphere (IH) out to Neptune's orbit. The outer boundary can be varied depending on the region of interest.

The SWMF/Geospace Model \cite[\cf][]{Haiducek:2017a, Welling:2020a} describes the tightly coupled basic elements of the magnetosphere-ionosphere system: the global magnetosphere (GM), the inner magnetosphere (IM), the ionospheric electrodynamics (IE).
An operational version of the SWMF/Geospace model has been running 24/7 at SWPC since 2016.


\subsection{AWSoM/AWSoM-R Configuration}
\label{subsec:awsomr}
It is commonly accepted that the gradient of the \alf wave pressure is the key driver for solar wind acceleration. Damping of \alf wave turbulence due to reflection from sharp pressure gradients in the solar wind is a critical component of coronal heating. For this reason, many numerical models explore the generation of reflected counter-propagating waves as the underlying cause of the turbulence energy cascade \citep[\eg][]{Cranmer:2010a}, which transports the energy of turbulence from the large-scale motions across the \textit{inertial range} of the turbulence spatial scale to short-wavelength perturbations. The latter can be efficiently damped due to wave-particle interaction. In this way, the turbulence energy is converted to random (thermal) energy \cite[\cf][]{Sokolov:2013a}.

\begin{figure}[tbh]
\centering
\includegraphics[width=1\textwidth]{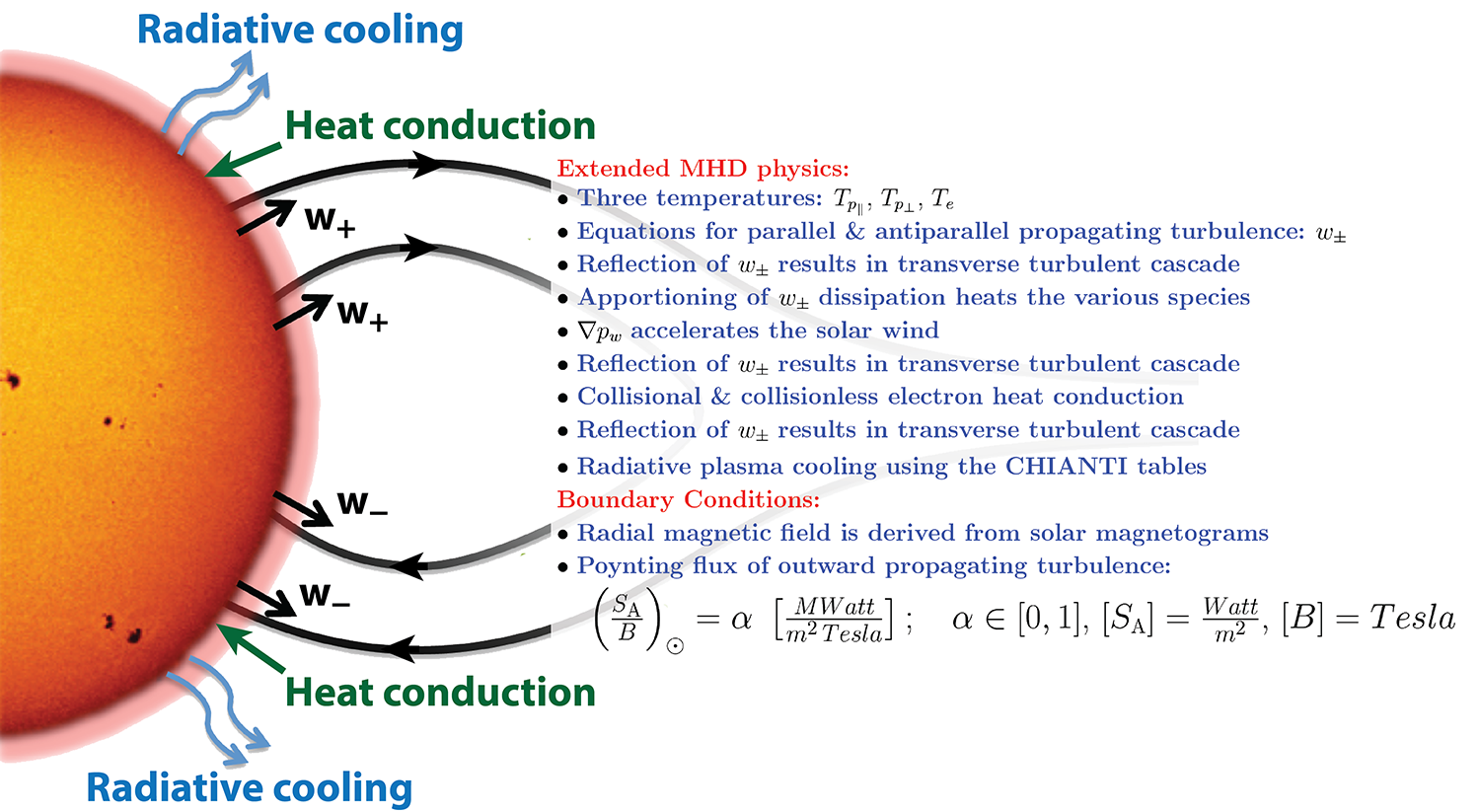}
\caption{Overview of the AWSoM and AWSoM-R physics. They solve XMHD equations with separate ion and electron temperatures. The energy densities of parallel and antiparallel propagating turbulence that are self-consistently coupled to each other and to the plasma are solved together with the XMHD equations. Heat conduction and radiative cooling are also taken into account. The turbulence is powered by the Poynting flux leaving the solar photosphere.}
\label{fig:awsom}
\end{figure}

\subsubsection{\textit{AWSoM}}
\label{subsec:plainawsom}
AWSoM \cite[]{vanderHolst:2014a, Sokolov:2013a, Gombosi:2018a, Sokolov:2021a} is a 3D global solar corona/solar wind model that self-consistently incorporates low-frequency \alf wave turbulence. The \alf waves are represented by the energy density distribution of two discrete populations propagating parallel and antiparallel to the magnetic field at the local \alf speed. The wave energy densities are imposed at the inner boundary with a Poynting flux of the outbound \alf waves assumed to be proportional to the magnetic field strength. In this model, outward propagating waves experience partial reflection on field-aligned \alf speed gradients and the vorticity of the background. In addition, the two populations counter-stream along closed field lines. The nonlinear interaction between oppositely propagating \alf waves results in an energy cascade from the large outer scale through the inertial range to the smaller perpendicular gyroradius scales, where the dissipation takes place. These processes are handled with analytic formulas that provide the resulting ion and electron heating. The solar wind is accelerated by the gradient of the \alf wave pressure. The main physics elements of the AWSoM model are shown in \figurename~\ref{fig:awsom}. 

The boundary conditions for the MHD quantities are obtained from the synoptic or synchronic photospheric magnetograms. The outward propagating Poynting flux at the solar surface ($S_\sA$) is measured in units of $W/m^2$ and it is taken to be proportional to the magnetic field magnitude, $B$ (measured in units of Tesla or T).  The proportionality constant $\alpha$ is measured in units of $MW/m^{2}/Tesla$, and its value varies between 0 and 1. The actual value of $\alpha$ depends on the phase of the solar cycle and on the choice of magnetogram (to account for the calibration differences between magnetograms).

The inner heliosphere (IH) component extends from about $20R_\sSun$ to anywhere between the orbits of the Earth and Neptune. It uses the BATS-R-US and it solves the same equations as the solar corona model, but on a Cartesian grid in either co-rotating or inertial frame. The IH model can propagate interplanetary CMEs (ICMEs) from the Sun to the planets. Adaptive mesh refinement is used to increase the grid resolution along the path of the CME
\cite[\cf][]{Manchester:2017b, Manchester:2014b, Roussev:2004a, vanderHolst:2009a}.

\subsubsection{\textit{Threaded-Field-Line Model and AWSoM-R}}
\label{subsec:tflm}
In the transition region the plasma temperature increases some two orders of magnitude over $\sim\!\!10^2\,$km, resulting in a temperature gradient of $\sim\!\!10^4\,$K/km. To resolve this gradient, 3D numerical simulations require sub-kilometer grid spacing, making these simulations computationally very expensive. AWSoM uses an artificial broadening of the transition region \cite[]{Lionello:2009a, Sokolov:2013a}.

An alternative approach is to reformulate the mathematical problem in the region between the chromosphere and the corona in a way that decreases the computational cost. Instead of solving a computationally expensive 3D problem on a very fine grid, one can reformulate it in terms of a multitude of much simpler 1D problems along \textit{threads} that allows us to map the boundary conditions from the the solar photosphere to the corona. This approach is called the \textit{Threaded-Field-Line Model} (TFLM) \cite[]{Gombosi:2018a, Sokolov:2021a}. 

The physics behind the reformulated problem is the assumption that between the solar surface and the top of the transition region ($R_\sSun \le r \le R_{b}$) the magnetic field is potential and varies slowly in time. Each \textit{thread} represents a field line and one can solve a 1D problem that describes evolution of the plasma in a magnetic flux tube around a given thread. The algorithm uses an implicit scheme to allow for large time steps. Using the TFLM methodology results in a significant speedup for time-dependent simulations. The AWSoM model with TFLM inner boundary conditions is called AWSoM-R, where the letter ``R'' implies that this version can run faster than real time on $\sim200$ cores at a moderate grid resolution (about 2$^\circ$ near the Sun).

\subsection{SWMF/Geospace Configuration}
\label{subsec:awsom}
While the BATS-R-US can model many of the dynamical plasma processes in the solar wind and magnetosphere, it is widely accepted that MHD alone cannot sufficiently describe the coupled solar wind -- magnetosphere -- ionosphere system. The ionosphere and space close to the Earth is not suited for MHD, and is beyond the numerical capabilities due to the high magnetic field intensity, which increases the wave speeds, thus requiring very small time steps and high spatial resolution. Furthermore, the inner magnetosphere ring current, which is an integral part of the storm dynamics, cannot be described by a temperature of a Maxwellian plasma population, which calls for separate treatment of the dynamics in the quasi-dipolar region. To that end, the SWMF/Geospace couples three different models describing these three domains. Furthermore, additional models can be coupled to tackle multiple plasma populations, kinetic physics, or other phenomena and processes (see Section \ref{sec:swapplications}.)

\begin{figure}[tbh]
    \floatbox[{\capbeside
        \thisfloatsetup{capbesideposition={left,top},
        capbesidewidth=0.45\textwidth}}]{figure}[\FBwidth]
    {\caption{Illustration of the models (components within SWMF) and couplings in the SWMF/Geospace configuration. Arrows denote the information that is passed between the components \cite[adapted from][]{Haiducek:2017a}.
    \label{fig:geospace}}}
    {\includegraphics[width=0.4\textwidth]{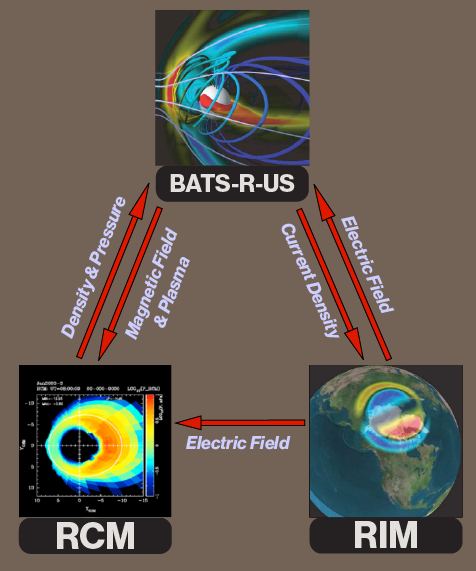}}
\end{figure}

The base SWMF/Geospace configuration is illustrated in \figurename~\ref{fig:geospace}. Under this setup, the global magnetosphere model BATS-R-US is coupled to the Ridley ionosphere electrodynamics model (RIM) \cite[]{Ridley:2004a} and the inner magnetosphere Rice Convection Model (RCM) \cite[]{Harel:1981a}. BATS-R-US supplies near-body field-aligned currents (FACs) to the RIM, which, using an empirical specification of conductance \cite[]{Ridley:2004a, Mukhopadhyay:2020a}, solves for the electric potential.  This electric potential is returned to BATS-R-US to set the plasma tangential velocity at the inner boundary. The RCM receives its initial and boundary field and plasma conditions from BATS-R-US as well as electric field from RIM. It returns total plasma pressure and density to BATS-R-US inside the closed field line region, significantly improving the inner magnetosphere results of the MHD solution \cite[]{DeZeeuw:2004a}, especially during geomagnetic storm times \cite[]{Liemohn:2018a}. 
In addition, the current configuration can include the Radiation Belt Environment (RBE) model \cite[]{Fok:2008a} that receives information from BATS-R-US and RIM and solves for the energetic electron population in the radiation belts.

The couplings default to 5-second (GM-IE) and 10-second (all other) frequency; faster coupling frequencies are required under extreme driving or when high-frequency output is produced \cite[]{Welling:2020a}. While the \emph{explicit} couplings are shown, the self-consistent nature of multi-model SWMF simulations produces \emph{implicit} couplings.  For example, while region-2 Birkeland currents are not explicitly passed from IM to GM physics modules, the improved pressure gradients in BATS-R-US due to the pressure coupling from RCM drives region-2 Birkeland currents \cite[]{Welling:2018a}. Under this model configuration, only upstream solar wind and IMF conditions, as well as F10.7 solar radio flux, are needed as inputs to the model.

The Geospace model is initialized by iterating GM and IE toward an approximate steady state solution  using the initial solar wind, IMF and F10.7 values for boundary conditions. Using a local time stepping mode, this is done very efficiently. Next the IM component is switched on and the Geospace model is run in time-dependent mode using the time varying boundary conditions. It takes about 5 hours for the ring current to build up to a realistic strength. After this point the model can be used for simulation and prediction. In operational use, the Geospace model is run continuously. The model is only reinitialized from scratch if there is a long (an hour or more) gap in the solar wind observations.

\begin{figure}[htb]
    \floatbox[{\capbeside
        \thisfloatsetup{capbesideposition={left,top},
        capbesidewidth=0.30\textwidth}}]{figure}[\FBwidth]
    {\caption{Grid configurations for BATS-R-US within the SWMF Geospace.  The left and right hand panels
    illustrate the grid configuration of the operational Geospace model versions 1 and 2, respectively  \cite[from][]{Haiducek:2017a}.
    \label{fig:mhdgrid}}}
    {\includegraphics[width=0.675\textwidth]{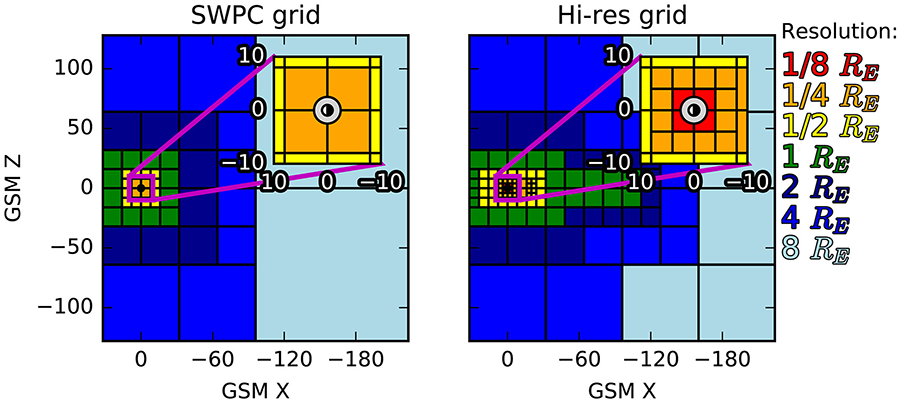}}
\end{figure}

In addition to the physics models and couplings, spatial resolution of the included models strongly affects the simulation results. RIM defaults to $2^{\circ}\times2^{\circ}$ grid spacing in geomagnetic longitude and latitude. BATS-R-US has no default grid, but the base SWMF/Geospace configurations are illustrated in \figurename~\ref{fig:mhdgrid} for version 1 and the more recent version 2. These configurations result in $\sim\!\!1$ million grid cells with a near-body resolution of $1/4\,R_\sE$ and $\sim\!\!2$ million grid cells with $1/8\,R_E$ maximum resolution, respectively.

While capable of running faster than real time on a modest number (about 100) of CPU cores, the operational SWMF/Geospace models can well reproduce large-scale features such as cross polar cap potential (CPCP) and Dst \citep{Haiducek:2017a, Mukhopadhyay:2021a}, and can also predict local ground magnetic perturbations with skill scores of practical value \cite[]{Pulkkinen:2013a, Toth:2014a}.

\subsubsection{\textit{Virtual Magnetic Observatories}}
\label{subsubsec:virtual}
The coupled-model approach of SWMF/Geospace allows for the production of virtual observatory simulations during code execution. The most widely used of these are virtual magnetometers. We use Biot-Savart integrals to find the total surface magnetic perturbation at an arbitrary point about the globe due to the simulated magnetospheric and ionospheric current systems \cite[]{Yu:2008a, Yu:2010a, Welling:2019a}. For a detailed description of the methodology see Appendix~\ref{subsec:groundmf}. While tools exist to create such outputs as part of post-processing \cite[]{Rasteatter:2014a}, the SWMF/Geospace combines information from the IE and GM models on-the-fly to provide continuous output during the simulation. A recently developed mathematical reformulation of the problem replacing the volume integrals with surface integrals speeds up the calculation by an order of magnitude (see Appendix~\ref{subsec:groundmf}). 

In a similar fashion, advanced virtual satellite observations are created by mapping kinetic distributions from the IM and optional RB modules along self-consistent global magnetic field lines obtained from GM. The net result is the ability to extract ring current and radiation belt flux distributions at arbitrary points about the inner magnetosphere. Virtual satellites have also been used to assess the simulation results through comparisons with in-situ spacecraft observations \cite[\cf][]{Welling:2012a, Glocer:2013a}.

\subsubsection{\textit{Operational Use at NOAA/SWPC and the CCMC}}
\label{subsubsec:swpc}

In 2015, NOAA's Space Weather Prediction Center (NOAA/SWPC) decided to transition a research model to operational space weather prediction. As part of this effort, a systematic study was undertaken to evaluate the performance of various physics-based and empirical models to predict ground magnetic perturbations \cite[]{Pulkkinen:2013a, Glocer:2016a}. The physics-based SWMF/Geospace model in particular was found to systematically be a top performing model using the selected metrics. That code has since been used for routine space weather prediction at NOAA/SWPC and at the Community Coordinated Modeling Center (CCMC) located at NASA GSFC. The operational codes run in the configuration illustrated in \figurename~\ref{fig:geospace}. In 2020, the NOAA/SWPC  upgraded to version 2 of the SWMF/Geospace model, which has a higher grid resolution near the Earth and better ionospheric conductance. 

\section{Growing Number of Space Weather Applications}
\label{sec:swapplications}
Space weather simulations using the SWMF have been carried out in multiple configurations and contexts, demonstrating that SWMF and its components are able to successfully simulate global-scale, meso-scale and micro-scale processes in a self-consistent manner, and integrate these processes to form a truly multi-scale space weather simulation capability. In addition, significant validation efforts have been made by a variety of comparisons with both in-situ and remote-sensing observations.

\subsection{Ambient Solar Wind}
\label{subsec:quietsw}
CMEs and ICMEs do not propagate and evolve in vacuum. They travel through the ambient interplanetary medium and interact with its plasma and magnetic field. Therefore, in order to simulate real space weather events, it is critical to have a validated ambient corona/solar wind model in which the CME/ICME will propagate and cause significant distortions. These distortions can include plasma pileup, shock fronts, magnetic field line distortion and many other phenomena \cite[\cf][]{Manchester:2004a, Manchester:2005a, Manchester:2008a, Manchester:2012a, Manchester:2014b}. The situation can be even more complicated when several CMEs are generated in rapid succession \cite[\cf][]{Lugaz:2005b, Lugaz:2008a, Lugaz:2009a}.

\begin{figure}[tbh]
\centering
\includegraphics[width=1\textwidth]{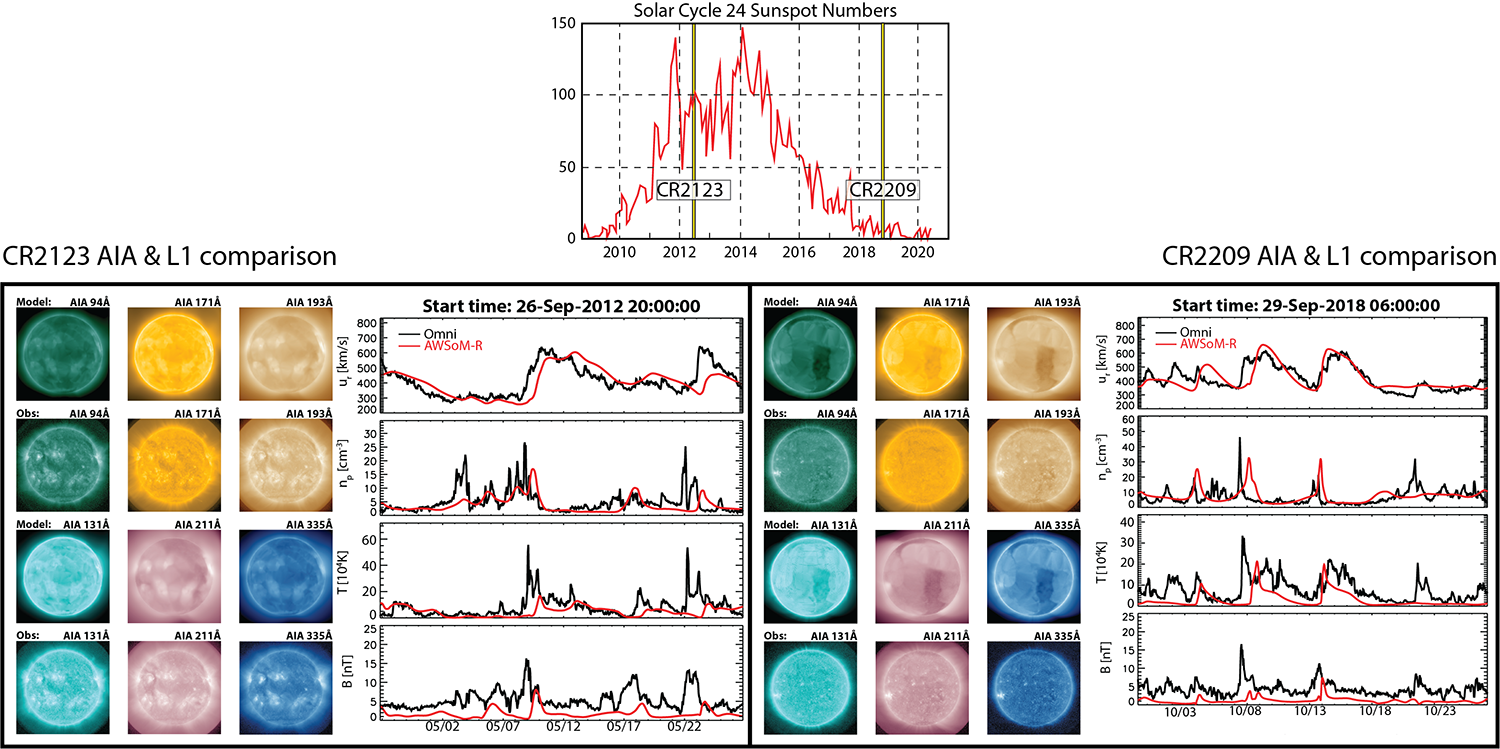}
\caption{Background corona and solar wind solutions with the AWSoM-R model for solar minimum and maximum conditions. The background solar wind is driven by an outward going photospheric turbulent energy flux per unit magnetic field of $1 \text{MJ}\, \text{m}^{-2}\, \text{s}^{-1}\, \text{Tesla}^{-1}$ (CR2209) and by $0.45 \text{MJ}\, \text{m}^{-2}\, \text{s}^{-1}\, \text{Tesla}^{-1}$ (CR2123).}
\label{fig:backgroundwind}
\end{figure}

\cite{Sachdeva:2019a} performed a detailed validation study of the AWSoM for the quiet-time solar wind for Carrington Rotations (CR) representative of the solar minimum conditions (CR2208 and CR2209). They compared simulation results with a comprehensive suite of observations extending from the solar corona to the heliosphere up to Earth's orbit. In the low corona ($r < 1.25 R_\sSun$), extreme ultraviolet (EUV) images from both the STEREO-A (Solar TErrestrial RElations Observatory Ahead) EUVI (extreme ultraviolet imaging) instrument and the SDO (Solar Dynamics Observatory) AIA (atmospheric imaging assembly) were compared to 3D tomographic reconstructions of the simulated electron temperature and density. Model results were also compared to tomographic reconstructions of the electron density from the SOHO (Solar and Heliospheric Observatory) LASCO (Large Angle and Spectrometric Coronagraph) observations in the $2.55 R_\sSun < r < 6.0 R_\sSun$ region. In the heliosphere, model predictions of solar wind speed were compared to velocity reconstructions from interplanetary scintillation observations. Simulation results at the first Lagrange point between the Sun and Earth (L1) were compared to OMNI data. The results of \cite{Sachdeva:2019a} show that the AWSoM performs well in quantitative agreement with the observations between the inner corona and 1 AU.

Recently AWSoM/AWSoM-R was also validated for solar maximum conditions. Using $S_\sSun/B_\sSun$ ($S_\sSun$ is the Poynting flux of outward propagating \alf waves at the solar surface) as an adjustable parameter, good agreement was found for CR2123 that characterizes solar maximum conditions for solar cycle 24 (see \figurename~\ref{fig:backgroundwind}). 
\figurename~\ref{fig:backgroundwind} shows the comparisons of AWSoM-R simulation results for CR2123 and CR2209 with AIA images and solar wind parameters at 1 AU. For both the rotations the AIA comparisons include six wavelengths (94, 171, 193, 131, 211 and 335 {\AA}). The L1 parameters include radial speed (U$_{r}$ in km/s), proton number density ($N_{p}$ in cm$^{-3}$) and temperature (T in K) and magnetic field magnitude (B in nT). Inspection of \figurename~\ref{fig:backgroundwind} reveals that we are able to match the observed slow/fast solar wind structure at 1 AU and, simultaneously, reproduce a number of optically thin coronal spectral observations. For AWSoM model results of CR2209 the reader is referred to \cite{Sachdeva:2019a}.
\subsection{CME Generation}
\label{subsec:cme}
The Eruptive Event (EE) generator algorithm of the SWMF is responsible for creating the initial conditions within the corona, which produces a CME eruption. This be done by inserting an unstable (or force imbalanced) flux rope into the steady solar corona solution, or inserting an arcade and applying shearing motion at the lower boundary of the corona model \cite[]{Antiochos:1999a, vanderHolst:2009a}. This approach offers a relatively simple, and inexpensive model for CME initiation based on empirical understanding of pre-event conditions.  We also have a SWMF component (EE), which is a physics-based extended MHD model (BATS-R-US) of the convection zone \cite[]{Fang:2012a, Fang:2012b}, where the domain is a localized wedge extending 30 Mm below the photosphere and hundreds of Mm into the corona. The wedge extends hundreds of Mm at the photosphere, sufficient to contain a large active region.  The model includes optically thin radiative loss terms appropriate for the corona and empirical cooling terms to approximate optically thick radiative transfer near the photosphere, which drives cellular convection \cite[]{Abbett:2003a, Abbett:2004a}.  In the environment, a CME may be initiated by the emergence of a flux rope from the convection zone \cite[]{Manchester:2004c}. Currently, the physics-based EE model only works in a stand-alone mode \cite[]{Fang:2012a, Fang:2012b}, and we use empirical models to generate CMEs in the SWMF \cite[]{Jin:2017b, Borovikov:2017b}.

\begin{figure}[tbh]
\centering
\includegraphics[width=0.85\textwidth]{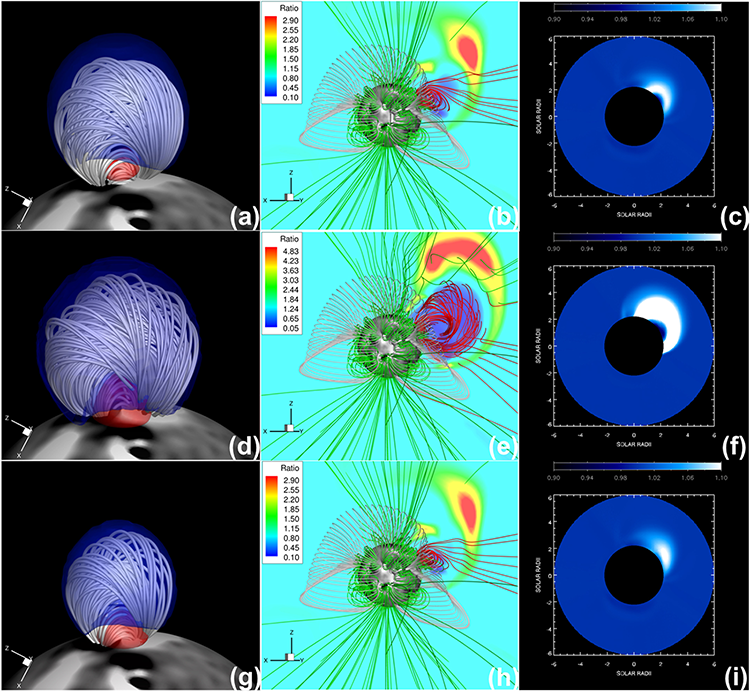}
{\caption{Three examples of \cite{Gibson:1998a} flux ropes with different size and magnetic strength parameters. Panels (a)-(f) and (g)-(i) show, respectively, flux ropes specified with radii of 0.8 and 0.6 $R_s$. Strength parameters are set to 0.6 for model run (a)-(c) and 2.25 for (d)-(i). The left column shows the initial configuration of the flux ropes with blue and red isosurfaces showing, respectively, the ratios of 0.3 and 2.5 of the mass density of the CME model divided by that of the pre-event corona. The middle column shows the resulting CME evolution at $t = 20$ minutes. Here, magnetic field lines are colored red, gray-shaded and green to illustrate the flux rope, large-scale helmet streamers, and magnetic fields surrounding active regions and open flux. Color contour images show the ratio of the mass density of the CME divided by that of the pre-event corona. The right column shows model-produced SOHO/LASCO white light images, where the total brightness is normalized by dividing by that of the pre-event background solar wind. \cite[from][]{Jin:2017b}}
    \label{fig:eeggl}}
\end{figure}

Magnetically-driven CMEs were first modeled with the SWMF suite in the early 2000s. First, the distorted spheromac-type \cite{Gibson:1998a} (GL) unstable flux-rope model was implemented \cite[]{Manchester:2004a, Manchester:2014a, Manchester:2014b, Lugaz:2005a, Lugaz:2005b}. Later, the \cite{Titov:1999a} (TD) twisted eruptive flux rope model was also added to the SWMF tool box as a CME initiation option \cite[]{Roussev:2003b, Roussev:2006a, Roussev:2007a}. The TD eruption model was used in the first physics-based Sun-to-Earth space weather simulation of two consecutive CMEs during the 2003 Halloween event \cite[]{Toth:2007a, Manchester:2008a} showing quantitative agreement with several observations including in-situ observations at 1 AU and coronagraph images from LASCO C2 and C3. An automated tool, the Eruptive Event Generator using Gibson-Low configuration (EEGGL) was developed \cite[]{Jin:2017b, Borovikov:2017b} and added to the SWMF suite to make CME simulations more widely available to the heliophysics community. In 2016, EEGGL was made available interactively through the CCMC's runs-on-request service to provide CME simulations.

Representative results from EEGGL-driven CME simulations are shown in \figurename~\ref{fig:eeggl} \cite[]{Jin:2017b} using a combination of  two flux rope sizes and two magnetic field strength parameters. The left panel shows the initial configuration of the flux ropes with two density isosurfaces. 
The middle panel depicts the resulting CME evolution at 20 minutes. The background color shows the density ratio between the CME solution steady background solar wind. The right panel shows the synthesized (model-derived) SOHO/LASCO white light images. The color scale shows the white light total brightness divided by that of the pre-event background solar wind. 
Comparing panels (a) and (d), we can see that with a higher magnetic field strength parameter, more plasma is added at the bottom of the flux rope (red isosurface). 
The second and third cases have the same magnetic field strength parameter but with different flux rope sizes. In this case, we can see the flux rope is considerably smaller at the beginning. With this smaller flux rope, the resulting CME speed is reduced and the morphology of CME in the synthesized white light image is quite different with narrower CME width angle. 

\subsection{ICME Simulation}
\label{subsec:icme}
\begin{figure}[ht]
    \floatbox[{\capbeside
        \thisfloatsetup{capbesideposition={left,top},
        capbesidewidth=0.35\textwidth}}]{figure}[\FBwidth]
    {\caption{
     CME-driven EUV waves in the simulation (left) and in the corresponding SDO/AIA observation (right). Both the simulation and observation images are produced by a tri-ratio running difference method. The tricolor channels are AIA 211 {\AA} (red), AIA 193 {\AA} (green), and AIA 171 {\AA} (blue). The ratio in each channel is identically scaled to $1 \pm 0.2$ for both observation and simulation.
    \cite[from][]{Jin:2017a}}
    \label{fig:Jin17_1}}
    {\includegraphics[width=0.625\textwidth]{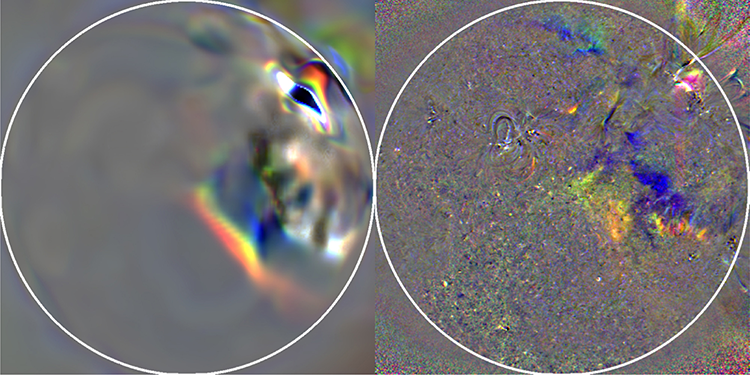}}
\end{figure}
The evolution of CMEs in the solar corona and interplanetary medium has been extensively simulated with the SWMF \cite[]{Manchester:2017b, Manchester:2014b, Manchester:2004a, Manchester:2012a, Manchester:2005a, Manchester:2008a, Manchester:2014a, Roussev:2004a, Roussev:2008a, vanderHolst:2009a, vanderHolst:2007a}. 
Current models (since 2014) start from the upper chromosphere with fixed temperature $T=5\times10^4$K and density $n=2\times10^{17} m^{-3}$. The \alf wave turbulence is launched at the inner boundary, with the Poynting flux scaling with the surface magnetic field. The electron and proton temperatures are solved separately. The smallest radial cell size is $\sim10^{-3}R_\sSun$ near the Sun to resolve the steep density and temperature gradients in the upper chromosphere. The initial condition for the radial magnetic field at the inner boundary is provided by synoptic/synchronic maps of the photospheric magnetic field using the Potential Field Source Surface (PFSS) model.

\begin{figure}[htb]
    \floatbox[{\capbeside
        \thisfloatsetup{capbesideposition={left,top},
        capbesidewidth=0.35\textwidth}}]{figure}[\FBwidth]
    {\caption{
     Comparison showing a general agreement between the white-light observations from SOHO LASCO C2 (top left) and STEREO-B COR1 (top right) and the respective synthesized white-light images from the simulation (bottom). The color contours show the relative total brightness changes compared to the pre-event background level.
    \cite[from][]{Jin:2018a}}
    \label{fig:Jin18}}
    {\includegraphics[width=0.625\textwidth]{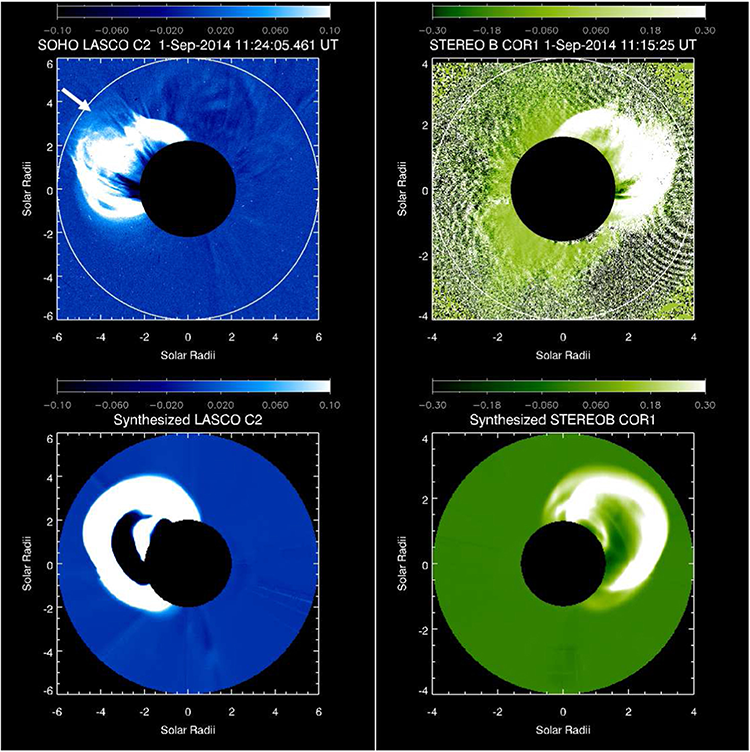}}
\end{figure}

The inclusion of the lower corona in our model allows us to produce synthesized extreme ultraviolet (EUV) images, which are then compared with the EUV observations from SDO/AIA \cite[]{Lemen:2012a} and STEREO/EUVI \cite[]{Howard:2008a}.  \figurename~\ref{fig:Jin17_1} shows an example of model results compared with observations of the 7 March 2011 CME event, which demonstrates enhanced emission from regions of the lower atmosphere compressed and heated by CME-driven shocks and compressional waves. 

In addition to EUV images, our model also allows us to make synthetic Thomson-scattered white light images of the corona. \figurename~\ref{fig:Jin18} shows a comparison between the observed white light images and the model synthesized images for the 7 March 2011 CME event \cite[]{Jin:2017a}. The synthesized running-difference images are able to reproduce the observed typical three-part CME structure comprising the bright core that represents the filament material, the dark cavity that corresponds to the flux rope, and the bright front that is due to the mass pile-up in front of the flux rope \cite[]{Illing:1985a}. Moreover, the model is also able to resolve the observed second faint front that is the outermost part of the increased intensity region associated with the CME-driven shock, as was first quantitatively demonstrated in  \cite{Manchester:2008a}. The white light comparison from three points of view confirms that the simulated CME propagates in the observed direction. The model results in \figurename~\ref{fig:Jin17_1} and \figurename~\ref{fig:Jin18} are produced by running the AWSoM with the magnetic field specified by GONG synoptic magnetograms for CR2107 and synchronous magnetograms for the month of September 2014, respectively.

\begin{figure}[ht]
    \floatbox[{\capbeside
        \thisfloatsetup{capbesideposition={left,top},
        capbesidewidth=0.35\textwidth}}]{figure}[\FBwidth]
    {\caption{1 AU results of the EEGGL simulation of the 12 July 2012 CME event simulated at CCMC. Shown are the simulated and observed plasma quantities plotted with dashed and solid lines, respectively. From top to bottom are the magnetic field component $B_x, B_y$ and $B_z$, the mass density, and the Earth-directed velocity $V_x$. Simulation results are shifted 10 hours to match the shock arrival.  We find good agreement, with the exception of the $B_x - B_y$ rotation and the excessive trailing velocity.}
    \label{fig:Chip12}}
    {\includegraphics[width=0.625\textwidth]{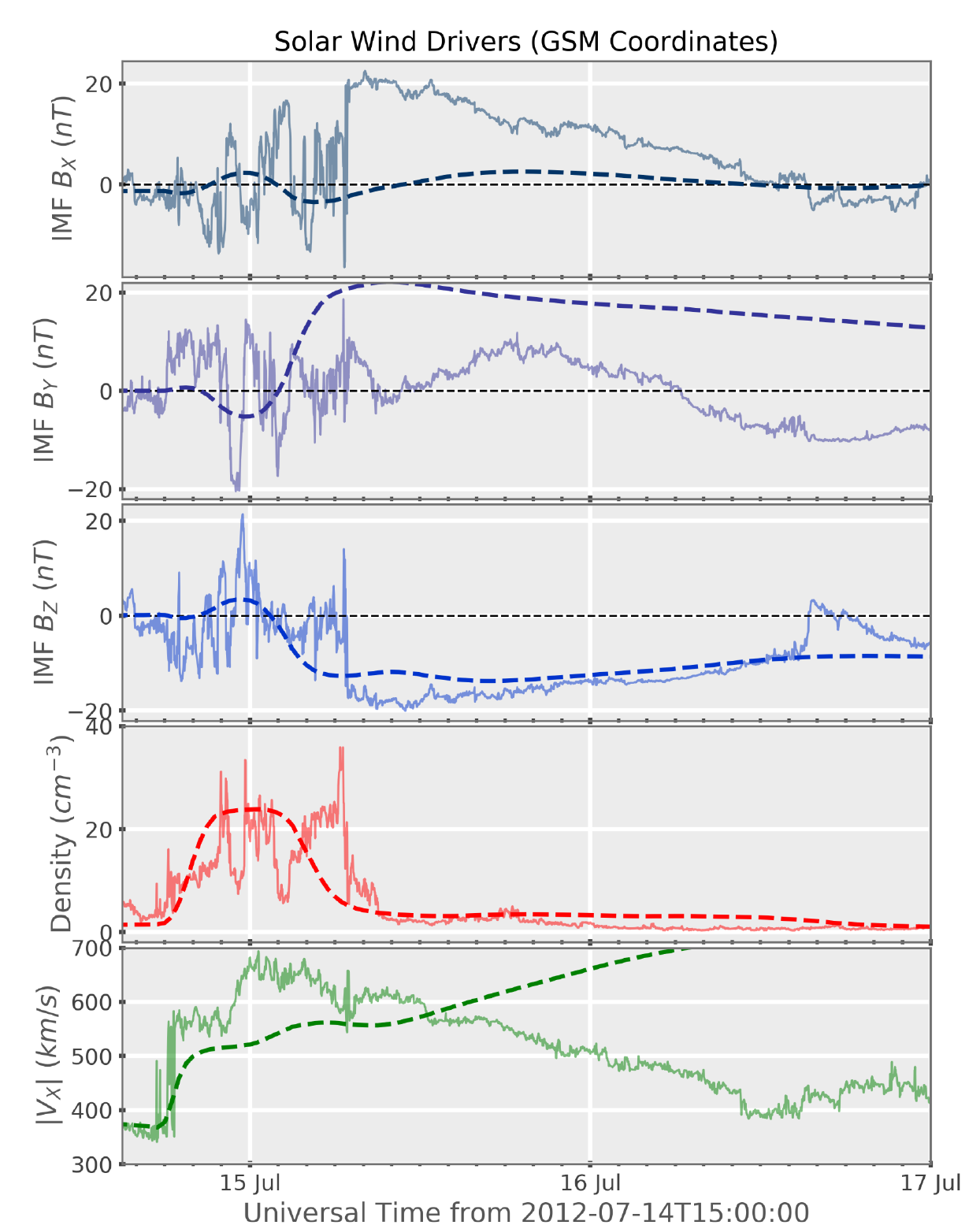}}
\end{figure}

EEGGL was designed to provide data-drive CME simulations that are capable of reproducing the solar wind disturbances at 1 AU that generate geomagnetic storms.  To achieve this goal, the model must capture the bulk plasma properties, in particular the plasma velocity, mass density and magnetic field. An example of this capability is shown in Figure \ref{fig:Chip12}, where we show the simulated (shown with dashed lines) and L1-observed plasma conditions (shown with solid lines) resulting from the Earth-directed CME that occurred on 12 July 2012. Here, time-series data are shown (top to bottom) for the Cartesian components of the magnetic field, mass density and Earth-directed velocity. We shift the simulated time by roughly 10 hours to provide a better comparison with observations.  We find that the magnetic $x$ and $y$ components appear to be miss-matched while the $z$ component very well matches the observed magnitude and time profile of the observations.  The velocity roughly matches the increase from the ambient background to the shocked value found in the sheath region, but then increases above observed values in the relaxation region. The model delivers mass density, early velocity and storm-driving $B_z$, which allows the model to successfully drive a magnetospheric simulation, while issues with flux rope rotation and stream-interaction remain to be addressed. This EEGGL-driven simulation was performed on demand at the CCMC where the model outputs are available to the public.

\subsection{Solar Energetic Particle Simulations}
\label{subsec:sep}
The acceleration of energetic particles in a CME-driven shock and the subsequent transport processes are modeled using the M-FLAMPA module in SWMF \cite[]{Sokolov:2004a,Borovikov:2018a}. The distribution function of energetic particles are solved on a multitude of extracted magnetic field lines advecting with the background plasma (Lagrangian grids) \cite[]{Sokolov:2004a}. M-FLAMPA is fully coupled with the solar corona (SC), inner heliosphere (IH), and the outer heliosphere (OH) components. The plasma and turbulence parameters along the magnetic field lines are extracted dynamically from the the BATS-R-US simulations.

\begin{figure}
\centering
\begin{subfigure}
  \centering
  \includegraphics[width=.45\linewidth]{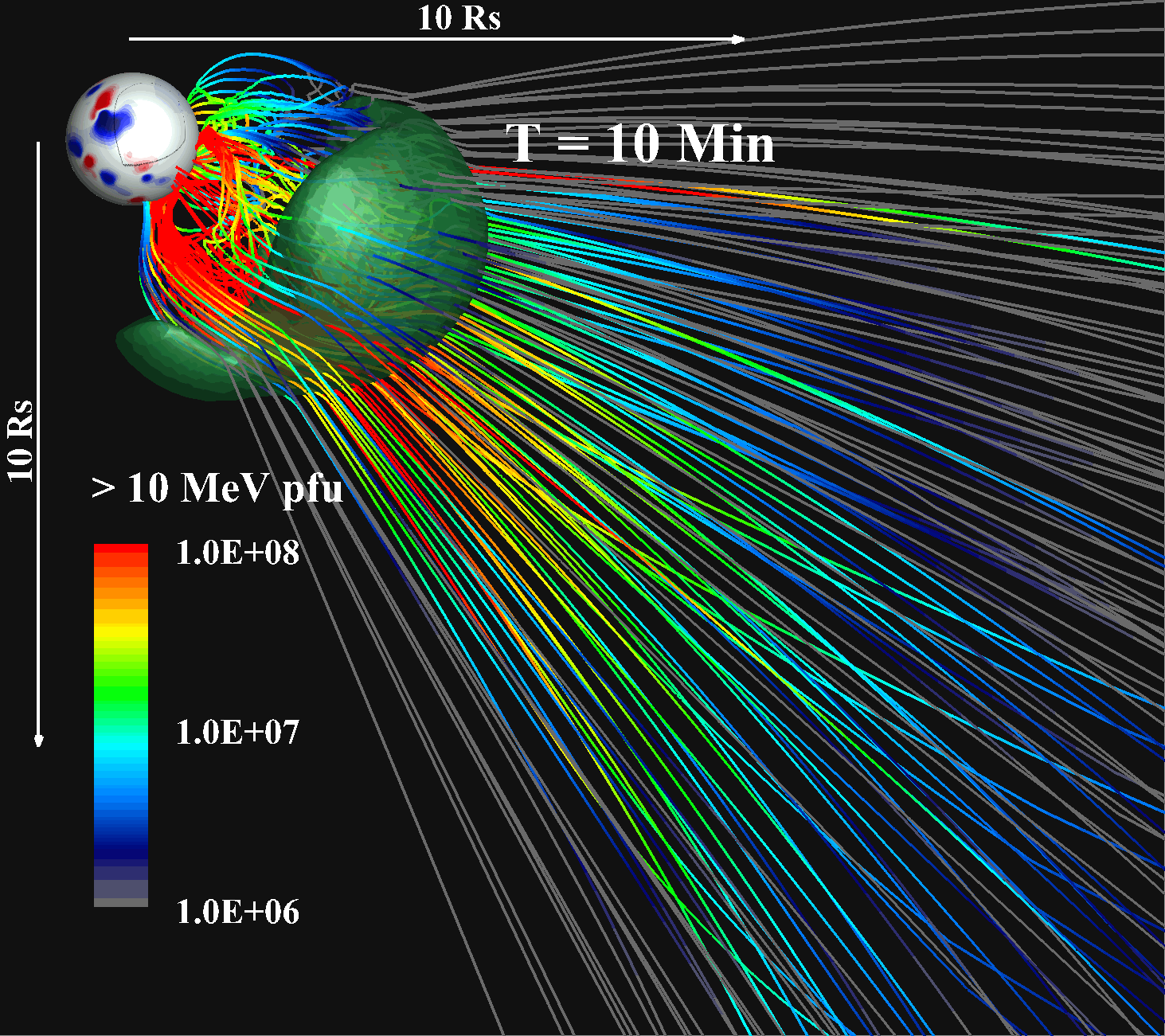}
\end{subfigure}%
\begin{subfigure}
  \centering
  \includegraphics[width=.45\linewidth]{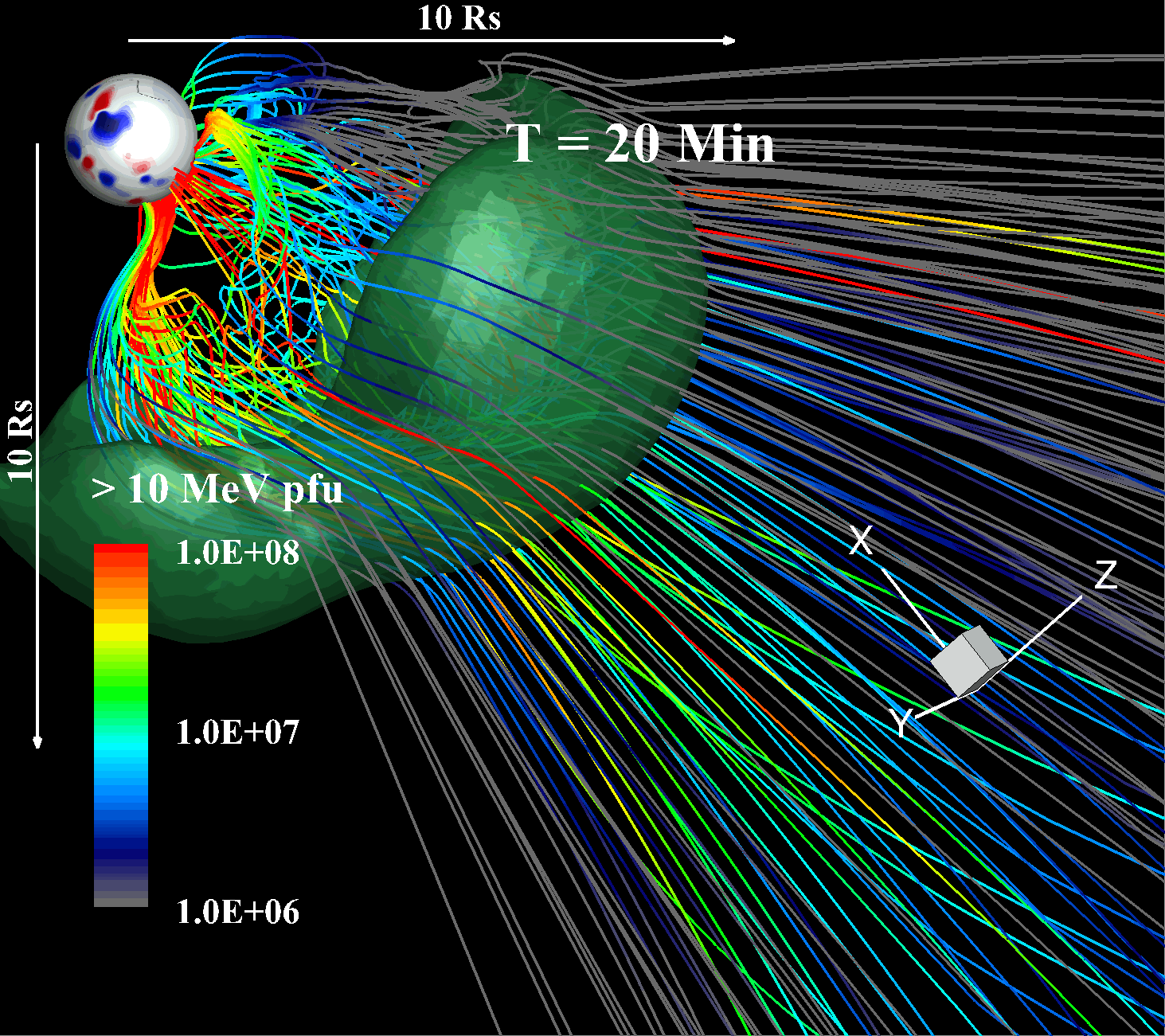}
\end{subfigure}
\caption{Distribution of the energetic particles ($>$ 10 MeV) along the extracted magnetic field lines at 10 min (left panel) and 20 min (right panel) after the eruption of CME. The flux is in the unit of particle flux unit (pfu, particles/cm$^2$/s/sr). The green isosurface represents the leading edge of the CME.}
\label{fig:mflampa}
\end{figure}

\figurename~\ref{fig:mflampa} shows the application of M-FLAMPA to model the acceleration and transport processes of energetic particles in an SEP event that occurred on 23 January 2012 \cite[]{Borovikov:2018a}.
The ambient solar corona and interplanetary steady-state solar wind background are obtained as discussed in \sectionname~\ref{subsec:quietsw} and the CME, which is the source of this SEP event, is simulated by inserting a flux-rope into the active region on the Sun using the EEGGL model (see \sectionname~\ref{subsec:cme}). 
In \figurename~\ref{fig:mflampa}, the green isosurface represents the leading edge of the CME. Hundreds of magnetic field lines whose footpoints on the solar surface are close to the active region are extracted using the coupled AWSoM-R, EEGGL, and M-FLAMPA modules.
Left and right panels are at 10 min and 20 min after the CME eruption, respectively. 
The colors on the magnetic field lines represent the flux, in the unit of particle flux unit (pfu, particles/cm$^2$/s/sr) of the energetic protons, whose energies are greater than 10 MeV.
Along single field lines, the proton flux is larger in the region close to the CME, where the acceleration takes place.
And the flux decreases away from the CME when the accelerated protons stream into interplanetary space.
The proton's flux is higher at the center of the CME than at the flank, indicating a stronger acceleration at the center where the compression is larger. 

\figurename~\ref{fig:mflampa} demonstrates the capability of using the self-consistent physics-based modules in SWMF to calculate the flux of the energetic particles at any location in the heliosphere, showing it to be a powerful tool to study the acceleration and transport processes of SEP events.

\subsection{Rigidity Cutoff Simulations}
\label{subsec:gcr}

Overall, the Earth's radiation environment is very dynamic. Such fluxes of the energetic ions (above 1 MeV per nucleon)  can be enhanced by several orders of magnitude during SEP events, which can last from a few hours to a week \cite[]{Baker:2008a}. SEPs are energetic particles ejected by the Sun in events that are correlated with coronal mass ejections (CMEs) and solar flares \cite[]{Reames:1999a}. The occurrence of SEPs is in positive correlation with ongoing solar activity.

\begin{figure}
\includegraphics[trim=0mm 0mm 0mm 0mm, clip,width=0.9\textwidth]{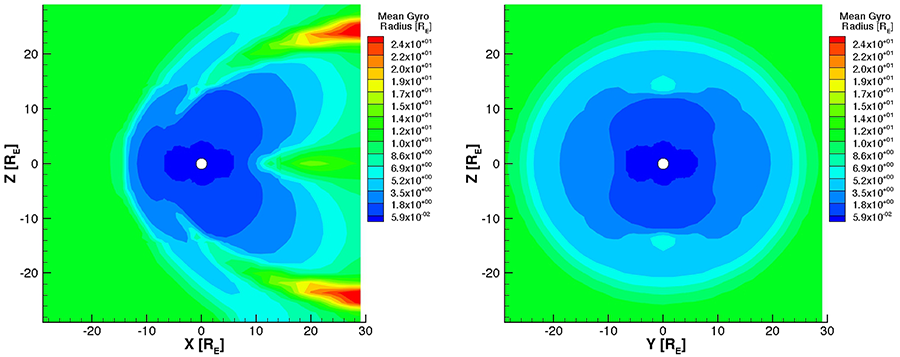}
\caption{
     Example of gyroradii of particles with 1 MeV   $ < E < 16$ MeV during quiet geomagnetic conditions. {\it Left:} Gyroradius map in the equatorial plane. {\it Right:} Gyroradius map in the meridional plane (X=0). The gyroradii of these particles can be as large as tens of $R_\sE$. Here, X-axis is directed toward the Sun, and Y-axis is in the equatorial plane, and Z-axis is such that the frame of reference is right-handed. The free-space energy spectrum of the simulated energetic particles is taken from \cite{Badavi:2011a}.}
    \label{gyro-radius-example-figure}
\end{figure}


The most stable component of the Earth's radiation environment, galactic cosmic rays (GCRs), varies by an order of magnitude at energies below a few hundred MeV per nucleon due to heliospheric modulation \cite[\cf][]{Vainio:2008a}.
Variability of GCRs observed in the Earth's magnetosphere is due to a combined effect of the IMF in the heliosphere and the geomagnetic field inside the magnetosphere on the GCR transport. 

\begin{figure}[thb]
    \floatbox[{\capbeside
        \thisfloatsetup{capbesideposition={left,top},
        capbesidewidth=0.45\textwidth}}]{figure}[\FBwidth]
    {\caption{
    Example of the calculated density of energetic protons with energies 1 MeV $<$ E $<$ 100 MeV) in geospace. Both the SEP's energy spectrum and geomagnetic parameters are taken for quiet conditions. The figure demonstrates that the topology of the SEPs population in the geospace is affected by the particles' gyro-motion. Here, X-axis is directed toward the Sun, Y-axis in the equatorial plane, and Z-axis is such that the coordinate frame is right-handed.}
    \label{density-topology-figure}}
    {\includegraphics[width=0.5\textwidth]{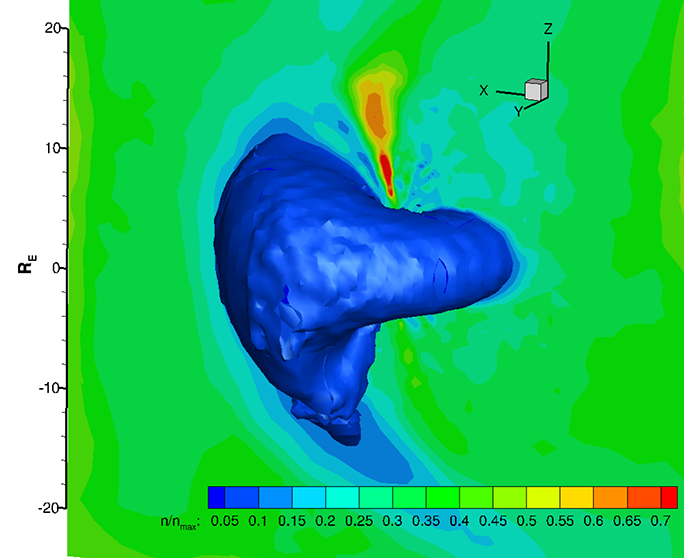}}
\end{figure}

The Earth's magnetosphere presents a shield against GCRs and SEPs. Those particles with energies below 100 MeV/n are effectively blocked by the Earth's magnetosphere \cite[]{Badavi:2011a}. Usually, the geomagnetic interaction of SEPs and GCRs is described in terms of rigidity, R (momentum/unit charge) rather than energy.
Transport of SEPs and GCRs in the geospace is a kinetic process due to a significant value of particles' gyroradius that can reach the value of tens of Earth's radii. An example of GCR's proton gyroradius calculated for quiet geomagnetic conditions is presented in Figure \ref{gyro-radius-example-figure}. 
One can see that even for particles that are on the lower end of the energetic spectrum of SEPs and GCRs penetrating in the geospace, the gyroradius can be as large as tens of Earth's radii, meaning that in practical calculations, kinetic methods that account for the gyro-motion of the energetic particles must be employed. 
The effect of the gyro-motion on the topology of the SEPs' population in geospace is illustrated in \figurename~\ref{density-topology-figure}, which shows the density of SEPs in the plane orthogonal to the equatorial plane and the SEP density's iso-surface in geospace.

\begin{figure}[bht]
\includegraphics[trim=0mm 0mm 0mm 0mm, clip,width=0.9\textwidth]{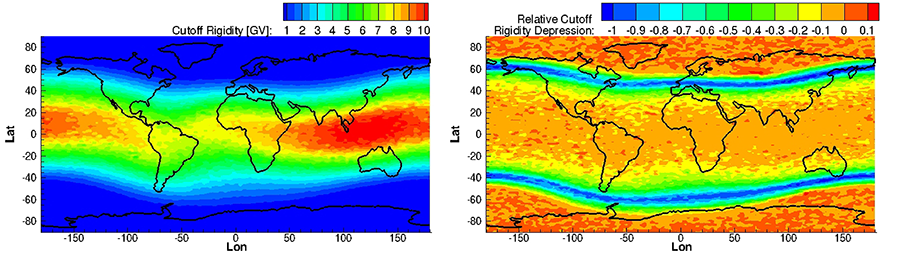}
\caption{Example of applying AMPS for rigidity cutoff calculation. The map is calculated for an altitude of 500 km. {\it Left:} Rigidity cutoff map calculated for quiet geomagnetic conditions. {\it Right:} Depression of the rigidity cutoff during a geomagnetic storm. The calculation was performed for conditions of the geomagnetic storm on 17 March 2015. One can see that the general rigidity cutoff patterns have changed mostly in the mid-latitude region. \cite[From][]{Tenishev:2021a}}
\label{cutoff-rigidity-example-figure}
\end{figure} 

An example of calculating cutoff rigidity detailed by \cite{Tenishev:2021a} is presented in \figurename ~\ref{cutoff-rigidity-example-figure}. 
The calculation is done using the Adaptive Mesh Particle Simulator (AMPS) employing 
 particle time-backward tracing starting from the altitude of 500 km. The calculations presented in the figure were performed for quiet geomagnetic conditions  ($p_\mathrm{SW} = 2\,$nPa, Dst$ = 1\,$nT, $B_y = -0.08\,$nT, and $B_z = 2\,$nT) and for the conditions during the geomagnetic storm on 17 March 2015 ($p_\mathrm{SW}= 10\,$nPa, Dst$ = -200\,$nT, $B_y = -7\,$nT, and $B_z = -10\,$nT). The left panel of \figurename~\ref{cutoff-rigidity-example-figure} shows the rigidity cutoff map before the storm. The right panel shows the relative depression during the storm. 
 The value shown in \figurename ~\ref{cutoff-rigidity-example-figure} is the ratio of the cutoff rigidity difference during the event to its original value. The relative depression of -1 means that the corresponding location becomes magnetically connected to the interplanetary magnetic field during the simulated geomagnetic storm.
 One can see that the general rigidity cutoff patterns have changed mainly in the mid-latitude region.

\subsection{Mesoscale Resolving Magnetosphere Simulations}
\label{subsec:mesoscale}
While the MHD plasma description has inherent restrictions in describing the microscale processes (see Section \ref{sec:epic} for treatment of kinetic processes), BATS-R-US, when run with high spatial resolution in key portions of the geospace, can easily resolve the Kelvin-Helmholtz instability and flux-transfer events (FTEs) \cite[]{Kuznetsova:2009a} found \eg at the magnetospheric boundary.  High-resolution MHD simulations in the magnetotail can reproduce intricate details of the interchange instability, bursty bulk flows, and other processes \citep{Yu:2017a}. The adaptive mesh refinement (AMR) guarantees that the run times, while higher for high resolution, remain manageable, as the increase in number of computational cells only increases by about a factor of a few. 

\begin{figure}[ht]
    \floatbox[{\capbeside
        \thisfloatsetup{capbesideposition={left,top},
        capbesidewidth=0.35\textwidth}}]{figure}[\FBwidth]
    {\caption{
    Results of high-resolution SWMF/BATS-R-US simulation with 1/16 $R_\sE$ grid resolution in the tail and magnetopause region in order to resolve small and mesoscale structures. It shows the filamentary current structures associated with the presence of both Kelvin-Helmholtz vortices at the flanks of the magnetopause and of bursty structures in the tail. Note that the associated flow velocities are not shown
    (Dorelli and Buzulukova, personal communications, 2020)}.
    \label{fig:hires}}
    {\includegraphics[width=0.625\textwidth]{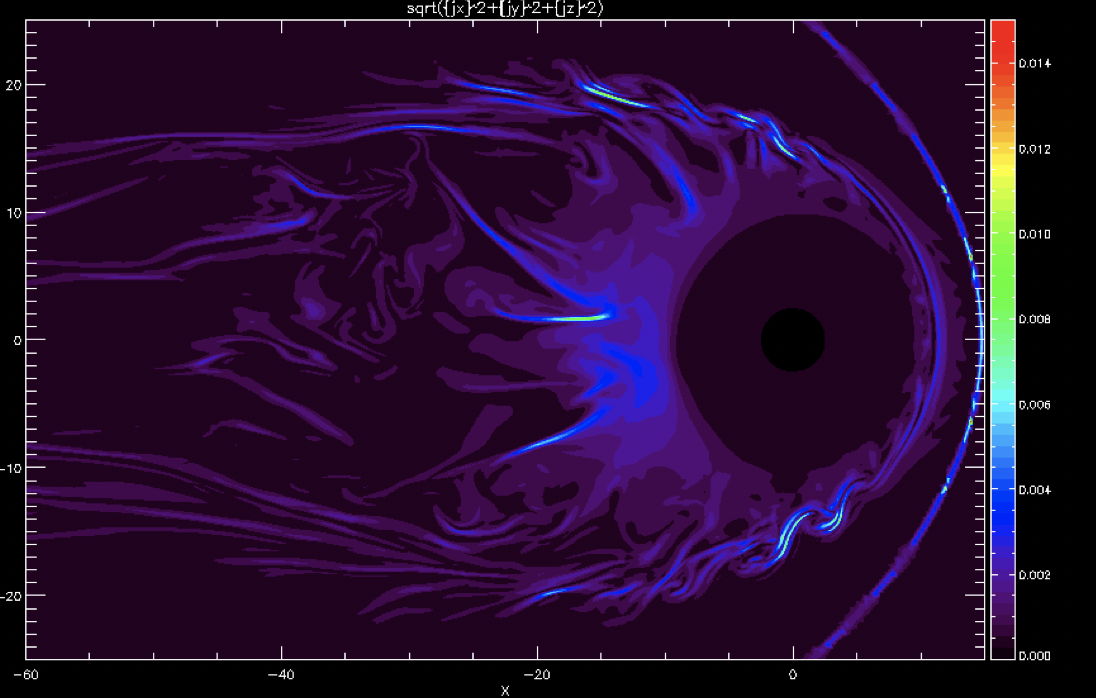}}
\end{figure}

An example of a very high-resolution simulation is shown in \figurename~\ref{fig:hires}. The SWMF/BATS-R-US simulation was run with 1/16 $R_\sE$ grid resolution in the tail and magnetopause region in order to resolve small and mesoscale structures in the magnetosphere. The results demonstrate the formation of Kelvin-Helmholtz vortices at the flanks of the magnetopause in response to the solar wind flow past the magnetic boundary. Furthermore, it was shown that reducing the resistivity in the model led to structuring of the reconnection in the magnetotail and the formation of narrow, elongated flow channels (or bursty bulk flows \citep{Angelopoulos:1994a}) throughout the width of the tail \citep{Haiducek:2020a}. 

\figurename~\ref{fig:hires} shows the current density in the equatorial plane during a geomagnetically active period. The filamentary current structures on the magnetopause and in the magnetotail are indicative of Kelvin-Helmholtz instability and mesoscale bursty bulk flows, respectively. The associated flow velocities for these structures are not shown. However, in this simulation it was found that while the main flow direction in the more distant magnetotail continues to be Earthward, the reconnection onset at the boundary of the quasidipolar and taillike magnetosphere creates tailward flows that strengthen at substorm onset (Dorelli and Buzulukova, personal communications, 2020). Such simulations are sufficiently accurate that they can be used to re-assess the substorm theories \citep[\eg][]{Baker:1996a, Angelopoulos:2008a}.

\subsection{Ionospheric Outflow Simulations}
\label{subsec:pwom}
Observations show that during geomagnetic storms O$^{+}$ can comprise as much as 40\% - 80\% of the ion density in the near-equatorial magnetosphere inside of 15 Re \cite[]{Lennartsson:1981a}. As O$^{+}$ can only originate in the ionosphere, its observed presence in the magnetosphere during storms is a clear indicator of the importance of the ionosphere in supplying magnetospheric plasma during space weather events. This is true not only of O$^{+}$, but it has been estimated that 65\% of the H$^{+}$ population during geomagnetic storms may also be of ionospheric origin \cite[]{Gloeckler:1987a}. 

\begin{figure}[tbh]
\centering
\includegraphics[width=1\textwidth]{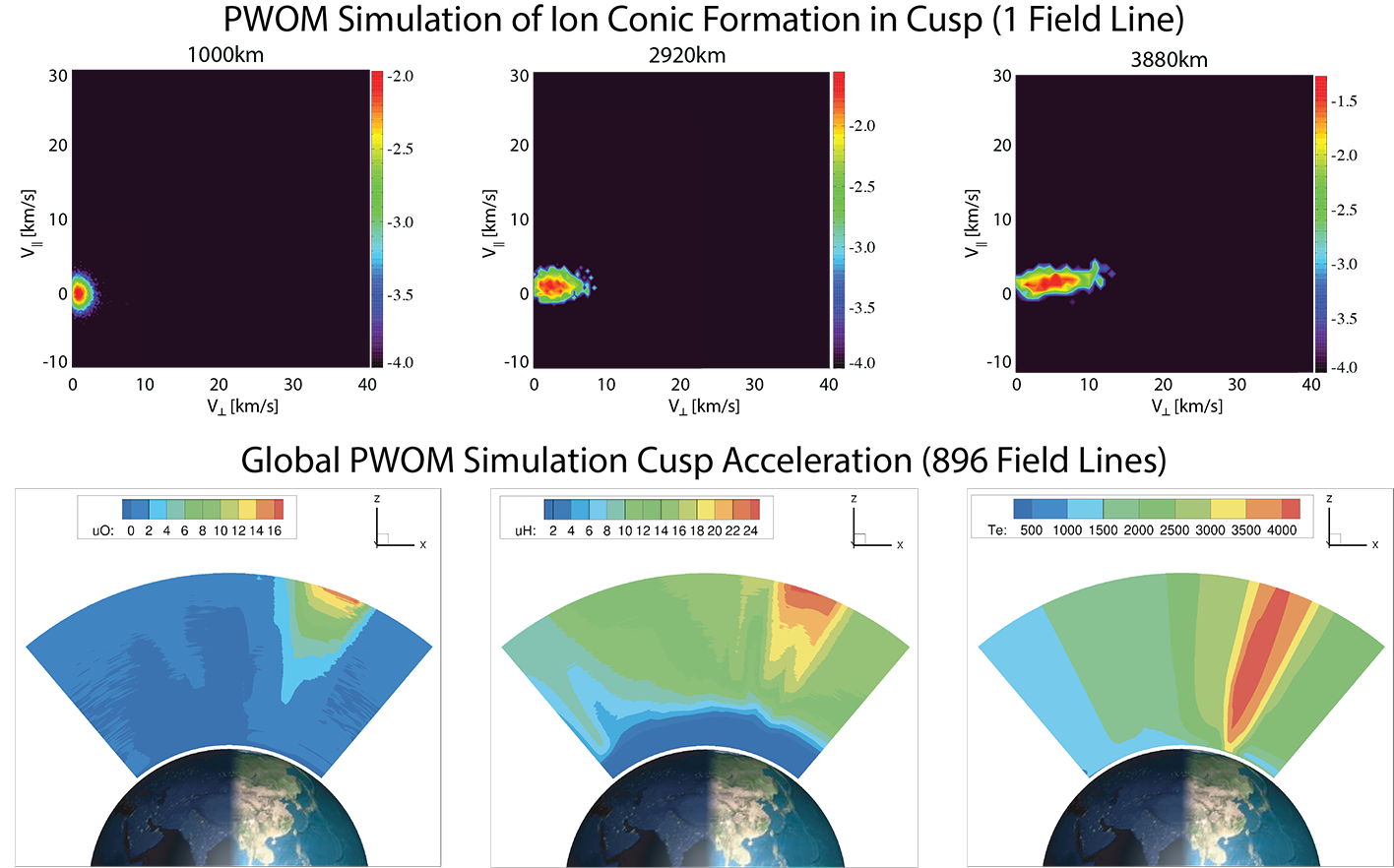}
\caption{An illustration of cusp accelerated ion outflow due to soft electron precipitation and wave-particle interactions as modeled by PWOM (adapted from  \cite[]{Glocer:2018a}). The top panel shows the PWOM computed ion distribution function along a single field line and demonstrates the ion conic evolution with altitude. Here $v_\spar$ and $v_\sperp$ are the parallel and perpendicular velocity and the color contour shows the log of phase space density in normalized, dimensionless, units.  The bottom panel shows the global simulation, including the kinetic processes in the cusp using 896 field lines. Here $u_\sO$ and $u_\sH$ are the bulk field aligned velocities for O$^{+}$ and H$^{+}$ in units of km/s, and $T_e$ is the electron temperature in Kelvin.}
\label{fig:pwom}
\end{figure}

The Polar Wind Outflow Model (PWOM) supplies the PW component in SWMF that calculates the transport of plasma from the ionosphere and sets the supply for the magnetosphere. This model solves the gyrotropic transport equations \cite[]{Gombosi:1989a} for multiple ion species from below the F2 peak to much higher altitudes. This model was expanded from solving a single field line to solving multiple field lines in order to reconstruct the global 3D outflow distribution \cite[]{Glocer:2009b}. The ability of the model to represent different critical drivers of ion outflow has also grown in recent years. The inclusion of various treatments of superthermal electron populations (photo, auroral, and secondary electrons) to PWOM has improved the model in comparison with observations \cite[]{Glocer:2012a, Glocer:2017a}. Most recently, PWOM has been expanded to move to a hybrid PIC description above 1000 km while maintaining a fluid description at lower altitudes \cite[see \figurename~\ref{fig:pwom}, adapted from][]{Glocer:2018a}. The latter expansion allows PWOM to treat wave-particle interactions due to processes like Ion Cyclotron Resonant Heating,  which is thought to be a major mechanism in creating ion conic distributions and energized O$^{+}$ escape. 

Simulations with SWMF are able to track the plasma calculated to escape the ionosphere throughout the magnetosphere using BATS-R-US when configured with multi-fluid or multi-species MHD. 
Using separate fluids or species for each constituent plasma population enables us to track the impact of ion outflow on the magnetosphere. 

SWMF in the configurations described above has been used to study the many impacts on ion outflow. The effect of ion outflow on the magnetic field at geosynchronous orbit and on the cross polar cap potential was studied and found to improve the prediction of the magnetic field as well as lower the cross polar cap potential \cite[]{Glocer:2009c,Welling:2012a}. The contribution of ion outflow to the ring current was examined and found to be a major contributor to the total ring current energy content during storms \cite[]{Glocer:2018a, Glocer:2020a, Welling:2015a, Ilie:2015a}. These codes, coupled together via the SWMF, have also been used to study ring current ENA observations from the TWINS mission \cite[]{Ilie:2013a}. These studies are only a subset of the total number of SWMF studies in this area and should not be taken as an exhaustive list. 

\subsection{Mesoscale Ionosphere Simulations}
\label{subsec:ionosphere}

The SWMF/Geospace is an important tool in the analysis of the polar region electrodynamics. The model's advantage is that it can produce superior spatial coverage for the magnetic disturbances whose observations are limited by the oceans and access to remote locations, and better spatial coverage for the field-aligned currents than those derived from spaceborne magnetic measurements by the AMPERE project \cite[]{Anderson:2000a}.

\begin{figure}[tbh]
\centering
\includegraphics[width=1\textwidth]{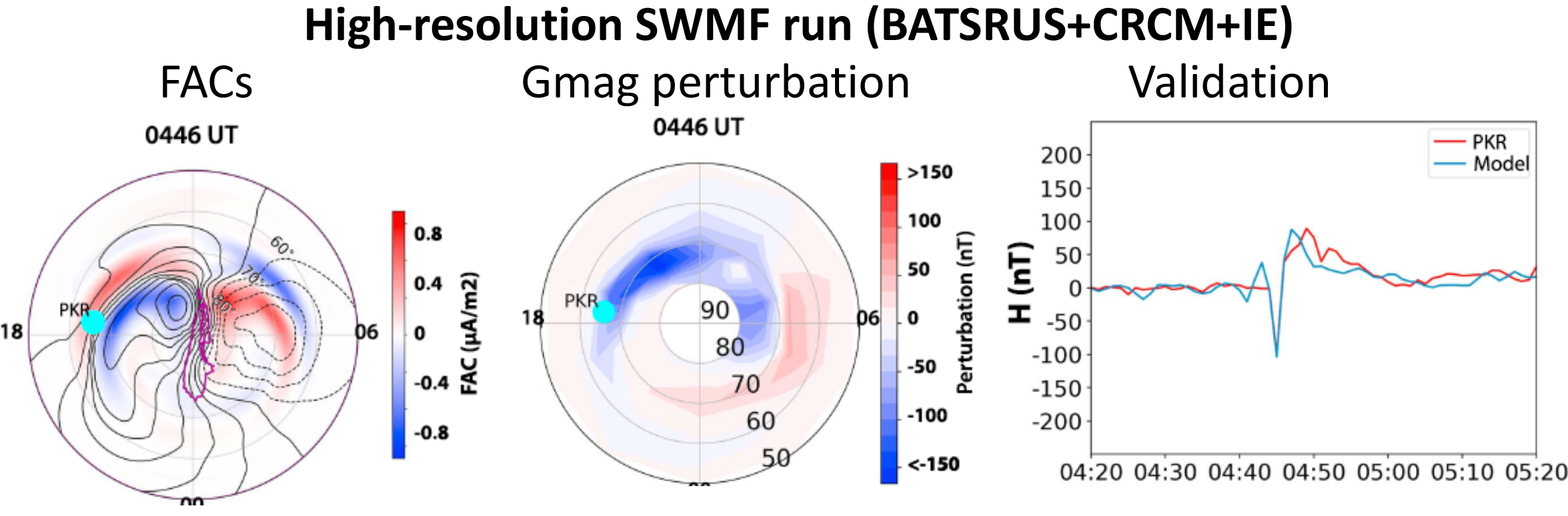}
\caption{FACs and ground magnetic perturbations due to shock compression on 17 March 2015 from a high-resolution BATSRUS run coupled with CRCM and IE. The simulated magnetic perturbations at Poker Flat were compared with the real magnetometer observations. The  results shown here illustrate the capacity of the SWMF to resolve mesoscale features of the magnetospheric dynamics in high-resolution MHD \cite[]{Zou:2017a}}
\label{fig:shasha}
\end{figure}

\figurename~\ref{fig:shasha} shows simulated field-aligned currents (FACs) and ground magnetic perturbations due to a solar wind pressure enhancement. The 2-way coupled BATSRUS, CRCM, and IE modules are utilized in this run with 1/8 $R_\sE$ resolution used from the dayside magnetopause to the near-Earth magnetotail to capture the magnetosphere reconfiguration due to compression and the subsequent relaxation. 
Several hundred virtual magnetometers have been included in the simulation at the locations of real magnetometers, and a uniformly distributed array covering the globe at a resolution of 4$^{\circ}$ in latitude and 12$^{\circ}$ in longitude. The left panel in \figurename~\ref{fig:shasha} shows the transient FACs during the Preliminary Impulse (PI) phase with the ionospheric convection contours superimposed on top. The location of the Poker Flat magnetometer is denoted by the cyan dot. The H component magnetic perturbation contours calculated from the uniform magnetometer array and a comparison of the time series of the simulated and observed H component perturbation at Poker Flat are shown in the middle and right panels of \figurename~\ref{fig:shasha}, respectively. 

The simulated magnetic perturbations matched both the polarity and the magnitude of the H component perturbation very well, suggesting the coupled models were able to capture the source, propagation, and closure of the compression-induced meso-scale field-aligned currents. The results of this coupled geospace run were then used to drive the Global Ionosphere and Thermosphere Model (GITM), and revealed a short-lived meso-scale fast flow channel in the ionosphere with intense Joule heating and sudden ion temperature enhancements \cite[]{Ozturk:2018a}. These simulation results provided a valuable explanation for a transient ion upflow event observed by the Poker Flat Incoherent Scatter Radar (PFISR) \cite[]{Zou:2017a}. 

Other important meso-scale features in the coupled geospace system have also been simulated by using the SWMF, including subauroral polarization streams (SAPS) \cite[]{Yu:2015a} and boundary flows between the Region 1 and Region 2 FACs \cite []{Wang:2019a}. 

\subsection{Geomagnetic Index Simulations}
\label{subsec:month}


For operational space weather forecasters, geomagnetic indices are a standard tool when distributing forecasts and warnings to users representing both spacecraft and ground system (e.g. power network) operators. These include K and Kp, \emph{Disturbance Storm Time} (Dst) and the related SYM-H index, and the Auroral Electrojet (AE, AU, AL, and AO) indexes \cite[]{Rostoker:1972a, Mayaud:1980a, Menvielle:2010a}.  The validity of the SWMF/Geospace suite can be checked by  reconstructing these indices from the simulation output data and comparing them with observations. 
Dst-equivalent outputs have long been a staple of SWMF simulations, 
serving as a quick-look diagnostic of inner magnetosphere performance \cite[\eg][]{Zhang:2007a, Welling:2011a}.
K, Kp, and AE indexes were added more recently and build off of internal virtual magnetometer capabilities.

\begin{figure}[!htb]
  \centering
    \includegraphics[width=0.45\textwidth]{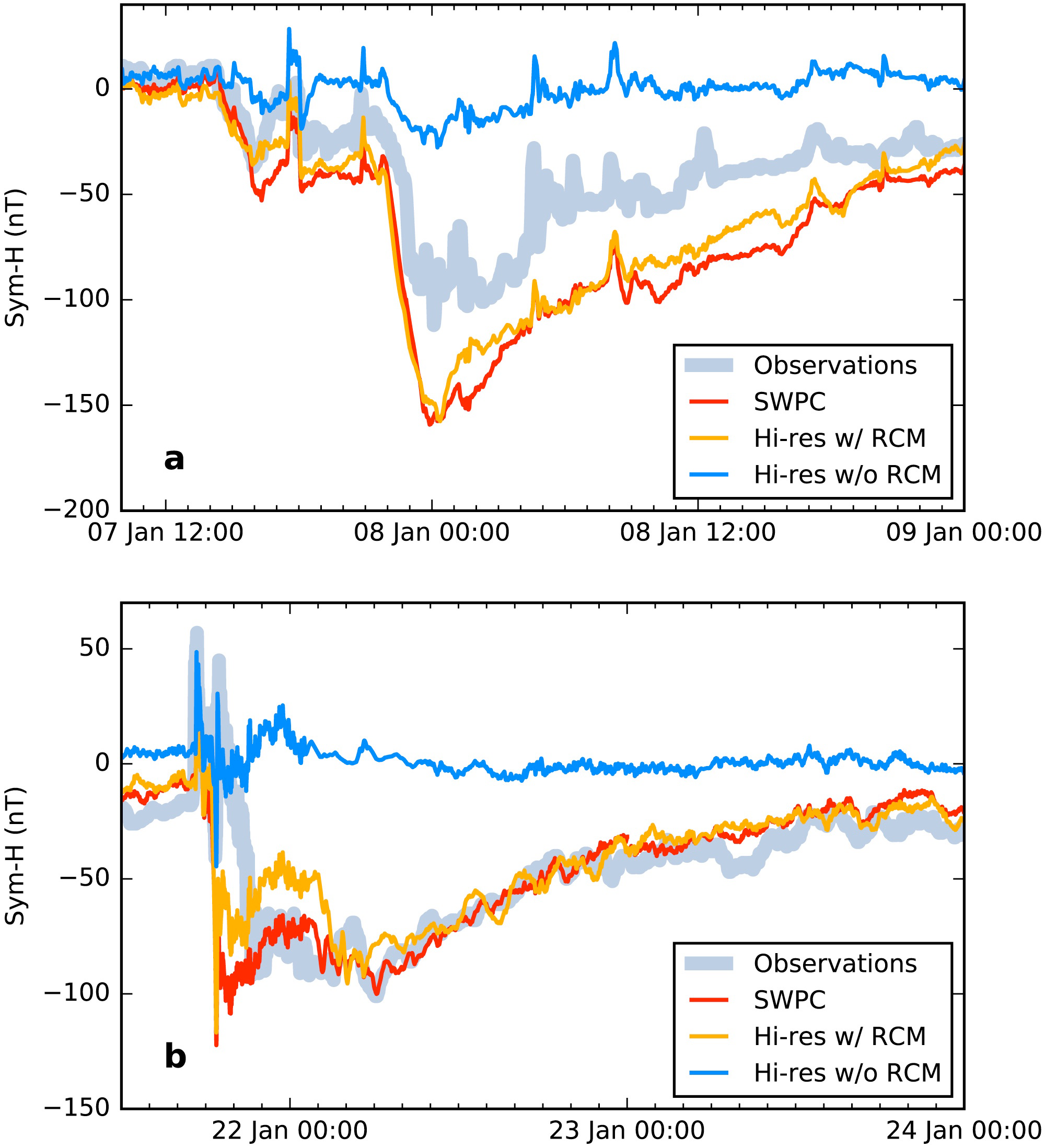}
    \hspace{1em}
    \includegraphics[width=0.5\textwidth]{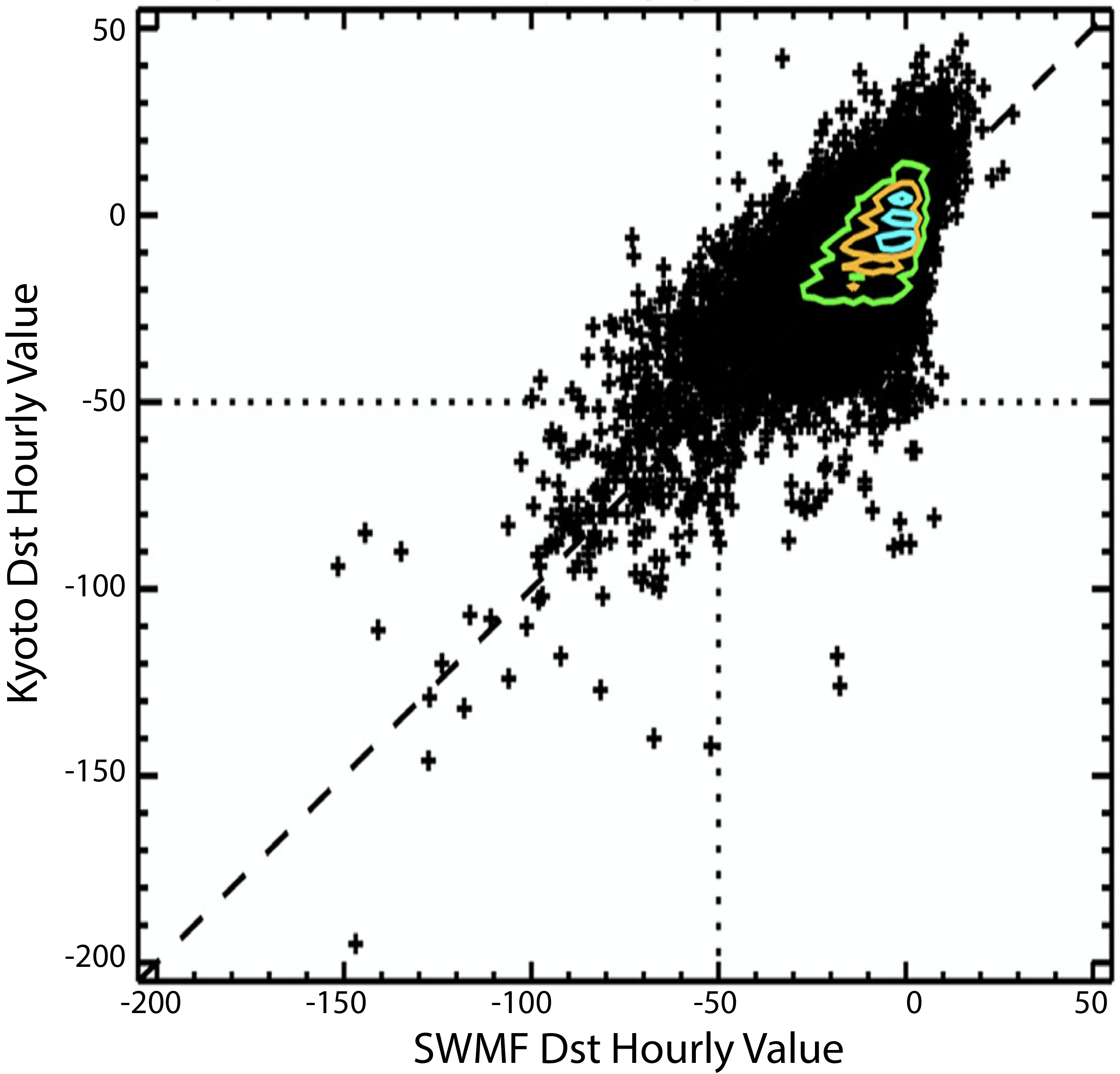}
    \caption{(left) Comparison of the storm SYM-H index for two storms from January 2005. The thick trace shows the observations, the red and orange traces show the normal resolution and high resolution SWMF/Geospace simulations, and the blue trace shows the SWMF/Geospace simulation run without the inner magnetosphere RCM component \cite[from][]{Haiducek:2017a}. (right) Scatter plot of the observed real‐time Dst time series (vertical axis values) against a SWMF prediction of the Dst (horizontal axis values). The green, orange, and cyan contour lines show the regions of 40, 80, and 120 values within a $2.6\times 2.6$-nT bin. The diagonal long-dashed line shows the ideal data-model relationship. The two dashed lines show the $-50$ nT threshold values  \cite[from][]{Liemohn:2018a}.}
  \label{fig:liemohn}
 \end{figure}

Several studies have focused on the performance of the virtual geomagnetic indexes produced by the SWMF.
\cite{Glocer:2016a} explored the local K-index predictive skill of the SWMF and other models, demonstrating the SWMF's strong capabilities and reproducing this value.
\cite{Haiducek:2017a} simulated the month of January 2005 using the observed solar wind data as input. The simulation was run with two different grid resolutions. The model was found to predict the ring current index SYM-H to good accuracy, with a root mean square error of less than 20 nT (see left panel of Figure \ref{fig:liemohn}). The geomagnetic index Kp performed well during storm time, but predicted larger than observed activity during quiet times. On the other hand, the auroral electrojet index AL was predicted reasonably well on average, but was systematically less negative than the observed values during high geomagnetic activity.  While the grid resolution caused only small variations to the results, runs without the inner magnetosphere component were not able to produce the storm dynamics \citep{Haiducek:2017a}.
\cite{Haiducek:2020a} further explored virtual AL performance during substorm activity, finding 
substorm-related perturbations to be weaker than observations.
\cite{Liemohn:2018a} continued along the same lines and used nearly three years of simulated geomagnetic index data from the experimental real-time SWMF runs at CCMC. The right panel of Figure \ref{fig:liemohn} shows scatter plots of observed and simulated Dst (hourly averaged SYM-H) values for the different simulation setups. They examined different metrics of success, and conclude that the correlation coefficient between the observed and model values was 0.69, the prediction efficiency was 0.41, and the Heidke skill score was 0.57 for an event threshold of $-50$ nT.

Overall, all these studies confirm that the Geospace model represents a reasonably accurate approximation to the real magnetosphere during a large variety of circumstances. This provides confidence in the more detailed predictions, such as local magnetic disturbance levels around the globe or the plasma parameters near the geosynchronous orbits. In CCMC-led modeling challenges that focus on geomagnetic index comparisons, the SWMF is consistently among the best of the global models \cite[\eg][]{Glocer:2016a, Pulkkinen:2013a, Rastaetter:2013a}. While more accurate models exist for predicting and forecasting geomagnetic indices, in particular those based on machine learning algorithms, the SWMF is, to-date, the most accurate reproduction of these indices from a solar-wind-to-ionosphere first-principles physics-based model of the full geospace system.

\section{Resolving Kinetic Scales in Global Simulations}
\label{sec:kineticscales}

\subsection{MHD-EPIC and MHD-AEPIC}
\label{subsec:aepic}
Kinetic models have been used for a long time to model the inner magnetosphere  \cite[\cf][]{Wolf:1982a, Buzulukova:2010a}, the radiation belts \cite[\cf][]{Fok:2008a} or the transport and diffusion of energetic particles along field lines \cite[\cf][]{Sokolov:2004a}. These kinetic models use some simplifying assumptions, such as restricting the motion of particles along field lines and ignoring feedback to the magnetic field, to drastically reduce the computational cost. 
Solving the full kinetic equations in 7 dimensions (one temporal, 3 spatial and 3 velocity) in a global simulation is simply not feasible on the current or even near future supercomputers. 

The idea of coupling MHD and kinetic PIC models has been around for a long time \cite[\cf][]{Sugiyama:2007a}, but making this work in 3D has been an elusive goal. In fact, many in the community argued that coupling fluid and kinetic models is impossible as they are simply not compatible with each other. The MHD and algorithm experts at Michigan initiated a collaboration with several PIC modelers, including Giovanni Lapenta, Stefano Markidis and Jeremiah Brackbill. It took years of working together to overcome all the obstacles. Some were seemingly simple, like converting units, and still took a long time. Others were much more complicated, such as keeping the PIC model stable and suppressing various instabilities, or avoiding discontinuities developing at the interface of the MHD and PIC regions. We found, for example, that using Hall MHD instead of ideal MHD improves stability, or using hyperbolic-parabolic cleaning in addition to the 8-wave scheme is necessary to eliminate accumulation of $\divB$ errors near the boundaries of the PIC region. 

Eventually our  work yielded results: the MHD with embedded PIC (MHD-EPIC) model became reality \cite[]{Daldorff:2014a}. 
It took a few more years to efficiently couple the models through the SWMF using a newly developed efficient parallel coupler, allow for different grids and different time steps, allowing for multiple PIC domains and generalizing the fluid model from single fluid (Hall) MHD to multi-species and multi-fluid MHD, as well as the five- and six-moment fluid equations \cite[\cf][]{Chen:2019a, Zhou:2020b}.

Running MHD-EPIC simulations for long simulation times also  revealed hidden issues with the PIC algorithms that could be avoided in stand-alone PIC simulations by careful tuning of various parameters, but were plaguing the more complicated MHD-EPIC simulations. 
We overcame these issues by developing the Gauss Law satisfying ECSIM (GL-ECSIM) algorithm \cite[]{Chen:2019a} that conserves energy and charge  at the same time. This new PIC algorithm, coupled with the extended MHD code, has finally delivered an accurate and reliable MHD-EPIC model. 

\cite{Toth:2017a} showed that the kinetic scales can be artificially changed by changing the mass per charge ratio of the ions and electrons and still obtain correct global solutions as well as correct, but scaled, kinetic solutions. The only limitation is that the modified kinetic scales should still be well separated from the global scales. For example, one can increase the kinetic scales by a factor of $f=16$ and thus reduce the computational cost of the PIC model by a factor of $f^4\sim65,000$. With such scaling it became possible to simulate Earth's magnetosphere with the MHD-EPIC model. \cite{Chen:2017a} modeled the kinetic reconnection process at the dayside magnetopause of Earth in a global simulation. The model correctly reproduced the properties of flux transfer events (FTEs) and revealed several new insights into the birth, development and final fate of FTEs starting from the kinetic scales and growing to the global scales. 

While MHD-EPIC with kinetic scaling opened the possibility of combining kinetic modeling with global simulations of Earth's magnetosphere dynamics, the simulations were still very expensive. This is especially true for the magnetotail, where the reconnection sites can move in a large volume due to the intrinsic dynamics of the reconnection X-lines, as well as to the flopping of the magnetotail caused by the changing solar wind. 

\begin{figure}[tbh]
\centering
\includegraphics[width=1\textwidth]{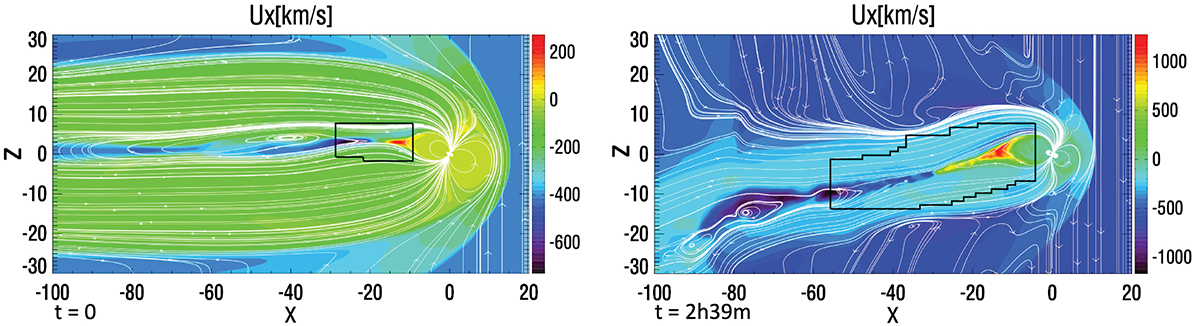}
\caption{MHD-AEPIC simulation of a geomagnetic storm. The 2D cuts display a part of the much larger 3D domain. Colors show the X component of the velocity and the white lines are traces of the magnetic field in the meridional plane. The black lines in the two snapshots, separated by 159 minutes of simulation time, indicate the edges of the active PIC regions. As the tail evolves, the active PIC region is continuously adapted to cover the reconnection sites of interest. }
\label{fig:MHD-EPIC-tail}
\end{figure}

To further improve the efficiency of the model, we have developed the MHD with an adaptively embedded PIC (MHD-AEPIC) algorithm. The main idea comes from the block-adaptive mesh and the hybrid schemes used in BATS-R-US: the PIC grid is decomposed into small blocks that can be activated and deactivated dynamically. 
While the idea is straightforward, the implementation is not. We had to abandon iPIC3D that uses a single grid, and implement the GL-ECSIM algorithm into the Adaptive Mesh Particle Simulator (AMPS) code \cite[]{Tenishev:2021a}. 
The resulting MHD-AEPIC model can achieve an order of magnitude or even more speed-up compared to the MHD-EPIC model that uses static PIC domains. 
We also developed a new PIC code, the Flexible Exascale Kinetic Simulator (FLEKS), to be used in MHD-AEPIC. FLEKS is based on the AMReX library \cite[]{Zhang:2019amrex, Zhang:2020amrex} and it was designed for flexibility and high performance with a state-of-the-art semi-implicit PIC algorithm. Novel particle splitting and merging algorithms have been designed for FLEKS to control the number of macro-particles per cell during long MHD-AEPIC simulations.

In general, the MHD-(A)EPIC model offers a powerful tool to study magnetospheric physics. The latest application is covering the tail reconnection site with an adaptive PIC region so that one can study geomagnetic storms and substorms in a more realistic way.  \figurename~\ref{fig:MHD-EPIC-tail} shows an example of an adaptive PIC region, which tracks the motion of the magnetotail reconnection site during a storm simulation.

\subsection{MHD-EPIC Results}
\label{subsec:epicresults}
MHD-EPIC has been used to simulate the terrestrial magnetosphere \cite[]{Chen:2017a,  Chen:2020a, Jordanova:2018a}, the interaction of Mercury \cite[]{Chen:2019b} and Mars \cite[]{Ma:2018b} with the solar wind and the mini-magnetosphere of Ganymede \cite[]{Toth:2016a, Zhou:2019b, Zhou:2020b}. To demonstrate the capabilities of MHD-EPIC, here we show some of the Earth magnetosphere simulation results by \cite{Chen:2017a}.

\begin{figure}[bht]
    \floatbox[{\capbeside
        \thisfloatsetup{capbesideposition={left,top},
        capbesidewidth=0.29\textwidth}}]{figure}[\FBwidth]
    {\caption{Snapshots showing $B_y$ strength (color) and the projected magnetic field lines in the meridional plane inside the PIC region. The color bar is different in each plot. \cite[from][]{Chen:2017a}}
    \label{fig:yuxi1}}
    {\includegraphics[width=0.7\textwidth]{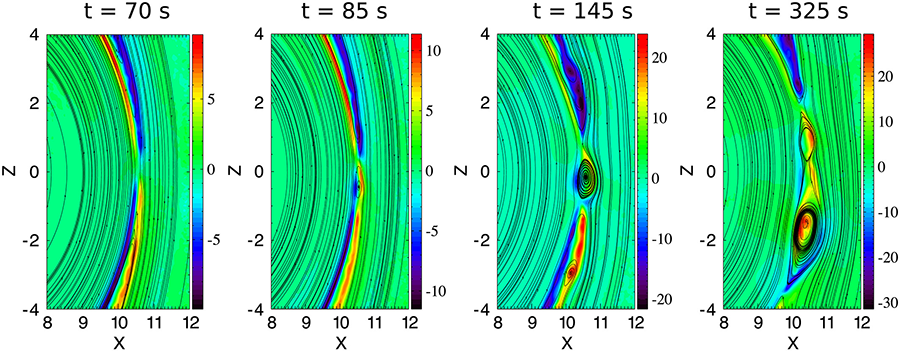}}
\end{figure}

An overview of the evolution of the dayside magnetopause is shown in \figurename~\ref{fig:yuxi1}, which contains the Hall magnetic field $B_y$ and the field lines at the meridional plane inside the PIC box. 
At $t = 70$ the Hall field extends far away from the X line with roughly the same field strength for each branch. 
Fifteen seconds later, south of the existing reconnection point, another X line starts to form. At $t = 145$ s, both X lines can be seen clearly, and a flux rope-like structure forms between the two X lines. 
At $t = 325$ s, the flux rope moves away from the top X line and the current sheet between them becomes unstable and a secondary flux rope is generated. During the one-hour simulation, flux ropes form near the subsolar point and move toward the poles quasi-periodically.

\begin{figure}[bht]
    \floatbox[{\capbeside
        \thisfloatsetup{capbesideposition={left,top},
        capbesidewidth=0.35\textwidth}}]{figure}[\FBwidth]
    {\caption{Crescent electron and ion phase space distributions 
    (a)  $E_x$ (mV/m) in the meridional plane at $t = 3,600$s; 
    (b) Normalized electron distribution in $V_y-V_x$ phase space; and 
    (c) Ion phase space distribution. 
    The phase-space density is normalized. \cite[from][]{Chen:2017a}}
    \label{fig:yuxi2}}
    {\includegraphics[width=0.625\textwidth]{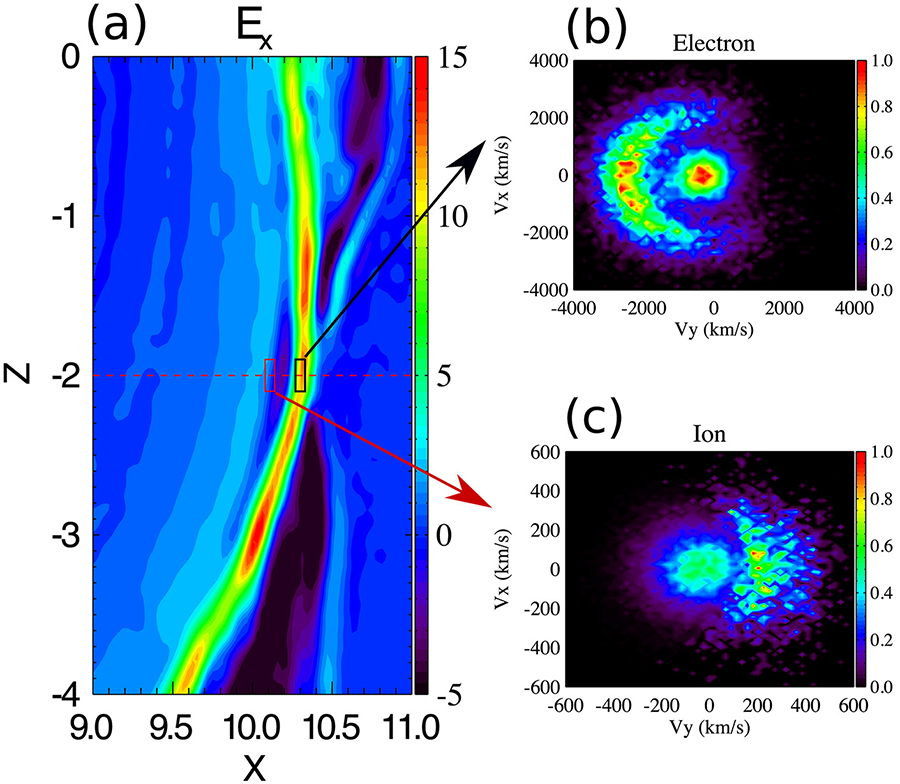}}
\end{figure}

Crescent shape electron phase space distribution has been observed near the electron diffusion region at the dayside magnetopause by MMS \cite{Burch:2016a}. The same distribution is also found in the 3D MHD-EPIC simulation (see \figurename~\ref{fig:yuxi2}). The phase space distribution of electrons inside a cube region on the dayside magnetopause
is shown in \figurename~\ref{fig:yuxi2}b. The crescent distribution is found in the $V_y-V_x$ plane, corresponding to the two velocity components perpendicular to the magnetic field. The crescent hot electrons are drifting along the negative $y$ direction with a speed close to 3,000 km/s. The direction of the flow is consistent with the $\bE \times \bB$ direction, and the velocity of the crescent particles is very close to the MMS observed by \cite{Burch:2016a}. Slightly farther away from the reconnection site, where the Larmor field appears, 
the ion phase space distribution also presents a crescent-like shape, as is shown in \figurename~\ref{fig:yuxi2}c. The crescent ions drift in a positive $y$ direction because $E_x$ is negative.

\section{Planetary environments and Solar Analogs}
\label{sec:planets}
Space weather phenomena at other solar system objects and at astropheres are the subject of increasing interest \cite[\cf,][]{Lilensten:2014a, Plainaki:2016a, Andre:2018a}. More recently, space weather phenomena in astropheres harboring extrasolar planets also became the focus of investigations \cite[\cf,][]{Pillet:2019a}.

The SWMF simulation suite has been used to simulate the space environment of most solar system planets, including Mercury \cite[\eg][]{Kabin:2000a, Kabin:2008a, Jia:2015a, Jia:2019a}, Venus \cite[\eg][]{Bauske:1998a, Ma:2013a}, Mars \cite[\eg][]{Liu:1999a, Bauske:2000a, Ma:2002a, Liemohn:2006b, Ma:2018a,  Regoli:2018a}, Jupiter \cite[\eg][]{Cravens:2003a, Sarkango:2019a}, Saturn \cite[\eg][]{Hansen:2000a, Gombosi:2005a, Hansen:2005a, Glocer:2007a, Gombosi:2010a, Zieger:2010a, Jia:2012a, Jia:2012b}, and Uranus \cite[]{Toth:2004a}. 

Moreover, the SWMF has been applied to comets \cite[\eg][]{Gombosi:1996a, Haeberli:1997a, Gombosi:1999a, Hansen:2007a, Rubin:2012a, Gombosi:2015a, Huang:2016a, Huang:2016c}, and planetary moons including Io \cite[\eg][]{Combi:1998a, Kabin:2001a}, Europa \cite[\eg][]{Kabin:1999b, Liu:2000a, Rubin:2015a, Jia:2018a, Harris:2021a}, Ganymede \cite[\eg][]{Toth:2016a, Zhou:2019b, Zhou:2020b}, Titan \cite[\eg][]{Kabin:1999a, Kabin:2000b, Nagy:2001a, Ma:2007a}, and Enceladus \cite[\eg][]{Jia:2010a, Jia:2010b, Jia:2010c}.

Finally, the SWMF suite of models has also been applied to the outer heliosphere \cite[\eg][]{Opher:2007a, Opher:2009a, Opher:2016a, Opher:2017a} and astrospheres \cite[\eg][]{Alvarado-Gomez:2020a, Cohen:2020a, Cohen:2010a, Cohen:2015a}.

As discussed above, the core MHD code within the SWMF, BATSRUS, can be configured to solve the governing equations of ideal MHD, resistive MHD, semi-relativistic MHD, multi-fluid MHD, MHD with anisotropic pressure, or high-order moment MHD. In addition to the basic equations, there are various source and loss terms included in BATSRUS that change from application to application (see details in Appendix~\ref{subsubsec:sourceterms}). Most relevant to our applications to the giant planet magnetospheres (e.g., Jupiter and Saturn) is the capability of including various mass-loading processes (ionization, charge-exchange, dissociative recombination, etc.) arising from the internal plasma sources associated with planetary moons [Jupiter: \cite{Sarkango:2019a}; Saturn: \cite{Zieger:2010a, Jia:2012a, Jia:2012b}]. In modeling planetary magnetospheres with an ionosphere, BATSRUS is normally coupled to the Ionospheric Electrodynamics (IE) module to simulate magnetosphere-ionosphere coupling. For planetary objects that do not possess a significant atmosphere/ionosphere, such as Mercury and Ganymede, we have extended the BATSRUS MHD model to include the planetary interior as part of the simulation domain such that the influence of the electrical conductivity of the planetary interior on the space environment can be directly modeled. We have applied such a model successfully to the magnetospheres of Mercury \cite[]{Jia:2015a, Jia:2019a, Chen:2019b} and Ganymede \cite[]{Zhou:2019b, Zhou:2020b}, where the induction effect of the subsurface conducting region (conducting core in the case of Mercury, and subsurface ocean in the case of Ganymede) plays an important role in the global magnetospheric interaction.

Here we show the results from two Mercury simulations that demonstrate the flexibility and capabilities of SWMF. The first is an extended MHD simulation that takes into account the finite conductivity of Mercury's interior \cite[]{Jia:2015a,Jia:2019a}, while the second is an MHD-EPIC simulation that takes into account kinetic effects \cite[]{Chen:2019b}.

A unique aspect of Mercury's interaction system arises from the large ratio of the scale of the planet to the scale of the magnetosphere and the presence of a large‐size core composed of highly conducting material. Consequently, there is strong feedback between the planetary interior and the magnetosphere, especially under conditions of strong external forcing. In applying the SWMF simulation suite to Mercury, \cite{Jia:2015a} used the resistive MHD version of BATSRUS to develop a global magnetosphere model in which Mercury's interior is modeled as layers of different electrical conductivities that electromagnetically couple to the surrounding plasma environment. One particular advantage of this model is its ability to characterize the dynamical response of Mercury to time‐varying external conditions in a self‐consistent manner, such as the induction effect at the planetary core. To demonstrate this capability, we have performed a series of idealized simulations as well as simulations of MESSENGER events for a wide range of upstream solar wind conditions \cite[]{Jia:2015a, Jia:2019a}. Our results show that, due to the induction effect, Mercury's core exerts strong global influences on the way Mercury responds to changes in the external environment, including modifying the global magnetospheric structure and current systems as well as affecting the extent to which the solar wind directly impacts the surface. There results have important implications for understanding the role of space weathering in generating Mercury's tenuous exosphere \cite[\eg][]{Jia:2019a}.

\begin{figure}[tbh]
\centering
\includegraphics[width=1\textwidth]{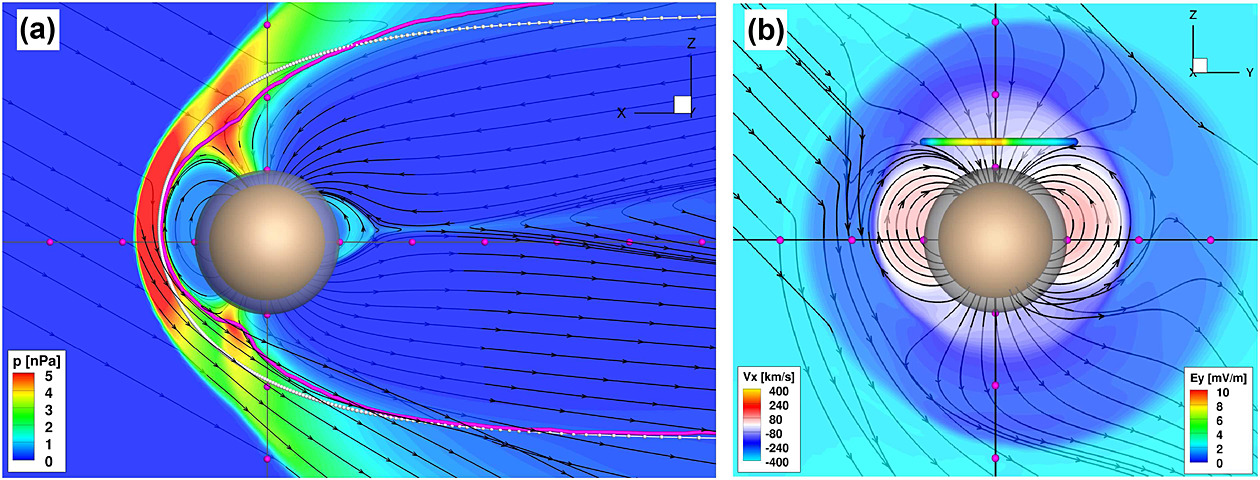}
\caption{Cuts through the simulation of the MESSENGER M2 flyby. (a) XZ cut at $Y = 0$ (noon‐midnight meridian) with color contours of plasma thermal pressure. The thick magenta line shows the modeled magnetopause boundary identified based on the current density, while the white dotted line represents the data‐based empirical magnetopause model of \cite{Winslow:2013a}. (b) YZ cut at $X = 0$ (terminator plane) with color contours of the x component of the flow velocity ($V_x$). The horizontal line at $Z = 1.3R_\sM$ is color coded by the y component of the convectional electric field ($E_y$), from which the cross‐polar cap potential is calculated. \cite[from][]{Jia:2015a}}
\label{fig:xianzhe1}
\end{figure}

As an example, we present results from the MESSENGER M2 flyby simulation by \cite{Jia:2015a} to illustrate the global configuration of the modeled magnetosphere. \figurename~\ref{fig:xianzhe1}a shows the model results in the noon‐midnight meridional plane, where color contours of plasma pressure, along with projections of sampled magnetic field lines, are plotted to delineate the magnetospheric configuration. One notable feature of the modeled magnetosphere is the pronounced asymmetry with respect to the planet's equatorial plane, which arises due to the northward offset of the internal dipole as well as the presence of a strong IMF $B_x$, which is typical at Mercury's orbit. Some general features of the modeled magnetosphere can be compared directly with MESSENGER observations, such as the location and shape of various important boundaries. For the upstream conditions used in this simulation, the magnetopause and bow shock standoff distances are about $1.5R_\sM$ ($R_\sM$ = 2440 km is Mercury's mean radius) and $1.9R_\sM$, respectively, in accordance with MESSENGER observations of the magnetosphere under similar upstream conditions \cite[]{Winslow:2013a}. Also plotted in \figurename~\ref{fig:xianzhe1}a is the modeled magnetopause boundary (magenta line) identified based on the total current density. For comparison, the empirical magnetopause model of \cite{Winslow:2013a}, constructed based on MESSENGER data, is also plotted. As can be seen, the overall shape of the magnetopause boundary in our model is in general agreement with the data‐based model.

\figurename~\ref{fig:xianzhe1}b shows a YZ cut at $X=0$ through the simulation in which the color contours represent the x-component of the modeled plasma flow velocity ($V_x$) and the lines with arrows are sampled field lines. Key regions of the interaction can be readily identified based on the $V_x$ contours. The transition from the ambient solar wind speed of $\sim400$ km/s to $\sim200$ km/s, which is characteristic of the magnetosheath flow at the terminator, marks the boundary of the bow shock, whereas further inward the transition from the sheath flows to convection flows with much smaller speeds ($<\sim100$ km/s) marks the magnetopause boundary. Inside the magnetosphere, flows with negative $V_x$ at high latitudes are the cross‐polar cap flows moving in the anti-sunward direction, while the flows with positive $V_x$ at low latitudes are those return flows convecting from the night side to the dayside. From the results shown in \figurename~\ref{fig:xianzhe1}b we can obtain the cross‐polar cap potential in the model, which provides a global measure of the strength of the coupling between the magnetosphere and the solar wind. The total potential drop in our model is about 25 kV, in reasonable agreement with the 30 kV estimated by \cite{Slavin:2009a} for this flyby based on MESSENGER observations.

\begin{figure}[tbh]
\centering
\includegraphics[width=0.9\textwidth]{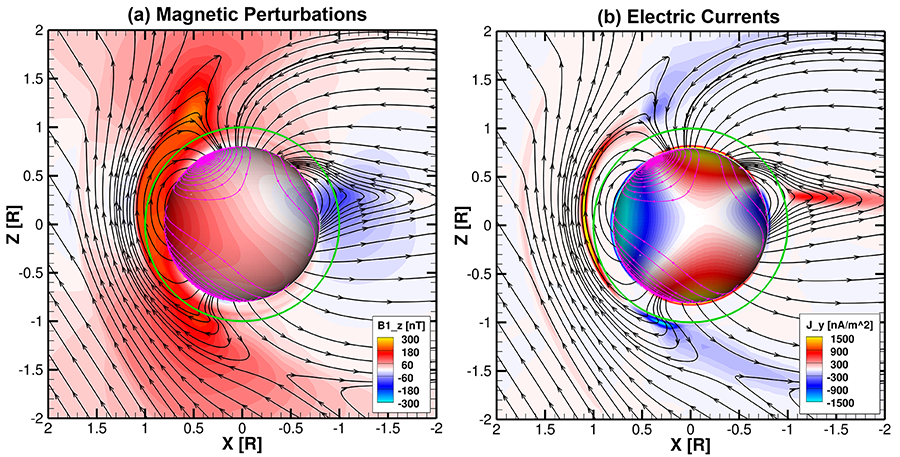}
\caption{A close-up view of the simulated magnetospheric configuration for the Highly Compressed Magnetosphere (HCM) event observed by MESSEGER on 23 November 2011. (a) Color contours of the z-component of the magnetic field perturbation ($B_{1z}$ in nT) shown in the XZ plane and also on a 3D sphere of radius $\sim$ 0.8$\rm R_{M}$ corresponding to the core-mantle boundary. (b) Same as (a) but for the current density in the y-direction ($J_{y}$ in $\rm nA/m^{2}$), indicating enhanced magnetopause and tail currents as well as the induction currents at the core in response to solar wind compression. In both panels, the black lines with arrows show projections of sampled magnetic field lines onto the XZ plane and the magenta lines with arrows show sampled streamlines of the induction currents generated at the core. The green circle of radius 1 $\rm R_{M}$ represents the planetary surface. \cite[from][]{Jia:2019a}}
\label{fig:xianzhe2}
\end{figure}

To demonstrate the model's ability to simulate the induction effect, we examine the global response of Mercury's magnetosphere to solar wind compression. \figurename~\ref{fig:xianzhe2} shows the results extracted from a time-dependent simulation for the Highly Compressed Magnetosphere (HCM) event observed by MESSENGER on 23 November 2011, which was produced by the passage of a CME \cite[]{Jia:2019a}. \figurename~\ref{fig:xianzhe2}a shows the z-component of the magnetic field perturbations, $B_{1z}$, which result from various current systems, including the Chapman-Ferraro currents, the tail current sheet, and the induction currents in the core, all of which are discernible in \figurename~\ref{fig:xianzhe2}b. As expected for this HCM event, both the magnetopause and the tail current sheet are displaced very close to the surface. The subsolar magnetopause stand-off distance in the simulation is $\sim$ 1.12 $\rm R_M$, which is in excellent agreement with the distance of 1.13 $\rm R_M$ determined by  \cite{Slavin:2014a} for this event based on MESSENGER observations. For this CME event, the current sheet on the night side almost reaches the planetary surface  with its inner edge at only $\sim$ 1.1 $\rm R_{M}$. As illustrated by the cyan and yellow colors at the core boundary in \figurename~\ref{fig:xianzhe2}b, strong currents flowing in the direction as indicated by the magenta arrows are induced to prevent the external magnetic perturbations from penetrating into the conducting core. On the day side, the induced currents, together with the intensified Chapman-Ferraro currents, produce the strong positive $B_{1z}$ perturbations, which are present throughout Mercury's resistive mantle (\figurename~\ref{fig:xianzhe2}a). The positive $B_{1z}$ perturbations in the dayside magnetosphere and inside the mantle reach amplitudes between $\sim$ 150 and 200 nT. On the night side, the cross-tail currents generate negative $B_{1z}$ perturbations planetward of the current sheet with an average amplitude of -100 nT. The intensification and displacement of the tail current sheet toward the planet also induce currents flowing at the core boundary that act to negate the effect of external variations. By driving the simulation with different upstream solar wind conditions, \cite{Jia:2019a} conducted a systematic numerical experiment to establish global context for interpreting the HCM events observed by MESSENGER during its entire mission. Their results also provide a quantitative assessment of the relative importance of the shielding effect from induction and the erosion effect from magnetopause reconnection.

\begin{figure}[tbh]
\centering
\includegraphics[width=0.9\textwidth]{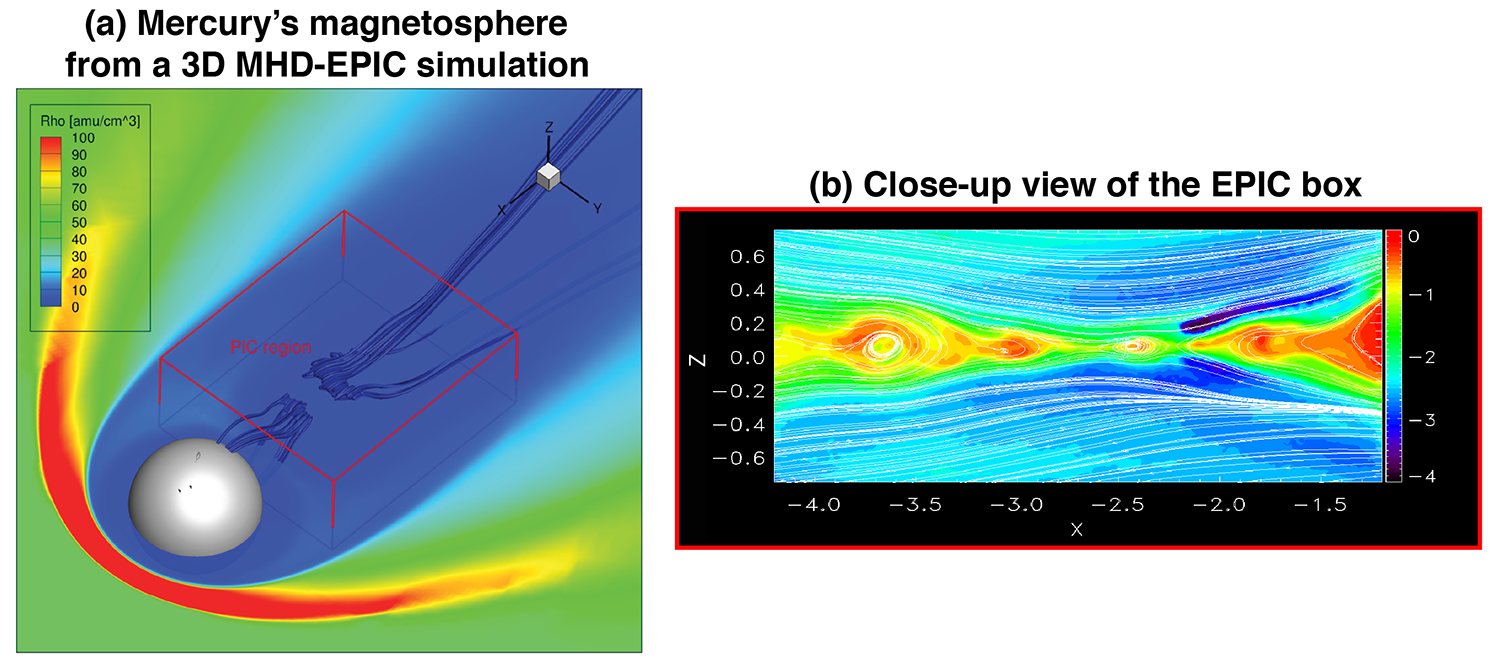}
\caption{Results from a 3D MHD-EPIC simulation of Mercury’s magnetosphere.
(a) 3D view of the simulated magnetosphere. Colors in the equatorial plane represent plasma density (in $amu/cm^3$), whereas solid lines indicate field lines. The red box shows the embedded PIC region in the magnetotail. (b) A close-up view of the PIC solution showing contours of electron pressure (in $Log_{10}$ nPa) and field lines in the noon-midnight meridian.\cite[from][]{Chen:2019b}}
\label{fig:Chen2019}
\end{figure}

As established by numerous MESSENGER observations, the dynamics of Mercury's magnetosphere is driven predominantly by magnetic reconnection, largely due to the IMF and solar wind conditions typically present in the inner heliosphere. Recognizing the importance of reconnection at Mercury, we have adapted the MHD-EPIC model to simulate Mercury's magnetosphere so that reconnection can be treated using a physics-based model \cite[]{Chen:2019b}, rather than through numerical or ad hoc resistivity as done in MHD models. \figurename~\ref{fig:Chen2019} shows the results from the MHD-EPIC Mercury simulation by \cite{Chen:2019b}, where a PIC box is embedded in the magnetotail to study tail dynamics and various asymmetries as observed by MESSENGER. While the upstream solar wind is held steady during this simulation, the magnetotail as simulated by the PIC code exhibits very dynamic behavior. A series of plasmoids of varying size form in the tail, and both tailward-moving and planetward-moving plasmoids are found in the simulation (\figurename~\ref{fig:Chen2019}b). The modeled plasmoids in the near-Mercury tail ($|X|$= 2 – 3$\rm R_{M}$) have an average size of ~ 0.3$\rm R_{M}$ in diameter, which is in accordance with the estimate based on MESSENGER observations in this region \cite[]{Sun:2020c}. 

While the imposed solar wind flow is symmetric about the Sun-Mercury line, due to kinetic effects significant dawn-dusk asymmetries develop in the magnetotail in the MHD-EPIC simulation, such as a thicker tail current sheet and higher plasma density and pressure on the dawn side. The occurrence and onset location of tail reconnection in the MHD-EPIC simulation also exhibit a strong preference for the dawn side, such that almost all dipolarization fronts and high-speed plasma flows arising from tail reconnection concentrate in the dawn sector. These simulation predictions are remarkably consistent with MESSENGER observations\cite[\eg][]{Sun:2016a, Poh:2017a}. The Mercury simulation presented here along with other published applications \cite[\eg][]{Toth:2016a, Chen:2017a, Zhou:2019b, Zhou:2020b} have demonstrated that MHD-EPIC is capable of capturing both local and global physics of the magnetosphere, and, therefore, provides an excellent tool for studying solar wind-magnetosphere coupling in planetary magnetospheres and for interpreting observations from spacecraft missions.

\section{Current and Future Directions}
\label{sec:directions}

\subsection{Machine Learning}
\label{subsec:ml}
The emergence of computational space physics at the turn of the 21st century was made possible by a collaboration between space physicists, applied mathematicians, computer and computational scientists.  At the beginning of the 2020s, we are again witnessing a scientific revolution, similarly to the one a quarter century ago. A new scientific discipline is emerging to offer unprecedented opportunities for advancing many research fields, including space weather: Artificial Intelligence (AI), including Machine Learning (ML) and Computer Vision (CV), can help computers ``learn'' how to find a needle in the haystack, and help us identify new connections between seemingly unrelated phenomena. Moreover, AI can greatly improve the assimilation of observations into computational models and to quantify the uncertainty of model results. The new challenge is how to integrate this new methodology with our existing space weather modeling capabilities.

A number of user-friendly and free libraries make it possible to apply  ML tools to a large variety of problems. Software such as TensorFlow from Google \cite[]{tensorflow:2015a}, AWS from Amazon \cite[]{Herrero:2011a_AWS}, Azure from Microsoft \cite[]{Dudley:1010a_azure}, or PyTorch from Facebook's AI Research Lab \cite[]{Paszke:2019a_pytorch} have allowed a proliferation of ML applications to space physics problems. 
A couple of years ago the SWMF team reached out to data science, machine learning and computer vision experts at the University of Michigan and we forged a new collaboration to bring advanced data science methods to space weather modeling. In this section we show two specific examples that demonstrate the potential of these new approaches.

\subsubsection{Total Electron Content (TEC) Maps}
\label{subsubsec:tec}
\begin{figure}[tbh]
    \floatbox[{\capbeside
        \thisfloatsetup{capbesideposition={left,top},
        capbesidewidth=0.35\textwidth}}]{figure}[\FBwidth]
    {\caption{Global TEC maps with 6‐hr interval under storm (13 October) conditions in 2016. The predicted TEC maps are from our LSTM/NN and corresponding ones from the CODE GIM, along with their differences, are, respectively, given in the left, middle, and right panels of each figure. The predictions shown are 1-hour ahead. The unit of the color contour is TECU ($10^16 \text{electrons/m}^2$). \cite[from][]{Liu:2020a}.
    \label{fig:TEC}}}
    {\includegraphics[width=0.6\textwidth]{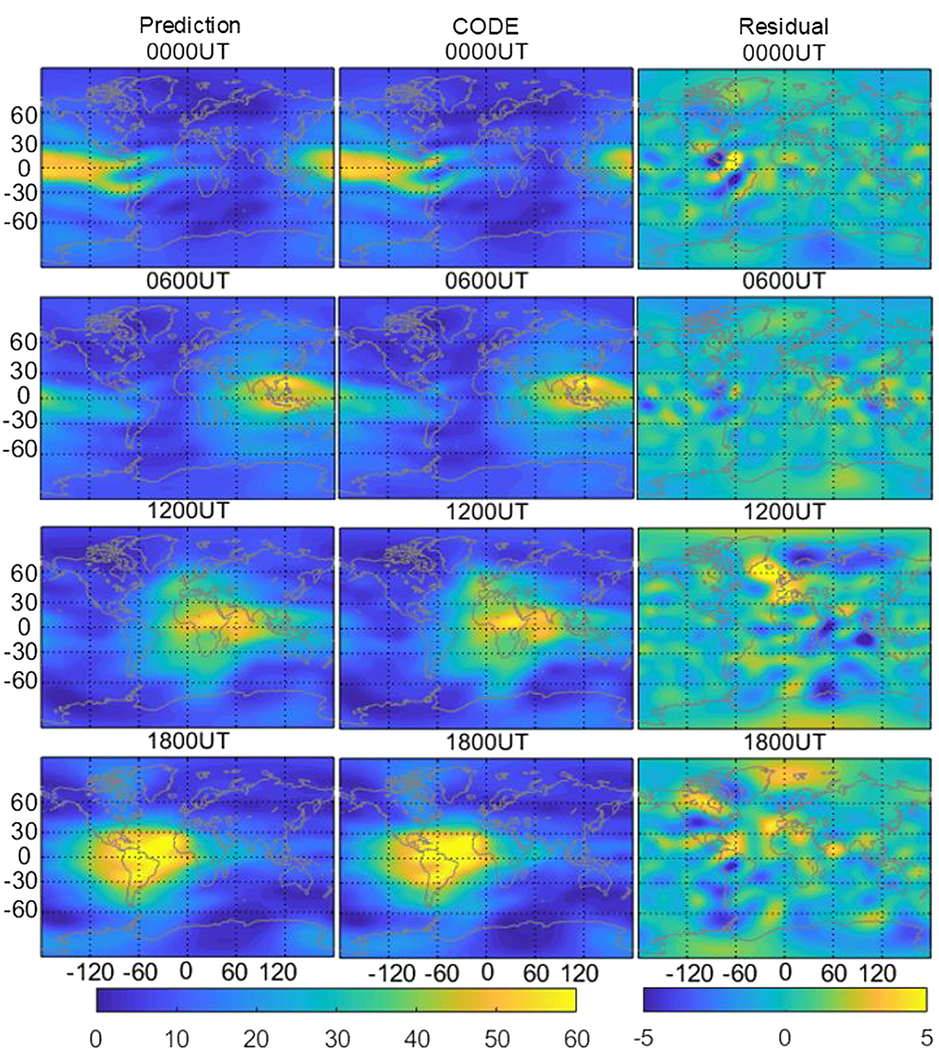}}
\end{figure}

The ionospheric total electron content (TEC) is arguably the most utilized physical parameter in ionospheric research in the GNSS era. The TEC maps provide us information about the ionospheric density structures and their evolution, and are also of practical importance since they can be used to estimate the GNSS signal delay due to the ionospheric plasma content between a receiver and a GNSS satellite. Recently, we applied advanced machine learning methods to the forecast of global ionospheric total electron content (TEC) maps (GIM) using maps from one of the International GNSS Services (IGS) centers, i.e., the Center for Orbit Determination in Europe (CODE). Spherical harmonic (SH) fitting is often used in constructing the GIM map. We applied an LSTM neural network method \cite[LSTM/NN,][]{Hochreiter:1997a} to forecast the 256 SH coefficients, which are then used to construct the GIM maps \cite[]{Liu:2020a}. The model results show that the first/second hour TEC root mean square error (RMSE) is 1.27/2.20 TECU during storm time and 0.86/1.51 TECU during quiet time. Comparing with the CODE GIMS, the RMSE of the LSTM prediction is 1.06/1.84 TECU for the 1st /2nd hour, while the RMSE errors from the IRI-2016 and NeQuick-2 models are around 9.21/5.5 TECU, respectively \cite[]{Liu:2020a}. Moreover, typical large-scale ionospheric structures, such as equatorial ionization anomaly and storm‐enhanced density are well reproduced in the predicted TEC maps during storm time. The ML model performs well in predicting global TEC when compared to two empirical models (IRI‐2016 and NeQuick‐2, see Figure \ref{fig:TEC}). 
        
The IGS GIM maps are constructed based on a few hundred IGS stations with limited spatial resolution. Critical meso-scale ionospheric structures that cause the most severe GNSS scintillations, i.e., the equatorial plasma bubbles, are smoothed out during the SH fitting procedure. Therefore, we also developed an innovative GIM construction method called VISTA (Video Imputation with SoftImpute, Temporal smoothing and Auxiliary data) \cite[]{Sun:2021aa}. Several extensions of existing matrix completion algorithms have been utilized to achieve TEC map reconstruction, accounting for spatial smoothness and temporal consistency while preserving important multi-scale structures of the TEC maps. This method utilizes the Madrigal TEC data based on over 5000 GNSS receivers and is able to overcome large missing data over the oceans. This newly proposed algorithm is targeted but not restricted to the temporal TEC map reconstruction. In the future we will use VISTA to process all historical Madrigal TEC data and then use the fully reconstructed maps to perform TEC forecast and data assimilation. 

\subsubsection{Neural Network Predictions of Solar Flares}
\label{subsubsec:flares}
Forecasting large solar flares with machine learning (ML) is at the heart of space weather prediction: Flare radiation and any information about the occurrence of a solar eruption event is carried at the speed of light, hence true forecasting is required. The first significant increase of solar energetic particle fluxes can take place within an hour after the flare. The US national space weather forecasting goal is to provide physics-based hourly space weather forecasts for validity periods of up to 120 hours. This goal can only be achieved if we can forecast solar flares and CMEs.

\begin{figure}[htb]
    \floatbox[{\capbeside
        \thisfloatsetup{capbesideposition={left,top},
        capbesidewidth=0.3\textwidth}}]{figure}[\FBwidth]
    {\caption{The ML strong flare (M/X class) probability index exhibits a dramatic increase about a day before the strong flare occurs on the Sun \cite[]{Chen:2019a_ML, Wang:2020a}. The ML methodology also forecasts the GOES x-ray peak intensity of the first strong flare \cite[]{Jiao:2020a}.}
    \label{fig:flarepredict}}
    {\includegraphics[width=0.675\textwidth]{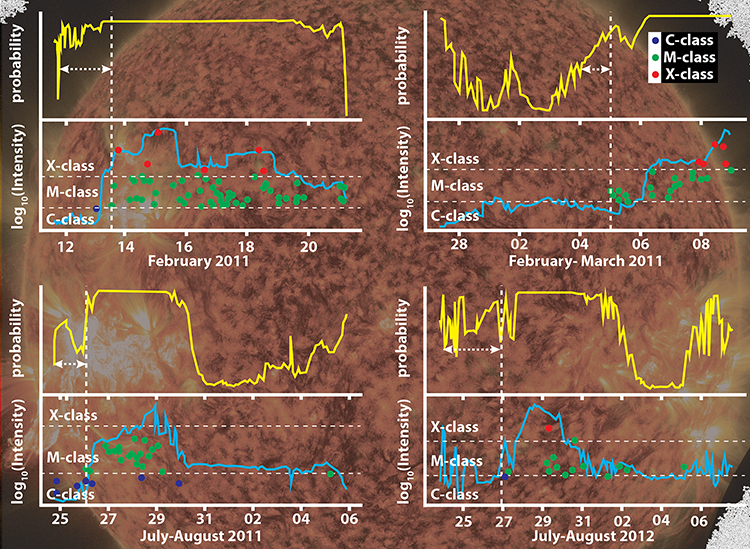}}
\end{figure}

Our approach is to break down the flare forecasting problem into a series of increasingly challenging ML/computer vision steps. First, we applied cutting-edge classification methodology to obtain a flare probability index that jumps from small values ($<0.3$) to near unity about one day before a large (M/X class) solar flare takes place \cite[]{Chen:2019a_ML} (see the yellow curves in \figurename~\ref{fig:flarepredict}). The ML algorithm for the flare probability index utilizes {\it SDO}/HMI-based SHARP parameters, physically-insightful summaries of active region photospheric magnetic fields. Thus, while this ML-based approach adds no new physics, it represents SOLSTICE's first step toward improving event forecasts with cross-disciplinary efforts.

Second, we developed an innovative mixed LSTM regression model, which combines binary classification of flaring and LSTM regression for flare intensity, that was used to predict the flare onset jointly with the maximum solar flare intensity observed by the GOES satellites within a 24-hr time window \cite[]{Jiao:2020a}. The predicted intensity peaks are shown by light blue curves in \figurename~\ref{fig:flarepredict}. While the results are quite encouraging, the method needs to be trained on larger data sets and improved in terms of computational efficiency and model interpretability. 

\subsection{Data Digestion and Assimilation, Ensemble Modeling and Uncertainty Quantification}
\label{subsec:data}

Physics-based modeling, while very successful in many applications, has several inherent limitations. Often the initial and boundary conditions are not fully known, so the equations to solve are not fully specified. The computational cost of the model may become prohibitive for high fidelity and/or well-resolved models. Here we describe a few of the challenges and future directions.

Solar simulations are driven by the boundary conditions applied near the surface of the Sun. While we have reasonable estimates of density, temperature and radial magnetic field at a large part of the surface, the other plasma quantities are not well understood. The line-of-sight velocity, for example, can be measured with Doppler shifts, but the other components are not known. The radial magnetic field on the back side of the Sun and near the poles is not well observed in general. The transverse components of the magnetic field can be obtained from vector magnetograms, however, there is an ambiguity for the sign of these components, and the observational errors in the transverse components are very significant. Using the ``observed'' magnetic field vector as a direct boundary condition for the AWSoM results in substantial unphysical flows due to the errors in the magnetic field. We are currently working on improving the data ingestion algorithm for the vector magnetogram information so that we will, hopefully, be able to initiate CMEs directly from the evolving magnetogram instead of inserting flux ropes, which is our current method.

Another way of using data in physics-based models is data assimilation. Data assimilation has the potential to significantly improve model performance, as it has been successfully done in terrestrial weather forecasting. To allow for the sparsity of observations of the Sun-Earth system, however, a different data assimilation method needs to be employed than the typical ensemble Kalman filter used in terrestrial forecasting. 
Presently we are developing a model that combines physics based modeling with data assimilation and uncertainty quantification (UQ). The model will start from the Sun with an ensemble of simulations that span the uncertain observational and model parameters based on a comprehensive UQ analysis. 
At the end the model will provide a probabilistic forecast of the space weather impacts. While the concept is simple, finding the optimal algorithm that produces the best prediction with minimal uncertainty is a complex and very challenging task that requires developing, implementing and perfecting novel data assimilation and uncertainty quantification methods. 

\subsection{Open-source Development}
\label{subsec:open}
Earlier in this paper we described how a large interdisciplinary group of researchers have developed, with sustained effort, the first-principles SWMF that is capable of modeling and forecasting space weather and other space physics phenomena. To make the SWMF impactful, it needs to be used by the space physics community. 

 From the beginning, we have made the SWMF \cite[]{Toth:2005swmf, Toth:2012a} available to the whole community via a user license. Users can obtain the full source code at \url{http://csem.engin.umich.edu/tools/swmf} with all scripts and documentation and use it for their research with minimal restrictions. In addition, BATS-R-US and the SWMF have been available for runs-on-request through the Community Coordinated Modeling Center (CCMC) at NASA Goddard through \url{https://ccmc.gsfc.nasa.gov/index.php}. CCMC made the SWMF, and many other models, accessible to a wide user community who may not have access to large computer resources and/or are not expert users who can configure and run a complex model.  

Some parts of the SWMF and related software have been truly open-source for a while. The Global Ionosphere Thermosphere Model (GITM) has been open-source at \url{https://github.com/~aaronjridley/} since 2012. The Space Science library for Python (spacepy), available from \url{https://github.com/spacepy/spacepy} since 2012, has become one of the best free visualization and analysis tools for the SWMF output. More recently, VisAnaMatlab at \url{https://github.com/henry2004y/VisAnaMatlab} and VisAnaJulia at \url{https://github.com/henry2004y/VisAnaJulia}, visualization and analysis tools written by Hongyang Zhou for the Matlab and Julia languages, respectively, have been made open-source too. 

Finally, the core SWMF model was also released in 2020 under a non-commercial open-source license at \url{https://github.com/mstem-quda}. MSTEM-QUDA contains the full core of the SWMF and the BATS-R-US, RIM, RCM, and RBE models. In addition, it contains a new Python library, swmfpy. We expect to add the CIMI inner magnetosphere model in the near future. The MSTEM-QUDA repository is an up-to-date mirror of the repositories developed at Michigan. Both the master and stable branches are available. 

Making a major part of the SWMF truly open-source opens a new era in the use and development of space physics and space weather model development. We are hopeful that it will lead to more use, faster and more reliable model development and productive collaboration in the community.

\section{Concluding Remarks}
\label{sec:conclude}
Over the last decades most scientific disciplines have undergone a major revolution, and the science behind space weather is no exception. A few decades ago, observations and theory were the two pillars of scientific discovery. Since then, the explosive advancements in computer hardware, software, numerical algorithms and data assimilation methods have made computational space physics a third pillar of space weather science.

The emergence of computational space physics was made possible by a close collaboration between space physicists, applied mathematicians, computer and computational scientists. But the formation of tightly integrated research efforts did not happen overnight: It takes years to educate researchers from diverse disciplines to understand each other's terminology, basic concepts, and methodology well enough to create a breakthrough product.

The SWMF development was started in the 1990s. Over the last three decades funding agencies and the University of Michigan invested over \$50M and about 200 person-year effort in this project. Maintaining and developing a world-class modeling framework requires collaboration of space scientists, mathematicians, numerical and computer scientists, and the space weather user community. Such large research environments take time to build and require continued investments both in intellectual and computational capabilities.

In this paper we described the present state of the SWMF and its main component, BATS-R-US. We also outlined its history and future directions. Today, SWMF, BATS-R-US and all their simulation and analysis tools constitute a cutting edge capability that is available to the space weather community.

\addcontentsline{toc}{section}{Acknowledgements}

\begin{acknowledgements}
		This work was supported by NASA DRIVE Science Center grant 80NSSC20K0600 and NASA MMS grant 80NSSC19K0564, NASA LWS grants 80NSSC20K0185, 80NSSC20K0190, 80NSSC20K1778 and 80NSSC17K0681, NASA SSW grant 80NSSC20K0854, NASA HSR 80NSSC20K1313, NASA 80NSSC21K0047, the NSF PRE-EVENTS grant 1663800 and NSF SWQU grant PHY-2027555. The authors thank Drs.\ John Dorelli and Natalia Buzulukova for providing their unpublished result to be shown in the paper (\figurename~\ref{fig:hires}).	The authors also thank Ms.\ Deborah Eddy for making the paper more readable.	
\end{acknowledgements}

	\bibliography{tig-new}

\appendix

\section{Fundamentals of BATS-R-US}
\label{sec:creation}

The BATS-R-US code was originally developed in the mid 1990s when there was a major national initiative to utilize the new transition from vector machines to massively parallel architectures. There were three principles guiding this development: (i) apply the latest advances in computational fluid dynamics to MHD, (ii) utilize the emerging adaptive mesh refinement (AMR) technology and (iii) create a data structure that is truly scalable to a very large number of CPU cores.

The emergence of computational space physics at the turn of the 21st century was made possible by a close collaboration between space physicists, applied mathematicians, computer and computational scientists. But the formation of tightly integrated research efforts did not happen overnight: It takes years to educate researchers from diverse disciplines to understand each other's terminology, basic concepts, and methodology well enough to create a breakthrough product.

\subsection{8-Wave Riemann Solver}
\label{subsec:8wave}
The first step of the BATS-R-US development was to attack a fundamental roadblock to high-resolution magnetohydrodynamics (MHD) simulations. High-resolution schemes are based on the \textit{conservative form} of the governing equations:

\begin{equation}
\frac{\partial\overline{\bU}}{\partial t} + \left(\divg\bar{\bar{\bF}}\right)^{T}=\overline{\bS}
\label{eq:DivForm}
\end{equation}

\noindent where $\overline{\bU}$ is the vector of conserved quantities defined by

\begin{equation}
\overline{\bU} = \left(\rho, \rho u_x, \rho u_y, \rho u_z, B_x, B_y, B_z, \varepsilon \right)^T
\label{eq:conserved}
\end{equation}

\noindent where $\rho$ is mass density, $u_x, u_y$ and $u_z$ are the three components of the plasma bulk flow velocity vector, $\overline{\bu}$, while $\overline{\bB}=\{B_x,B_y,B_z\}$ is the magnetic field vector and $\varepsilon$ is the total energy density

\begin{equation}
\varepsilon = \varepsilon_{hd} + \frac{\overline{\bB} \cdot \overline{\bB}}{2\muo} = \frac{p}{\gamma -1} + \rho \frac{\overline{\bu} \cdot \overline{\bu}}{2} + \frac{\overline{\bB} \cdot \overline{\bB}}{2\muo} 
\label{eq:energy}
\end{equation}

\noindent Here $\varepsilon_{hd}$ denotes the hydrodynamic energy density, while $\gamma$ is the specific heat ratio and $\muo$ is the permeability of vacuum. The flux tensor, $\bar{\bar{\bF}}$, can be written as

\begin{equation}
    \renewcommand{\arraystretch}{1.5}
\bar{\bar{\bF}} = \left( \begin{array}{c}
\rho \overline{\bu}  \\
\rho \overline{\bu} \overline{\bu} + \left( p + \frac{\overline{\bB} \cdot \overline{\bB}}{2\muo}\right) \unitI - \frac{\overline{\bB} \overline{\bB}}{\muo} \\
\overline{\bu} \overline{\bB} - \overline{\bB} \overline{\bu} \\
\overline{\bu} \left(\varepsilon + p + \frac{\overline{\bB}\cdot \overline{\bB}}{2\muo} \right) - \frac{\left(\overline{\bu} \cdot \overline{\bB} \right) \overline{\bB}}{\muo}
\end{array} \right)^{T}
\label{eq:flux}
\end{equation}

\noindent Finally, $\overline{\bS}$ is a ``source'' vector, containing the terms that cannot be expressed in divergence form:

\begin{equation}
\overline{\bS} = -\divB
    \left( \begin{array}{c} 
    0 \\ \overline{\bB} \\ \overline{\bu} \\ 
    \overline{\bu} \cdot \overline{\bB} 
    \end{array} \right)
\label{eq:source}
\end{equation}

The ``source term'' given by \equationname~(\ref{eq:source}) can be handled two different ways. One can directly apply Maxwell's equation that expresses the absence of monopoles resulting in a $\overline{\bS}\equiv0$ identity. However, setting $\overline{\bS}$ to zero results in a degenerate eigensystem for \equationname~(\ref{eq:DivForm}) \cite[\cf][]{Roe:1996b}. Due to this degeneration even advanced MHD codes solve only the hydrodynamic part of the MHD equations with high-resolution methods and advance the magnetic field separately \cite[\cf][]{Clarke:2010a, Lyon:2004a, Zhang:2019b}. In a groundbreaking paper, \cite{Powell:1997a} proposed a Riemann solver that formally keeps the $\divB$ term in \equationname~(\ref{eq:source}) and makes it a passively convected quantity. This method resolves the degeneracy of the eigensystem of \equationname~(\ref{eq:DivForm}) and results in an 8th wave that carries information about the discontinuity in the normal component of the magnetic field. This so-called \textit{8-wave scheme} ensures the solenoidity condition to truncation accuracy \cite[]{Powell:1997a} and it makes it possible to formulate the MHD problem in a way that makes it suitable for high-resolution schemes. \cite{Toth:2000a} carefully evaluated the various methods that constrain $\divB$ and concluded that the 8-wave approach performs as well as the alternative methods generally applied by the computational MHD community.

The complete numerical algorithm solving the MHD equations with the 8-wave scheme was published by \cite{Powell:1999a}. It gives a detailed description of the full eigensystem of the MHD equations together with a space physics example. The publication of the \cite{Powell:1999a} paper created an avalanche of negative reactions from the space physics MHD modeling community, once again proving George Barnard Shaw's sarcastic comment: ``All great truths begin as blasphemies.'' The criticism culminated in a paper by \cite{Raeder:2000a} that tried to discredit the 8-wave scheme. The subsequent comment and reply exchange \cite[]{Gombosi:2000comment, Raeder:2000a} stopped the open criticism, but the underlying skepticism from some competitors still lingers even today.

\subsection{Adaptive Mesh Refinement}
\label{subsec:amr}
BATS-R-US uses a simple and effective block-based adaptive mesh refinement (AMR) technique \cite[]{Stout:1997a}. The approach closely follows that first developed for two-dimensional gas dynamics calculations by \cite{Berger:1985a, Berger:1989a}. This block-based tree data structure is advantageous for several reasons. One of the primary advantages is the ease with which the grid can be adapted. If, at some point in the calculation, a particular region of the flow is deemed to be sufficiently interesting, better resolution of that region can be attained by refining a block, and inserting the eight finer blocks that result from this refinement into the data structure. Removing refinement in a region is equally easy. Decisions as to where to refine and coarsen are made based on either geometric considerations or on comparison of local flow quantities to threshold values.

The governing equations are integrated to obtain volume-averaged solution quantities within computational cells. The computational cells are embedded in regular structured blocks of equal-sized cells. The blocks are geometrically self-similar. Solution data associated with each block are stored in standard indexed array data structures, making it straightforward to obtain solution information from neighboring cells within a block. Note that the data on each block can be associated with any one of a multitude of coordinate systems including Cartesian, curvilinear, and more.  

\begin{figure}
  \begin{floatrow}
    \ffigbox[\FBwidth]
      {\includegraphics[width=0.9\linewidth]{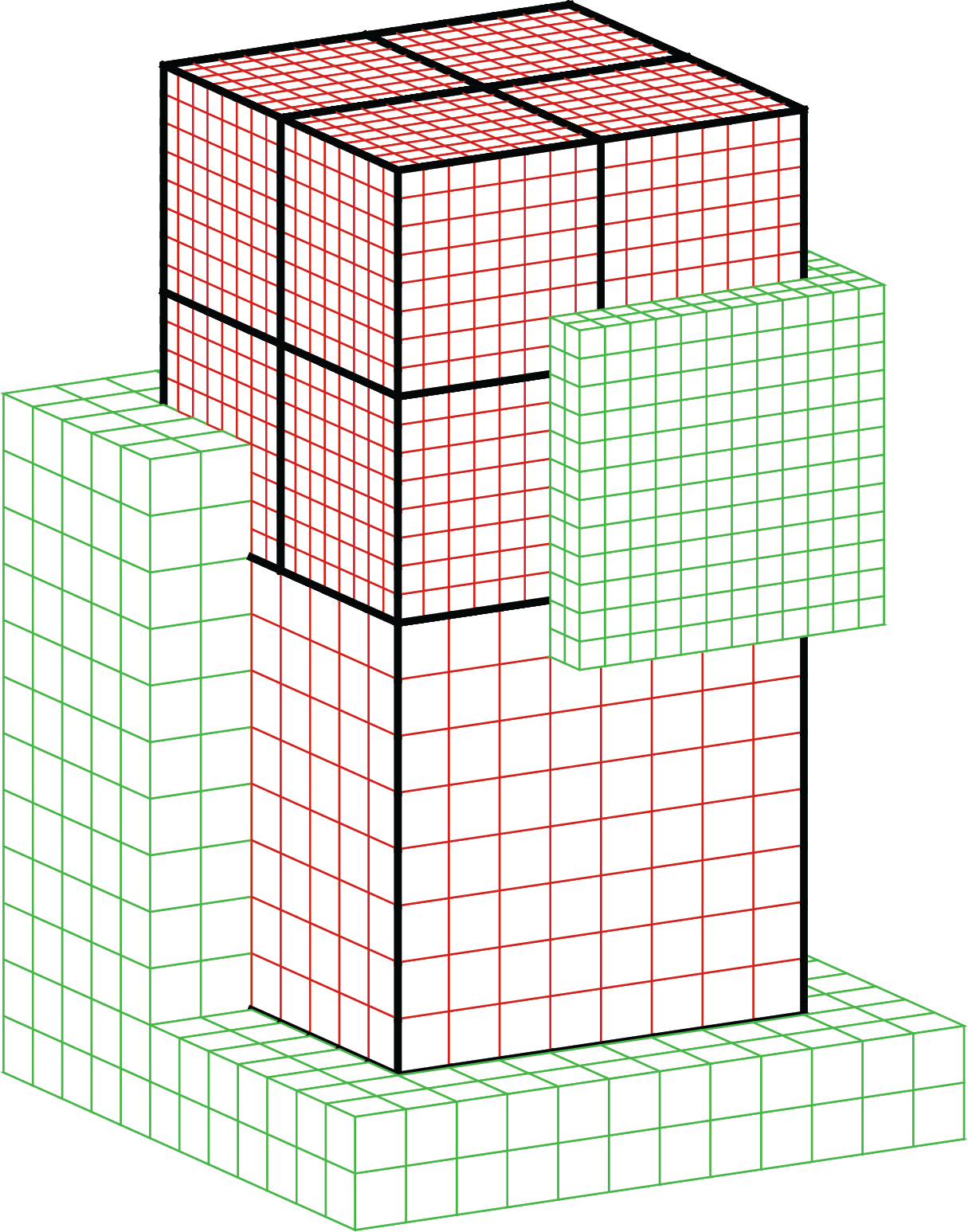}}
      {\caption{Self-similar blocks illustrating the double layer of ghost cells for both coarse and fine blocks. (from \cite{Gombosi:2003a})}
      \label{fig:amr-blocks}}
    \ffigbox[\Xhsize]
      {\includegraphics[width=0.75\linewidth]{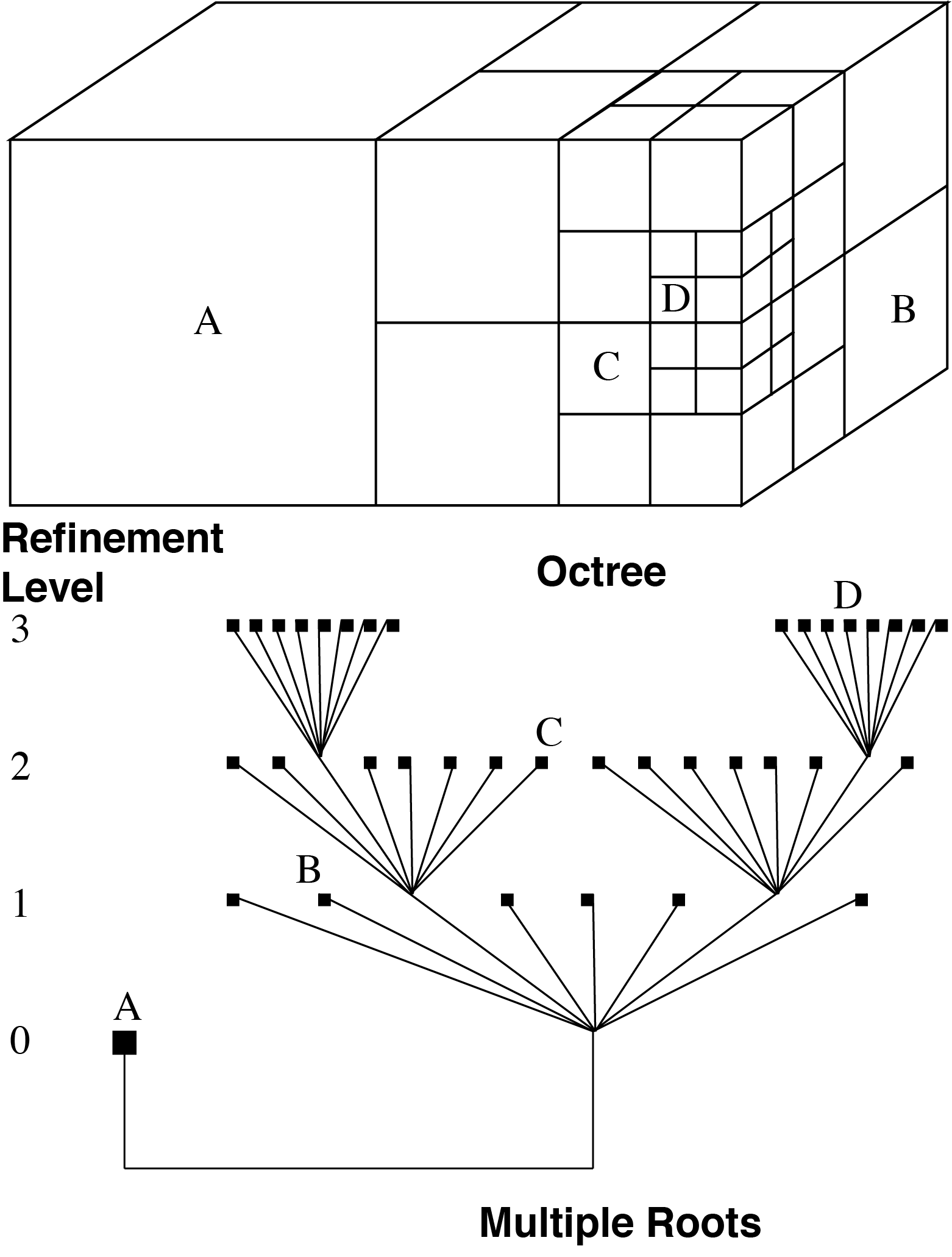}}
      {\caption{Solution blocks of the BATS-R-US computational mesh with three refinement levels.  (from \cite{Gombosi:2003a})}
      \label{fig:octree}}
    \end{floatrow}
\end{figure}

Computational grids are composed of many self-similar blocks. Although each block within a grid has the same data-storage requirements, blocks can be of different sizes in terms of the volume of physical space they occupy. Starting with an initial mesh consisting of blocks of equal size (that is, uniform resolution), spatial adaptation is performed by dividing and coarsening appropriate solution blocks. In regions requiring increased cell resolution, a parent block is refined by dividing itself into eight children, or offspring. Each of the eight octants of a parent block becomes a new block with the same number of cells as the parent, which doubles cell resolution in the region of interest. Conversely, in over-resolved regions, the refinement process reverses; eight children coarsen and coalesce into a single parent block. Thus, cell resolution reduces by a factor of 2. Multigrid-type restriction and prolongation operators are used to evaluate the solution on all blocks created by the coarsening and division processes, respectively.

When a 3D block is refined, it is split into eight octants (see \figurename~\ref{fig:amr-blocks}). Each octant forms a block with the same number of cells as the original block, but the resolution is increased by a factor of two. The resulting grid structure is an octree of blocks, and the equations are solved at the finest level only, i.e. on the leaves of the tree (see \figurename~\ref{fig:octree}). 

The hierarchical data structure and self-similar blocks simplify domain decomposition and enable good load-balancing, a crucial element for truly scalable computing. For explicit time stepping (all blocks use the same time-step) natural load-balancing occurs by distributing the blocks equally among the processors. For more complicated time-stepping schemes the load-balancing is more challenging, but it still can be done on a block by block basis. We achieve additional optimization by ordering the blocks using the Peano-Hilbert or Morton space-filling curves to minimize inter-processor communication. The self-similar nature of the solution blocks also means that serial performance enhancements apply to all blocks and that fine-grained algorithm parallelization is possible. The algorithm's parallel implementation is so pervasive that even the grid adaptation performs in parallel.

\subsection{BATS-R-US Performance}
\label{subsec:performance}
For most computational models that involve the solution of partial differential equations (PDEs), domain decomposition (i.e., partitioning the problem by dividing the computational domain into subdomains, and farming the subdomains off onto separate cores) is a natural and, in many cases, the most practical approach to parallelization. The block-based AMR solver was designed from the ground up with a view to achieving very high performance on massively parallel architectures \cite[]{Stout:1997a}. The hierarchical data structure and self-similar blocks make domain decomposition of the problem almost trivial and readily enable good load-balancing, a crucial element for truly scalable computing.  A natural load balancing is accomplished by simply distributing the blocks equally amongst the processors. The parallel implementation of the algorithm has been carried out to such an extent that even the grid adaptation is performed in parallel.

\begin{figure}[ht]
    \floatbox[{\capbeside
        \thisfloatsetup{capbesideposition={left,top},
        capbesidewidth=0.4\textwidth}}]{figure}[\FBwidth]
    {\caption{The cell update rates as a function of number of cores for the BAT-S-R-US model. The problem size scales in proportion to the number of parallel processes. The dotted lines represent linear scaling.}
    \label{fig:bats-scale}}
    {\includegraphics[width=0.5\textwidth]{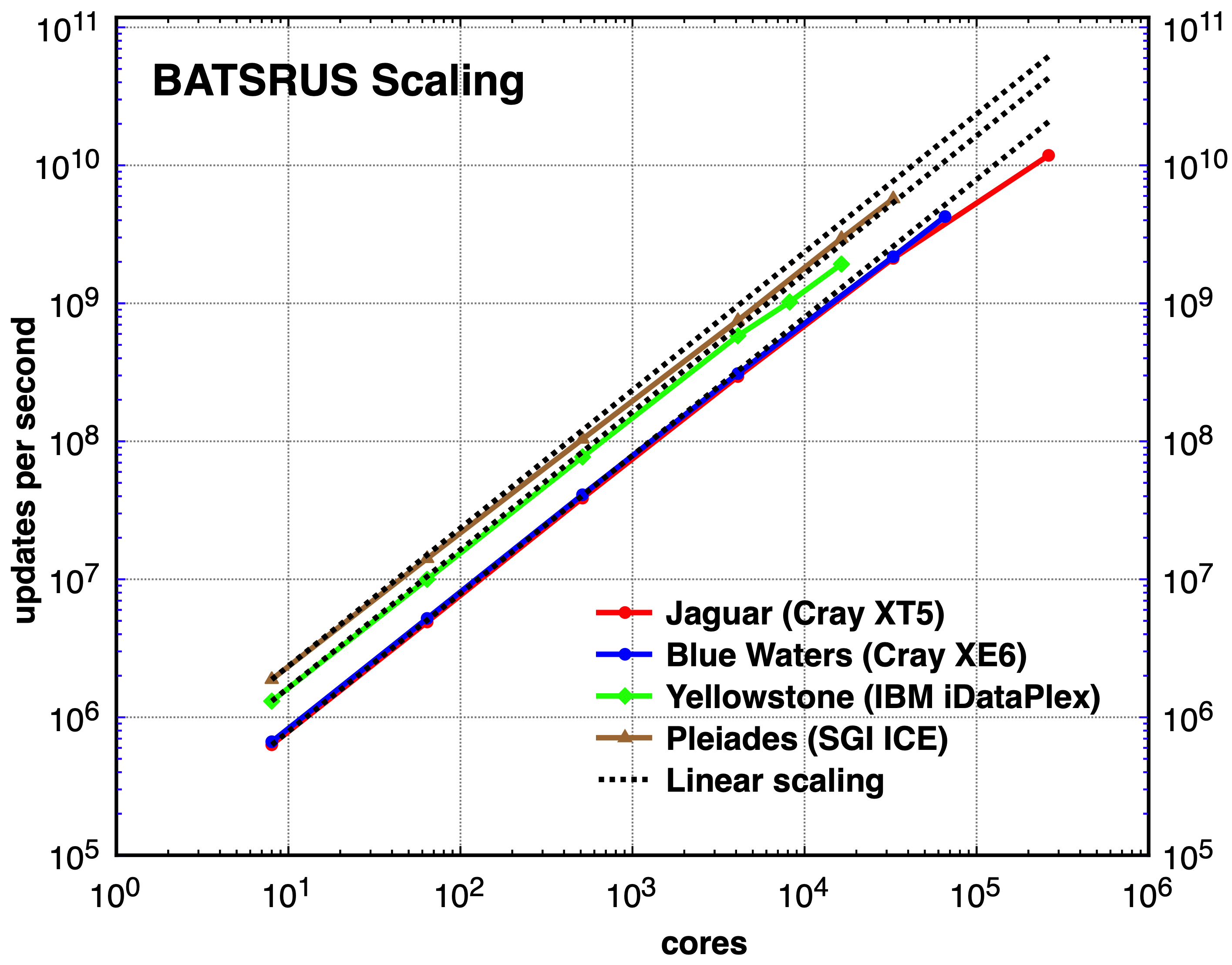}}
\end{figure}

\figurename~\ref{fig:bats-scale} shows the weak scaling (how the solution time varies with the number of processors for a fixed problem size per processor) of BATS-R-US on several supercomputers from 8 up to more than 200,000 cores. 
Recently, BATS-R-US has been further developed to use hybrid MPI and OpenMP parallelism that allows scaling to beyond 500,000 cores \cite[]{Zhou:2020a}.

\section{SWMF}
\label{sec:swmf2005}
The Space Weather Modeling Framework (SWMF) was developed in the early 2000s \cite[]{Toth:2005swmf} with a combination of support from the DoD MURI (Multidisciplinary University Research Initiatives) and NASA Earth and Space Sciences HPCC (High Performance Computing and Communications) programs. This development closely followed the development path of BATS-R-US: its intellectual leadership came from a tightly integrated team of senior university faculty in computer science, software engineering, applied mathematics and space plasma physics, while the actual development was carried out by a group of early-to-mid career scientists with help from postdocs and graduate students. The availability of significant stable funding for over a decade was a critical element of the success of the SWMF development (the funding came just as the BATS-R-US development resources were winding down).

\subsection{Structure}
\label{subsec:swmfstructure}
\begin{figure}[ht]
    \floatbox[{\capbeside
        \thisfloatsetup{capbesideposition={left,top},
        capbesidewidth=0.35\textwidth}}]
            {figure}[\FBwidth]
    {\caption{The original physics modules of the SWMF \cite[]{Toth:2005swmf}. This was the first successful coupling \cite[]{DeZeeuw:2004a} of a gyrokinetic ring current model \cite[]{Harel:1981a, Wolf:1991a, Toffoletto:2003a, Sazykin:2002a} to a global MHD model describing the magnetosphere.}
    \label{fig:swmf2005}}
    {\includegraphics[width=0.6\textwidth]{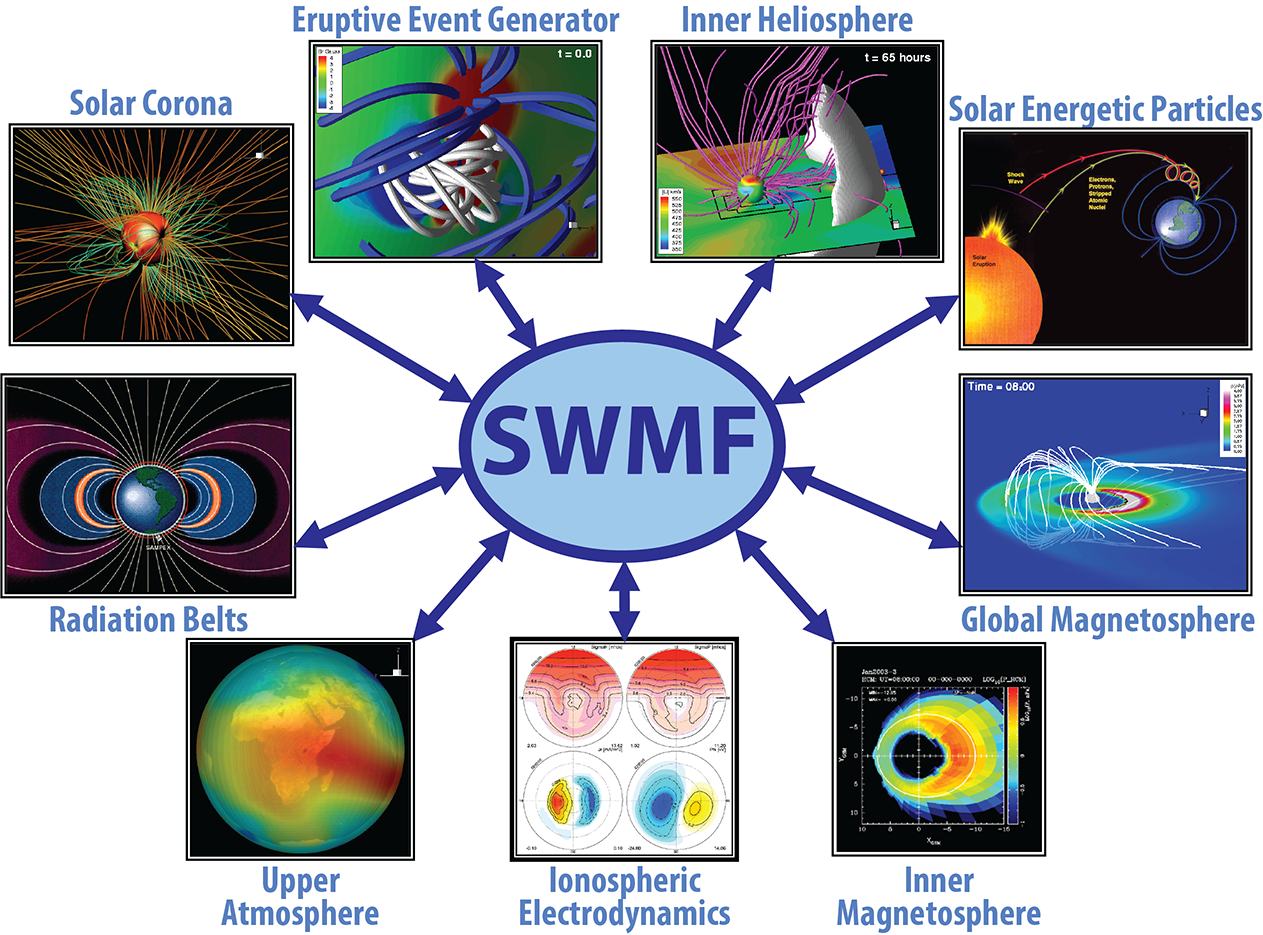}}
\end{figure}

\figurename~\ref{fig:swmf2005} shows the structure of the SWMF. There are over a dozen components or physics domains. In an actual simulation one can use any meaningful subset of the components. 

If the simulation starts from the Sun, it is typically driven by solar magnetogram data and flare/CME observations. Simulations restricted to magnetospheric components are usually driven by the solar wind data obtained by satellites upstream of the Earth, for example ACE, Wind or Geotail. We also use the F10.7 solar flux for some of the empirical relationships in the ionosphere and thermosphere models.

\begin{figure}[ht]
    \floatbox[{\capbeside
        \thisfloatsetup{capbesideposition={left,top},
        capbesidewidth=0.4\textwidth}}]
            {figure}[\FBwidth]
    {\caption{The layered architecture of the SWMF. (from \cite{Toth:2012a})}
    \label{fig:swmf-layer}}
    {\includegraphics[width=0.4\textwidth]{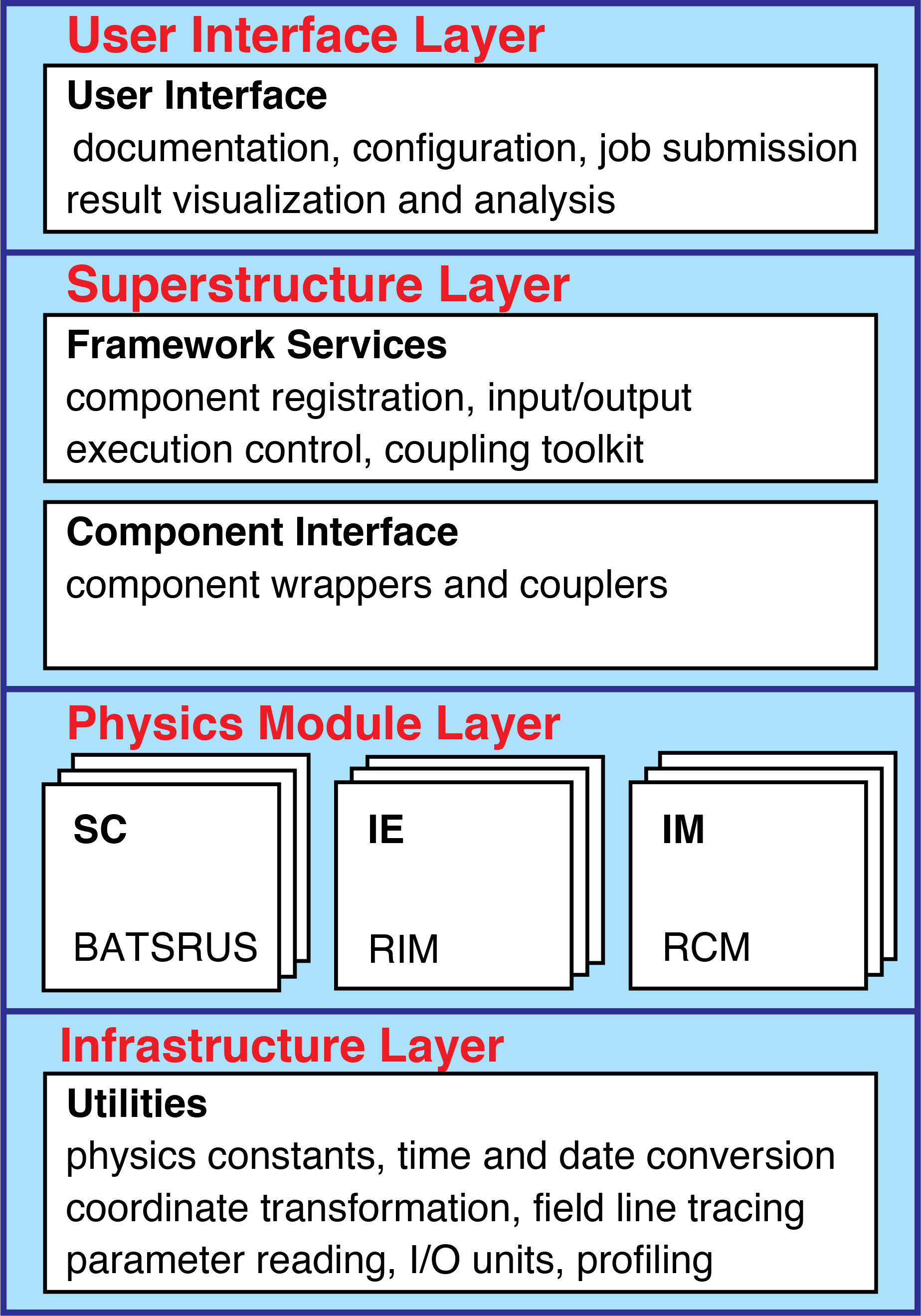}}
\end{figure}

The SWMF has a layered architecture (see \figurename~\ref{fig:swmf-layer}). The top layer is the user interface. The second layer contains the control module, which is responsible for distributing the active components over the parallel machine, executing the models, and coupling them at the specified frequencies. The third layer contains the physics domain components. Each component can have multiple physics models. Each component version consists of a physics model with a wrapper and one or more couplers. The wrapper is an interface with the control module, while each coupler is an interface with another component. The physics models can also be compiled into stand-alone executables. The fourth and lowest layer contains the shared library and the utilities that can be used by the physics models as well as by the SWMF core.

\subsection{Couplers}
\label{subsec:swmfcouplers}
The SWMF couples together the various models at regular intervals, based on either simulation time or iteration number. The relevant physical quantities are passed with efficient MPI communication. In addition to transferring the data, SWMF has to transform between coordinate systems, take care of unit conversions, and interpolate between different grids. Often the models are moving or rotating relative to each other so that the mapping has to be recalculated every coupling time. A further complication arises for adaptive grids that may change between two couplings. SWMF includes utilities to take care of coordinate transformations and interpolation between various grids.

Since the models use widely different grids and time steps, coupling through a simple interface may be very challenging, especially when the flow is slower than the fast magnetosonic speed. A possible solution is to overlap the models. For example the inner boundary of the inner heliosphere model is provided by the solar corona model at 20 $R_\sSun$, while the solar corona obtains its outer boundary conditions from inner heliosphere module at 24 $R_\sSun$. The overlap serves as a buffer to suppress numerical artifacts due to the differences between the spatial and temporal resolutions.

In some cases the coupling between the physics models requires some complicated and expensive calculations. For example the inner magnetosphere and the radiation belt models require passing the magnetic field geometry and the plasma state along the closed magnetic field lines of the global magnetosphere model. Tracing magnetic field lines is challenging because the global magnetosphere grid is large and distributed over many processors.
SWMF uses highly parallel and efficient schemes for
tracing multiple field lines \cite[]{DeZeeuw:2004a, Glocer:2009c}
that provide mapping information, integrate quantities along the lines, or extract state variables and positions along the lines.

\subsection{Original SWMF Modules}
\label{subsec:modules2005}
\figurename~\ref{fig:swmf2005} shows the components of the original SWMF and the models that can represent these components. Several components were represented by the BATS-R-US code. Since the SWMF is compiled into a single library, the components cannot contain modules, external subroutines or functions with identical names. An automated script ensured that BATS-R-US codes representing various components could be compiled together and they could be configured and run with different parameters. The original models of SWMF were the following:

\subsubsection{\textit{Solar Corona (SC)}}
The Solar Corona (SC) (represented by BATS-R-US) domain started at the photosphere and extended to a few solar $R_\sSun$. The MHD equations were solved with empirical heating functions, heat conduction, and radiative cooling on a co-rotating spherical grid with highly stretched radial coordinates to capture the transition region \cite[]{Cohen:2007a, Downs:2010a}.

\subsubsection{\textit{Eruptive Events (EE)}}
The Eruptive Event generator component is responsible for creating a CME. This was achieved with empirical models that insert an unstable flux rope into the steady solar corona solution, or insert an arcade and apply shearing motion at the lower boundary of the corona model \cite[]{Roussev:2003b, Manchester:2004a}.

\subsubsection{\textit{Inner Heliosphere (IH)}}
The Inner Heliosphere model originally extended from about $20R_\sSun$ to the orbit of the Earth and has been later extended to include the planets. BATS-R-US solves the ideal or two-temperature MHD equations on a Cartesian grid in either co-rotating or inertial frame, and it can model the propagation of CMEs from the Sun to the Earth \cite[]{Toth:2005swmf, Lugaz:2005b, Manchester:2006b}.

\subsubsection{\textit{Global Magnetosphere (GM)}}
The Global Magnetosphere domain surrounds the Earth and it extends about $30R_\sE$ toward the Sun, a few hundred $R_\sE$ toward the magnetotail, and about $60R_\sE$ in the orthogonal directions. BATS-R-US solves the MHD equations on a Cartesian or spherical grid. As an alternative, the \cite{Tsyganenko:1989a} empirical model can provide the magnetic field as a function of observed solar wind parameters and planetary indexes.

\subsubsection{\textit{Inner Magnetosphere (IM)}}
The Inner Magnetosphere model consists of the closed magnetic field line region around the Earth. The Rice Convection Model (RCM) \cite[]{Harel:1981a, Wolf:1991a, Toffoletto:2003a, Sazykin:2002a} solves for the bounce averaged and isotropic but energy resolved particle distribution of electrons and various ions. This was the first successful coupling \cite[]{DeZeeuw:2004a} of a gyrokinetic ring current model  to a global MHD model describing the magnetosphere.

\subsubsection{\textit{Radiation Belts (RB)}}
The Radiation Belt domain overlaps with IM but it models the relativistic electrons. The RBE \cite[]{Fok:2008a} model solves the bounce-averaged Boltzmann equation.

\subsubsection{\textit{Ionospheric Electrodynamics (IE)}}
The Ionospheric Electrodynamics model is a two dimensional height-integrated spherical surface at a nominal ionospheric altitude (at around 110 km for the Earth). The Ridley Ionosphere Model (RIM) \cite[]{Ridley:2004a} code uses the field-aligned currents to calculate particle precipitation and conductances based on empirical relationships, and then it solves for the electric potential on a 2D spherical grid.

\subsubsection{\textit{Upper Atmosphere (UA)}}
The Upper Atmosphere contains the thermosphere and the ionosphere extending from around 90 km to about 600 km altitude for the Earth. The GITM \cite[]{Ridley:2006a} code solves the equations of multi-species hydrodynamics with viscosity, thermal conduction, chemical reactions, ion-neutral friction, source terms due to solar radiation, etc. on a spherical grid in a corotating frame. The MSIS \cite[]{Hedin:1991a} and IRI \cite{Bilitza:2001a} empirical models provide statistical average states for the upper atmosphere and ionosphere, respectively.

\subsection{Additional Simulation Tools}
\label{subsec:simtools}
In addition to BATS-R-US, the SWMF includes several world-class models that provide one of the most advanced space weather simulation capabilities ranging from kinetic and meso-scales to global description of the space environment. Here we briefly summarize the most important simulation/postprocessing tools included in the SWMF suite.

\subsubsection{\textit{EEGGL}}
The Eruptive Event Generator using the Gibson-Low configuration tool \cite[EEGGL,][]{Jin:2017b} is the first community model (available at the CCMC) to simulate magnetically driven CMEs. It is an automated tool for finding the flux rope parameters \cite{Gibson:1998a} to reproduce observed CME events \cite[e.g.,][]{Jin:2017b, Borovikov:2017a}. The solar magnetogram is first used to specify the inner boundary condition of the magnetic field for AWSoM(-R), which is then employed to generate an ambient solar wind solution. Simultaneously, the input magnetogram and the observed CME speed are used by EEGGL to determine the \cite{Gibson:1998a} flux rope parameters. With the derived parameters, a \cite{Gibson:1998a} flux rope is inserted into the ambient solar wind to initiate the CME event. The various parameters of the \cite{Gibson:1998a} model are carefully selected to reproduce the observed CME source region and speed. The user can change parameters, such as helicity and initial orientation, to experiment with the properties of the resulting eruption. Presently, we are working on more eruptive event generator tools, which employ different physical processes (such as the \cite{Titov:1999a} mechanism) to initiate solar eruptions.

\subsubsection{\textit{M-FLAMPA}}
The Multiple Field Line Advection Model for Particle Acceleration solves the kinetic equation for solar energetic particles along a multitude of interplanetary magnetic field lines originating from the Sun \cite[]{Sokolov:2004a, Borovikov:2018a}. It is seamlessly coupled to AWSoM-R and EEGGL and can therefore account for the temporal evolution of field lines as the CME moves outward from the Sun. The diffusion coefficient used in M-FLAMPA is self-consistently calculated from the energy densities of the Alfv{\'e}nic turbulence simulated by AWSoM-R. Together, M-FLAMPA, AWSoM-R and EEGGL provide a high-performance, self-consistent description of the solar corona, inner heliosphere, and solar energetic particle distribution to study solar storms and their impact on the inner heliosphere.

\subsubsection{\textit{AMPS}}
The Adaptive Mesh Particle Simulator (AMPS) is a high-performance kinetic Monte Carlo code originally developed for modeling neutral planetary environments, where it was used to solve the Boltzmann equation accounting for particle collisions, internal degrees of freedom, and chemical reactions \cite[]{Tenishev:2008a, Tenishev:2021a}. 
AMPS employs AMR mesh with cut-cells for discretizing the simulated domain. The implemented cut-cells methodology allows the code to simulate gas flows around objects with arbitrarily complex surface geometry like, \eg nuclei of comets \cite[]{Tenishev:2016a}.

The distinct feature of AMPS is the ability to model two-phase environments, where a dust phase is simulated concurrently with the ambient gas or plasma. In such a simulation, the electric charge of the dust grains varies according to the ambient plasma conditions, and the size and/or chemical composition of a dust grain can change affected by, \eg sublimation of volatiles carried the grain \cite[]{Tenishev:2011a}. 

AMPS was extended to simulating energetic charged particle transport in the inner heliosphere and the Earth’s magnetosphere \cite[]{Tenishev:2005a,Tenishev:2018a}. This model can be used to describe a broad range of suprathermal particle populations including magnetospheric particles with energies exceeding $\sim\!\!1$keV/nucleon, solar energetic particles in the MeV to GeV range, or galactic cosmic rays with energies above $\sim\!\!100$MeV/nucleon. 
Recently, AMPS was extended by adding an implicit PIC capability. Now it can be used for simulating various plasma phenomena either as a stand-alone modeling tool or coupled to other components of the SWMF. 

AMPS is coupled to several components of the SWMF, allowing  multi-scale and multi-physics simulations. Specifically, two-way coupling has been developed with the Global Magnetosphere (GM) and Outer Heliosphere (OH) modules. A one-way coupling procedure is implemented to couple AMPS to the Solar Corona (SC) and Inner Heliosphere (IH) components of the SWMF for SEP simulations. 

\subsubsection{\textit{iPiC3D}}
iPiC3D is a parallel high-performance implicit Particle-in-Cell (PIC) code \cite[cf.,][]{Markidis:2010a}. It solves the full set of Maxwell's equations for the electromagnetic fields coupled with the equations of motion for electrons and ions on 3D Cartesian grids. The discretization is based on the implicit moment PiC (IMPiC) method that employs an implicit time integration for the electric field, then the magnetic field is updated from the induction equation, finally the particles are moved with a simple iterative scheme \cite[]{Brackbill:1982a, Brackbill:1986a, Brackbill:2008a, Lapenta:2006a}. The main advantage of iPiC3D is that it is capable of taking larger grid cell sizes and time-steps and thus making the coupled simulation affordable on today's supercomputers. There are still open questions about the use of implicit PiC codes. The bottom line is that if one wants to resolve Debye scale phenomena the use of expensive explicit PiC codes are necessary. However, if one is mainly interested in reconnection and other space plasma phenomena the use of implicit PiC codes is not only justified, but also necessary \cite[cf.,][]{Ricci:2002a}.

\subsubsection{\textit{FLEKS}}
The FLexible Exascale Kinetic Simulator (FLEKS) is a new particle-in-cell (PIC) code that is designed for the MHD with adaptively embedded PIC (MHD-AEPIC) simulations. FLEKS uses the Gauss's law satisfying energy-conserving semi-implicit method (GL-ECSIM) \cite[]{Chen:2019a} as the base PIC solver. Novel particle splitting and merging algorithms have been designed to control the number of macro-particles per cell during a long MHD-AEPIC simulation. The particle splitting algorithm improves statistical representation and reduces noise in the cells with low macro-particle number, while the particle merging algorithm alleviates the load imbalance and speeds up simulations. The FLEKS grid is Cartesian, but the active PIC region is not limited to be a box anymore since any Cartesian cells can be switched on or off at any point of the simulation. FLEKS uses the high-performance parallel data structures provided by the AMReX library \cite[]{Zhang:2019amrex, Zhang:2020amrex} to store the fields and also the particles. 

\section{Physics}
\label{sec:physics}
BATS-R-US has a layered modular software architecture to handle several applications with a single base code (see \figurename~\ref{fig:bats-structure}). The state variables of the equation system are defined by the equation modules, while the rest of the application dependent details are implemented into user modules. A configuration script is used to select the equation and user modules that are compiled together with the code. There are currently dozens of equation and user modules (obviously not all combinations are possible) which means that BATS-R-US can be configured for quite a few different applications. In addition to the basic equations, there are various source terms that change from application to application: collisions, charge exchange, chemistry, photo-ionization, recombination, radiative losses, etc. The boundary and initial conditions vary greatly as well.

\subsection{Conservation Laws in BATS-R-US}
\label{subsec:xmhd}
BATS-R-US can be configured to solve the governing equations of ideal and resistive MHD \cite[]{Powell:1999a}, semi-relativistic \cite[]{Gombosi:2002a}, anisotropic \cite[]{Meng:2012a}, Hall \cite[]{Toth:2008a}, multispecies \cite[]{Ma:2002a} and multi-fluid \cite[]{Glocer:2009a} extended magnetofluid equations (XMHD) and more recently non-neutral multifluid plasmas \cite[]{Huang:2019a}. 

\begin{table}[htb]
\caption{Conservation laws in BATS-R-US.}
\centering
\small
\begin{tabular}{p{1.5in} p{1.in} p{0.75in} p{1.3in} p{1.25in}}
\hline
\textbf{Physics}  & \textbf{E\&M} & \textbf{Fluids} & \textbf{Resistivity}  & \textbf{Fastest wave} \\
\hline
ideal MHD$^{a}$  & Ohm's law & single/multi$^{b}$ & numerical & fast magnetosonic \\
resistive MHD$^{c}$  & Ohm's law & single/multi & numerical + Ohmic & fast magnetosonic \\
semi-relativistic MHD$^{d}$  & Ohm's law & single/multi & numerical & light (reduced) \\
anisotropic MHD$^{e}$ & Ohm's law & single/multi & numerical + Hall &  whistler \\
Hall MHD$^{f}$ & Ohm's law & single/multi & numerical + Hall &  whistler \\
5-, 6-moment transport$^{g}$ & Maxwell's eqs & multi & numerical + Hall & light (reduced) \\
\hline
\multicolumn{5}{l}{$^{a}$\cite{Powell:1999a}; $^{b}$\cite{Glocer:2009a}; $^{c}$\cite{Kuznetsova:2007a}; $^{d}$\cite{Gombosi:2002a};} \\
\multicolumn{5}{l}{$^{e}$\cite{Meng:2012a}; $^{f}$\cite{Toth:2008a}; $^g$\cite{Huang:2019a}}
\end{tabular}
\label{table:t1}
\end{table}

\tablename~\ref{table:t1} summarizes the various extended MHD conservation laws that can be solved by BATS-R-US.

\subsubsection{Extended MHD Equations}
\label{subsubsec:laws}
BATS-R-US can solve many approximations to the low-order velocity moments of the Boltzmann equations (we refer the interested readers to the literature \cite[\cf,][]{Burgers:1969a, Schunk:1980a, Gombosi:1991b, Gombosi:1998book, Shumlak:2003a, Huang:2019a}. The governing equations for species `s' can be written as
\begin{subequations}
\label{eqn:xmhd}
\begin{align}
& \frac{\partial \rho_s}{\partial t} + \divg (\rho_s \overline{\bu}_s) = 0	\\
& \frac{\partial \rho_s \overline{\bu}_s}{\partial t}  + \divg 
    \left[\rho_s \overline{\bu}_s \, \overline{\bu}_s + p_{s_\sperp} \unitI 
        + (p_{s_\spar} - p_{s_\sperp})\,\overline{\bb}\,\overline{\bb}\right] =
        \frac{q_s}{m_s}\rho_s(\overline{\bE}+\overline{\bu}_s \times\overline{\bB})	\\
& \frac{\partial p_{s_\spar}}{\partial t}  + \divg ( p_{s_\spar} \overline{\bu}_s) =
    - 2 p_{s_\spar} \overline{\bb} \cdot (\overline{\bb} \cdot \grad )\, \overline{\bu}_s 	\\
& \frac{\partial p_{s_\sperp}}{\partial t} + \divg(p_{s_\sperp}\overline{\bu}_s) = 
   - p_{s_\sperp} (\divg\overline{\bu}_s) 
   + p_{s_\sperp} \overline{\bb} \cdot (\overline{\bb} \cdot \grad)\, \overline{\bu}_s
\end{align}
\end{subequations}
where $\rho$ and $\overline{\bu}$ denote the mass density and the velocity vector, respectively, and $q$ and $m$ are the charges and masses of the particles. For the pressure tensor we used the CGL approximation \cite[]{Chew:1956a}: $\bar{\bar{\bP}} = p_{\sperp} \unitI + (p_{\spar} - p_{\sperp})\, \overline{\bb}\,\overline{\bb}$, where $\unitI$ is the identity matrix, $\overline{\bb}$ is the unit vector along the magnetic field direction, $p_{\spar}$ is the pressure along the parallel direction of the magnetic field and $p_{\sperp}$ is the pressure in the perpendicular direction. The scalar pressure can be written as $p = (p_{\spar} + 2 p_{\sperp})/3$. BATS-R-US has the capability to solve the full equation system (\ref{eqn:xmhd}) or reduce it and only solve for the scalar pressure, $p$.

Equation~(\ref{eqn:xmhd}) can be obtained from the Boltzmann equation by considering the infinite series of velocity moments, called \textit{Maxwell's equation of change} \cite[\cf,][]{Gombosi:1994book}.
\begin{equation}
\label{eqn:change}
\begin{split}
\frac{\partial\left<\mathcal{M}_s\mathcal{F}_s\right>}{\partial t}  
    & + \divg\left(\overline{\bu}_s\left<\mathcal{M}_s\mathcal{F}_s\right>\right)
    + \divg\left<\overline{\bc}_s\,\mathcal{M}_s\mathcal{F}_s\right> 
    + \left<\mathcal{F}_s\,[(\overline{\bc}_s\cdot\grad)\overline{\bu}_s]
    \cdot\grad_{c_s} \mathcal{M}_s\right>
    \\ &
+ \left(\frac{\partial\overline{\bu}_s}{\partial t}  
    + [(\overline{\bu}_s\cdot\grad)\overline{\bu}_s] \right) \cdot
    \left<\mathcal{F}_s\,\grad_{c_s} \mathcal{M}_s\right>
    - \left<\mathcal{F}_s\,(\overline{\ba}_s\cdot\grad_{c_s}) \mathcal{M}_s\right>  
    = \left<\mathcal{M}_s\left(\frac{\delta\mathcal{F}_s}
    {\delta t}\right)_{\text{coll}}\right>
\end{split}
\end{equation}
Here $\mathcal{F}_s$ is the velocity distribution function of species `s' (expressed in terms of the random velocity, $\overline{\bc}_s$), 
$\mathcal{M}_s$ is a physical quantity of a single particle of species `s' dependent on the random velocity. $\langle\rangle$ denotes averaging over the entire random velocity space. The order of $\mathcal{M}_s$ in the random velocity defines the order of the velocity moment equation. For instance, $\mathcal{M}_s=m_s$ (zeroth-order moment equation) results in the continuity equation, describing the conservation of mass. The first-order velocity moment equations are obtained by using $\mathcal{M}_s=m_s \overline{\bc}_s$ and they express the conservation of momentum. The second-order velocity moment equations are obtained by using $\mathcal{M}_s=m_s\,\overline{\bc}_s \,\overline{\bc}_s$. There is one zero-order, three first-order and six second-order moment equations (due to the symmetric nature of the $\overline{\bc}_s \,\overline{\bc}_s$ diad).

It is important to note that \equationname~(\ref{eqn:change}) leads to an infinite number of velocity moment equations. The ``villain'' is the third term on the left hand side of \equationname~(\ref{eqn:change}), $\divg\left<\overline{\bc}_s\,\mathcal{M}_s\mathcal{F}_s\right>$. If $\mathcal{M}_s$ is $n$-th order in velocity, the term  $\left<\overline{\bc}_s\,\mathcal{M}_s\mathcal{F}_s\right>$ is the $(n+1)$-th velocity moment. In other words, the transport equation for the $n$-th velocity moment contains the divergence of the $(n+1)$-th moment, resulting in an infinite series of partial differential equations.

The infinite series of velocity moment equations must be closed some way to obtain a closed set of differential equations. There are a number of closures in the literature \cite[\cf,][]{Chapman:1916a, Enskog:1917a, Grad:1949a, Levermore:1996a}. The simplest (and most popular) closures either neglect the third-order velocity moments (the heat flow), or express a high-order velocity moment in terms of lower moments \cite[cf,][]{Grad:1949a}. Equation~(\ref{eqn:xmhd}) was obtained by neglecting the heat flow tensor and using the CGL approximation \cite[]{Chew:1956a} for the pressure tensor. In this approximation there are six velocity moments we solve for: $\rho_s$, the three components of $\overline{\bu}_s$ and the two pressure components, $p_{\spar}$ and $p_{\sperp}$. For this reason this is called the \textit{six moment approximation}.

The electric ($\overline{\bE}$) and magnetic fields ($\overline{\bB}$) are obtained from Maxwell's equations:
\begin{subequations} 
\label{eqn:maxwell_orig}
\begin{align}
& \frac{\partial \overline{\bB}}{\partial t} + \curl \overline{\bE} = 0 \\
& \frac{\partial \overline{\bE}}{\partial t} - c^2\ \curl\overline{\bB} = -c^2\muo\,\overline{\bj} \\
& \divg \overline{\bE} = \frac{\rho_c}{\epso} \label{eqn:dive} \\
& \divB = 0 \label{eqn:divb}
\end{align}
\end{subequations}
where $\epso$ is the vacuum permittivity, $\muo$ is the vacuum permeability, $c=1/\sqrt{\epso \muo}$ is the speed of light, $\rho_c = \sum_s (q_s/m_s)\rho_s$ is the total charge density and $\overline{\bj} = \sum_s (q_s/m_s) \,\rho_s \overline{\bu_s}$ is the current density. Equations~(\ref{eqn:dive}) and (\ref{eqn:divb}) are constraints on the initial conditions and analytically these conditions are preserved. Numerically, however, this is not guaranteed to hold. BATS-R-US uses a variety of methods to enforce the solenoidal magnetic field condition (for more details see \cite{Toth:2012a}).

It is important to point out that in the multifluid formulation the electric current density depends on the charge averaged, and not the mass averaged, ion velocity. This can be seen by looking at the definition of $\overline{\bj}$:

\begin{equation}
\label{eqn:current}
\overline{\bj} = e\left(\sum\limits_{\text{s=ions}} Z_s  \,n_s \overline{\bu}_s - n_e \overline{\bu}_s \right) = e \, n_e\,(\overline{\bu}_+ - \overline{\bu}_e )
\end{equation}

\noindent where $Z_s$ is the ionization state of a given ion species and $\overline{\bu}_+ = \sum_{\text{s=ions}} Z_s (n_s/n_e) \,\overline{\bu}_s$ is the charge averaged ion velocity. Note that in general $\overline{\bu}_+ \ne \overline{\bu}$ and the two vectors can be quite different. BATS-R-US takes into account the full definition of $\overline{\bu}_+$ and thus self-consistently accounts for the different velocities of the various ion species \cite[]{Glocer:2009a}. This is different from the approximate solution applied in the LFM code \cite[\cf][]{Wiltberger:2010a, Merkin:2011b} that assumes that the macroscopic plasma velocity in the direction perpendicular to the magnetic field coincides with the electrical drift velocity and therefore is the same for all ion species.

Extended magnetohydrodynamics (XMHD) makes two fundamentally important assumptions: (i) electrons are assumed to be massless and (ii) charge neutrality is assumed at all scales. These two assumptions lead to the generalized Ohm's law:

\begin{equation}
\label{eqn:efield}
\overline{\bE} = -\bu_e \times \overline{\bB} - \frac{1}{e n_e} \divg [p_{e_\sperp} \unitI + (p_{e_\spar} - p_{e_\sperp})\overline{\bb}\,\overline{\bb}]
\end{equation}

In a single-ion plasma the electron velocity is $\overline{\bu}_e = \overline{\bu}_i - \overline{\bj}/(e n_e)$ resulting in the motional electric field plus the Hall term. The second term in \equationname~(\ref{eqn:efield}) is the ambipolar electric field. It is interesting to note that the parallel (field aligned) component of the electric field is

\vspace{-1em}
\begin{equation}
\label{eqn:efield-parallel}
E_\spar =  \overline{\bb} \cdot \overline{\bE} = 
  - \frac{\grad_\spar p_{e_\spar}}{e n_e} 
  + \frac{p_{e_\spar} - p_{e_\sperp}}{e n_e} \frac{\grad_\spar B}{B}
\end{equation}
where $\grad_\spar = \overline{\bb} \cdot \grad$ is the parallel gradient operator. In \equationname~(\ref{eqn:efield-parallel}) the first term describes the parallel ambipolar electric field while the second term represents adiabatic focusing. BATS-R-US has the capability to solve various XMHD approximations, from ideal MHD to resistive, Hall, anisotropic pressure, multispecies and multifluid limits. A more detailed description of these capabilities can be found in \cite{Toth:2012a}.

\subsubsection{Six Moment Equations}
\label{subsubsec:sixmoment}
A recent addition to the BATS-R-US equation set is the six-moment approximation \cite[]{Huang:2019a}. This approximation solves the full set of \equationname~(\ref{eqn:xmhd}) and \equationname~(\ref{eqn:maxwell_orig}) without neglecting the electron mass and assuming charge neutrality. Consequently there is no Ohm's law to express the electric field and we need to solve the full set of electron fluid equations and the full set of Maxwell's equations. In this approximation the fastest wave mode is the light wave. Since the speed of light typically well exceeds the typical MHD wave speeds, one can artificially reduce it to allow larger time steps and more efficient computation. An additional benefit is that the whistler wave speed is also limited by this reduced speed of light. The six-moment equations describe several phenomena that are not captured by simpler MHD equations, Hall physics, relativistic limit of fast and whistler waves, net charge, anisotropy of both electron and ion pressures, etc. An additional benefit is that one can have multiple species with positive and negative charges, including multiple electron fluids or negatively charged dust.  See \cite{Huang:2019a} for details.

\subsubsection{Source Terms}
\label{subsubsec:sourceterms}
The collision terms in the transport equations describe the various physical processes that transfer mass, momentum and energy between various ionized or neutral species. These terms represent the underlying physics that enable us to model the interaction of space plasma flows with planets \cite[\cf][]{Ma:2004a, Ma:2013a, Ma:2018b, Sarkango:2019a}, planetary moons \cite[\cf]{Rubin:2015a, Jia:2018a, Harris:2021a}, comets \cite[\cf]{Gombosi:1996a, Huang:2016a} and other objects of interest.
The collision term describes the rate of change of the distribution function due to interaction between various species. In BATS-R-US we consider the following processes:
\begin{outline}
	\setlist{itemsep=-1ex, topsep=0ex, labelwidth = 1ex, labelsep= 1ex,leftmargin=5ex,listparindent=2ex}
\1 elastic collisions
\1 photoionization and impact ionization (using the Beer-Lamber law),
\1 charge transfer, and
\1 recombination.
\end{outline}

Next, we discuss the contributions of these processes to the collision
term. We make the following simplifying
asssumptions:
\begin{outline}
	\setlist{itemsep=-1ex, topsep=0ex, labelwidth = 1ex, labelsep= 1ex,leftmargin=5ex,listparindent=2ex}
\1 
All particles are assumed to lack any internal degrees of freedom,
\1
Energy thresholds of various processes (such as chemical reactions,
ionization thresholds, etc.) will be neglected,
\1
All neutral species are considered cold ($T_n=0$) and are assumed to
move with the same bulk velocity, ${\overline{\bu}}_n$.
\end{outline}
These simplifications limit the scope of our approximations, but our methodology still provides useful insights into collisional effects in space plasmas.

In the present approximation all particles are assumed to possess no intrinsic
degrees of freedom, therefore all inelastic collisions change the
identity of a particle. These reactions result in ionization, charge
transfer, or recombination.

\paragraph{\textit{Elastic Collisions.}}
Elastic collisions do not change the identity of particles, but do
change the momentum and energy of individual particles. The effects of these collisions is described in the general framework of the
relaxation-time approximation \cite[\cf][]{Bhatnagar:1954a, Burgers:1969a, Gombosi:1994book}. The main idea behind this approximation is the
recognition that collisions drive all gas components toward
equilibrium. Since equilibrium phase-space distributions are
Maxwellians, the cumulative effect of elastic collisions can be
formally described by gradually replacing the present distribution
function ($F_s$) with the appropriate Maxwellian, $F_{s(st)}$ \cite[\cf][]{Gombosi:1994book}:

\begin{equation}
   \left(\frac{\delta F_s}{\delta t}\right)_{el} =
       \sum_{t=all} \frac{F_{s(st)} - F_s}{\tau_{st}}
\label{eq:BGK.1}
\end{equation}

In expression~(\ref{eq:BGK.1}) the subscript ``$t$'' refers to all
species other than ``$s$'', and $\tau_{st}$ is a ``relaxation time''
characterizing how fast the distribution function $F_s$ approaches
equilibrium due to collisions between particles of types ``$s$'' and
``$t$''. Equation~(\ref{eq:BGK.1}) means that ``$st$'' and ``$st'$''
collisions may drive particles ``$s$'' toward two different
equilibria: however, in steady-state equilibrium all species will reach
the same bulk velocity and temperature.

The relaxation timescale, $\tau_{st}$, can be different for different species. For instance, electrons relax toward equilibrium faster than ions in ion-electron collisions. In practice, the \textit{momentum transfer collision frequency}, $\bar{\nu}_{st}$ is used instead of the relaxation time. The momentum transfer collision frequency includes a mass-dependent factor that accounts for the efficiency of momentum transfer in an elastic collision:

\begin{equation}
   \frac{1}{\tau_{st}} = \frac{m_s+m_t}{m_t} \bar{\nu}_{st}
\label{eq:nubar}
\end{equation}

The parameters of the Maxwellian, $F_{s(st)}$, are chosen in a way that mass, momentum and energy are conserved while the gas is driven toward equilibrium \cite[]{Burgers:1969a, Gombosi:1994book}:

\begin{equation}
   F_{s(st)} = n_s \left(\frac{m_s}{2\pi k_\sB T_{s(st)}}\right)^{3/2}
   \times \exp\left[-\frac{m_s}{2k_\sB T_{s(st)}}\left(\overline{\Bv}_s + \overline{\bu}_s-\overline{\bu}_{st}\right)^2\right]
\label{eq:BGK.2}
\end{equation}

\noindent where

\begin{equation}
   n_s={\int\hspace{-0.1in}\int\limits_\infty \hspace{-0.1in}\int} 
      F_s(t,\overline{\br},\overline{\Bv}_s)\,d^3v_s
\label{eq:BGK.3}
\end{equation}
\begin{equation}
   \overline{\bu}_{st}=\frac{m_t\overline{\bu}_t + m_s\overline{\bu}_s}{m_s+m_t}
\label{eq:BGK.4}
\end{equation}
\begin{equation}
    T_{s(st)} = T_s + \frac{m_s m_t}{(m_s+m_t)^2} \left[2(T_t-T_s) + \frac{m_t}{3k_\sB} (\overline{\bu}_t-\overline{\bu}_s)^2 \right]
\label{eq:BGK.5}
\end{equation}

\noindent In these expressions, $k_\sB$ is the Boltzmann constant and the kinetic
temperature is defined as
\begin{equation}
   T_s = \frac{m_s}{3n_s k_\sB}{\int\hspace{-0.1in}\int\limits_\infty
         \hspace{-0.1in}\int} 
         v_s^2 F_s(t,\overline{\br},\overline{\Bv}_s)\,d^3v_s
\label{eq:BGK.6}
\end{equation}

Equations~(\ref{eq:BGK.3}) through ~(\ref{eq:BGK.5})  describe the
number density of species ``$s$'', the drift velocity of species
``$s$'' with respect to the center mass of fluids ``$s$'' and ``$t$'',
and the ``stagnation temperature'' of species ``$s$''. It should be
noted that ${\bf u}_{st}={\bf u}_{ts}$ and in general $T_{s(st)}\neq
T_{t(ts)}$.

\paragraph{\textit{Ionization.}}
There are four primary ionization processes to be considered:
photoionization, impact ionization by superthermal electrons, impact
ionization by energetic ions, and finally impact ionization by
energetic neutrals. These ionization processes create new charge,
therefore we consider them separately from the charge transfer
reactions.

The ionization process converts a particle from the thermal neutral
population to one of the charged particle species. Since the neutral
gas is assumed to be cold ($T_n=0$) the net ionization source can be
approximated by the following expression:

\begin{equation}
   \left(\frac{\delta F_s}{\delta t}\right)_{ion} =
    \nu^{io}_{s'}
    n_{s'} \  \delta^3 \left(\overline{\bu}_s + \overline{\Bv}_s - \overline{\bu}_n \right) 
\label{eq:source.1}
\end{equation}

\noindent where $\nu^{io}_{s'}$ is the sum of the photoionization and impact ionization frequencies of species ``$s'$'',  $n_{s'}$ is the density of particles
producing charged particles of type ``$s$''. Throughout this paper the
charge state of particles ``$s'$'' is one less than the charge state
of particles ``$s$''.

\paragraph{\textit{Charge Exchange.}}
Charge exchange transfers an electron from one particle to an other
(an example is the accidentally resonant O$^+$+H$\rightleftharpoons$O+H$^+$ reaction). Although there is a transfer of electrons between two heavy particles, in most cases each particle tends to retain its original kinetic energy. Here we limit our consideration to singly charged ions and we consider the following general charge exchange reaction: ${\rm S} + {\rm M}^+ \rightarrow {\rm S}^+ + {\rm M}$. The ion, ${\rm S}^+$, is referred to as species ``$s$'', while particles ${\rm S}$ are species ``$s'$''. In our approximation the neutral
particles form a cold gas, therefore one can write the net rate of
change of the phase-space distribution function of particles ``$s$''
is the following:

\begin{equation}
     \left(\frac{\delta F_s}{\delta t}\right)_{cx} =
     - F_s \left(\sum_{t'=neutrals} k_{st'} n_{t'}\right) 
    + \left(\sum_{t=ions} k_{ts'}\  n_{t} \right)\,n_{s'} 
      \delta^3 \left(\overline{\bu}_s + \overline{\Bv}_s - \overline{\bu}_n \right) 
\label{eq:chargeX.2}
\end{equation}

\noindent Here $k_{ts'}$ and $k_{ts}$ are charge exchange rates. The first term
describes the loss of particles ``$s$'' due to charge exchange with
all neutral species, while the second term describes the creation of new
``$s$'' particles by charge exchange with ``$s'$'' type particles.

\paragraph{\textit{Recombination.}}
Recombination removes a positive and a negative charge from the
system. It represents a sink for electrons and for particles ``$s$''
and a source for particles ``$s'$''. This  leads to the following loss
rate for ions ``$s$'':

\begin{equation}
   \left(\frac{\delta F_s}{\delta t}\right)_{rec} = - \alpha_s n_e F_s
\label{eq:recombination.1}
\end{equation}

\noindent where $\alpha_s$ is the recombination coefficient and $n_e$ is the
electron density. Equation~(\ref{eq:recombination.1}) also gives the source term for species ``$s'$'' (naturally with positive sign).

\paragraph{\textit{Combined Collision Term.}}
Next, we combine the collision terms for all processes discussed above
and combine equations (\ref{eq:BGK.1}), (\ref{eq:source.1}),
(\ref{eq:chargeX.2}) and (\ref{eq:recombination.1}) to obtain:

\begin{align}
\frac{\delta F_s}{\delta t} &= 
       - \left(\sum_{t=all} \frac{m_s+m_t}{m_t} \bar{\nu}_{st}
       + \sum_{t'=neutrals} k_{st'} n_{t'}  
       + \alpha_s n_e \right) F_s 
       \nonumber \\ &
  + \left(\nu^{io}_{s'} + \sum_{t=ions} k_{ts'}\ n_{t}\right) n_{s'} 
      \delta^3 \left(\overline{\bu}_s + \overline{\Bv}_s - \overline{\bu}_n \right)
  + \sum_{t=all} \frac{m_s+m_t}{m_t} \bar{\nu}_{st} F_{s(st)}
\label{eq:combined}
\end{align}

In BATS-R-US we take the appropriate velocity moments of \equationname~(\ref{eq:combined}) (corresponding to the actual approximation used for the governing equations). 

\subsection{Coupled MHD Turbulence}
\label{subsec:turbulence}
The \textit{ad hoc} elements can be eliminated from the solar corona model by assuming that the coronal plasma is heated by the dissipation of \alf wave turbulence \cite[\cf][]{Sokolov:2013a}. The dissipation itself is caused by the nonlinear interaction between oppositely propagating waves \cite[\eg][]{Hollweg:1986a}. 

Within coronal holes, there are no closed magnetic field lines, hence, there are no oppositely propagating waves. Instead, a weak reflection of the outward propagating waves locally generates sunward propagating waves as quantified by \cite{vanderHolst:2014a}. The small power in these locally generated (and almost immediately dissipated) inward propagating waves leads to a reduced turbulence dissipation rate in coronal holes, naturally resulting in the bimodal solar wind structure. Another consequence is that coronal holes look like cold black spots in the EUV and X-ray images,  while closed field regions are  hot and bright. Active regions, where the wave reflection is particularly strong, are the brightest in this model \cite[see][]{Sokolov:2013a, Oran:2013a, vanderHolst:2014a}.

As has been shown by \citet{Jacques:1977a}, the \alf waves exert an \textit{isotropic} pressure on the plasma. The relation between the wave pressure and wave energy density is $p_\sA=(w_++w_-)/2$, where $w_\pm$ are the energy densities for the turbulent waves propagating along the magnetic field vector ($w_+$) or in the opposite direction ($w_-$). The \cite{Wentzel:1926a, Kramers:1926a, Brillouin:1926a} approximation (WKB) is used to derive the equations that govern the transport of \alf waves, which may be reformulated in terms of the wave energy densities. Dissipation of \alf waves is the physical process that drives the solar wind and heats the coronal plasma.

\alf wave dissipation occurs when two counter-propagating waves interact. \alf wave reflection from steep density gradients is the physical process that results in local wave reflection, thus maintaining a source of both types of waves.
In order to describe this wave reflection we go beyond the WKB approximation that assumes that the wavelength is much smaller than spatial scales of the background variations.

The equation describing the propagation, dissipation, and reflection of \alf turbulence has been derived by \cite{vanderHolst:2014a}:

\begin{equation}
\label{eq:w_pm}
    \frac{\partial w_\pm}{\partial t} + \divg\left[ (\overline{\bu} \pm \overline{\bV}_\sA) w_\pm\right] + \frac{w_\pm}{2} \left(\divg\overline{\bu}\right) = - \Gamma_\pm w_\pm \mp \mathcal{R} \sqrt{w_-w_+}
\end{equation}

\noindent where $\mathbf{V}_A$ is the \alf velocity, while $\Gamma_\pm$ and ${\cal R}$ are the reflection coefficient and the dissipation rate, respectively. Finally, with the help of  the dissipation rate of \alf turbulence one can express the ion and electron heating rates \cite[]{vanderHolst:2014a, Gombosi:2018a}.

In this model there are only two free parameters: (i) the Poynting flux of \alf waves leaving the photosphere ($P_{\sA_\sSun}$), and (ii) the transverse correlation length of \alf turbulence ($L_\perp$). Our solar corona model assumes that $P_{\sA_\sSun}\sim B_\sSun$ and $L_\perp\sim B_\sSun^{-1/2}$ \cite[\cf][]{Gombosi:2018a}.

\subsection{Gyrokinetic Models}
\label{subsec:gyrokinetic}

\subsubsection{Kinetic PWOM}
\label{subsubsec:k-pwom}

The original polar wind model that PWOM is based on solved the field aligned gyroptropic transport equations for each ion species as described by \cite{Gombosi:1989a}. 
After PWOM was incorporated at the PW component of the SWMF it expanded to a global polar wind model, but retained it's fluid nature \cite[]{Glocer:2009b}. 
Given the importance of kinetic processes to many ionospheric outflow mechanisms beyond the polar wind, multiple steps were taken to include these processes in PWOM. 

The initial expansion to kinetic processes came with the inclusion of superthermal electrons whose energy is much greater than the thermal energy of ionospheric electrons (0.3 eV). 
These superthermal electrons are either photoelectrons generated by the photoionization of the neutral atmosphere, precipitating electrons of magnetospheric origin (auroral electrons/polar rain), or secondary electrons generated by other energetic electrons impacting neutral particles. 
To encorporate these superthermal electrons into PWOM, we split the electron population into thermal and superthermal components that must statisify charge neutrality and current conservation \citep{Glocer:2012a}:

\begin{equation}
n_{e}+n_{\alpha}=\sum_{i}n_{i} 
\end{equation}
\begin{equation}
    n_{e}u_{e}+n_{\alpha}u_{\alpha}=\left( \sum_{i}n_{i}u_{i} -\frac{j}{e}\right)
\end{equation}

\noindent Here subscripts $e$ and $\alpha$ indicate the thermal and superthermal electrons, respectively.
Once the superthermal electron population is known, the thermal population is determined from the above equations as well as the thermal electron energy equation (not shown), which also includes the energy deposition due to collisions between the thermal and superthermal populations. 
At that point the ambipolar field is determined and the ion solution can be updated, including the effect of additional sources due to impact ionization. 
In PWOM we have used three approaches to specifying the superthermal electron population. First, in \citet{Glocer:2012a} we used the output of a two-stream calculation of the photoelectron source together with a collisionless kinetic mapping. Second, in \citep{Glocer:2017a} we coupled PWOM to the kinetic STET code and thereby obtained the superthermal electron solution by solving the Boltzmann equation presented by \citet{Khazanov:1994a} as:

\begin{equation}
\frac{\beta}{\sqrt{E}}\frac{\partial \phi}{\partial t}+\mu\frac{\partial \phi}{\partial s} - \frac{1-\mu^{2}}{2}\left(-\frac{F}{E}+\frac{1}{B}\frac{\partial B}{\partial s}\right)\frac{\partial \phi}{\partial \mu} + EF \mu \frac{\partial}{\partial E} \left( \frac{\phi}{E} \right) = q+\langle S \rangle \label{stet1}
\end{equation}

\noindent where the constant $\beta=1.7\times10^{-8}eV^{1/2}cm^{-1}s$, the superthermal differential flux is given by $\phi=\phi(t,E,\mu,s)$, the kinetic energy is $E$, and cosine of the local pitch angle is provide by $\mu$. We also have $s$ defined as the distance along the field line, the magnetic field is given by $B$ and $F$ is the force associated with the parallel electric field and $Q$ is production rate of superthermal electrons, and  $\langle S \rangle$ represents the collision operators. 

The approach to including superthermal electrons as a true kinetic population as described by \cite{Glocer:2017a} is the most complete and physical approach, 
but it can be computationally expensive. Therefore, \cite{Glocer:2018a} 
included the option to use the two-stream approach  \cite[\cf,][]{Nagy:1970a, Solomon:1988a, Barakat:2001a, Lummerzheim:1994a, Schunk:2009a}. 
The two-stream approach has energy dependence and transport/collisional effects, but does not include the effects of pitch angle diffusion or trapping.
It is however dramatically faster than the fully kinetic approach and therefore represents an acceptable compromise between physical completeness and computational efficiency for many problems. 
However, specific problems that rely on detailed kinetic electron effects such as trapping will still require the more comprehensive treatment. 

The inclusion of superthermal electrons allows PWOM to treat only some of the outflow mechanisms, but many other processes require the inclusion of kinetic ions. 
Most prominant of these processes are wave-particle interactions, which drive ion acceleration in the cusp and auroral zones. 
Motivated by this, \cite{Glocer:2018a} expanded the PWOM code to include kinetic ions based on a gyroaveraged PIC approach at high altitudes while keeping the fluid approach at low altitudes for computational efficiency. 
In the high altitude PIC region each macro particle in PWOM for a species `\textit{i}' is advanced by solving the gyro-averaged particle equation of motion given by:

\begin{equation}
m_{i}\frac{\partial v_{i_\spar}}{\partial t} - q_{i}E_{\spar} + \frac{G \,m_{i}M_{planet}}{r^{2}} + \mu_{a_i}\frac{\partial B}{\partial s} = 0
\end{equation}

\noindent where $m_{i}$ specifies the mass, $v_{i}$ specifies the velocity, $t$ specifies the time, and $q_{i}$ is the ion charge. The external forces are given by the parallel electric field $E_{\spar}$, and gravity. In this equation, $\mu_{a_i}$ is the particle's first adiabatic invariant specified by 

\begin{equation}
    \mu_{a_i}=\frac{m_{i}v_{\sperp}^2}{2B}
\end{equation}

At the interface between the low altitude fluid region and the high altitude PIC region information is exchanged. PWOM uses the fluid solution in the last fluid computational cell to sample particles for the PIC region, and the first computational cell of the PIC region is used to compute moments and set boundary conditions for the fluid domain. 
Collisions are included in both the fluid and PIC regions with the fluid collisional terms provided by Burger's fully linear approximation \cite[]{Burgers:1969a}, while in the PIC region collisions are included using the Monte Carlo approach described by \cite{Takizuka:1977a} and modified by \cite{Nanbu:1998a} to allow for particles with variable statistical weights.
Wave particle interactions in the PIC region are implemented using the approach described by \cite{Retterer:1987a}. This method includes the heating by randomly perturbing the perpendicular velocity of the macroparticles with the variance determined by a diffusion coefficient, which depends on the wave power. 
Formulations of these diffusion coefficients are given by \cite{Crew:1990a}, \cite{Barakat:1994a}, and others. 

\subsubsection{Dynamic Global Core Plasma Model}
\label{subsubsec:dgcpm}

Being the coldest ($\sim 1 eV$) magnetospheric population within the magnetosphere, the plasmasphere's evolution is dominated by advection via $\overline{\bE}\times\overline{\bB}$ drift and refilling via ionospheric outflow at mid- and low-latitudes.  
The Dynamic Global Core Plasma Model (DGCPM, \cite{Rasmussen:1993a, Ober:1997a, Liemohn:2004a, Borovsky:2014a}) captures these dynamics by solving a continuity equation for the total flux tube content, $N$:

\begin{equation}
    \frac{\partial N}{\partial t} = \mathcal{S} - \mathcal{L} - \overline{\bu}_{\sperp}\cdot\grad N 
    \label{dgcpm:cont}
\end{equation}

\noindent $\overline{\bu}_{\sperp}$ is the horizontal bulk velocity of the cold plasmasphere fluid (set by the 
local $\overline{\bE}\times\overline{\bB}$ drift).
A dipole magnetic field is assumed, electric potential is a required input and typically obtained via an empirical model.
$\mathcal{S}$ and $\mathcal{L}$ represent the net source and loss of plasma from/into a given flux tube, respectively.
On the day side, ionospheric plasma is assumed to fill flux tubes until saturation density, $N_S$, is reached:

\begin{equation}
    \mathcal{S} = \frac{N_S - N(t)}{\tau_{fill}}
    \label{dgcpm:source}
\end{equation}

\noindent Saturation values are a function of radial distance and determined empirically \citep{Carpenter:1992a}.
The filling time constant, $\tau_{fill}$, has a configurable value but defaults to 6.7 days.
The loss term includes simple loss into the ionosphere at either end of the flux tube:

\begin{equation}
    \mathcal{L} = \frac{N(t)}{\tau_{loss}}
    \label{dgcpm:loss}
\end{equation}

\noindent The loss time constant, $\tau_{loss}$, is set to 3 days.

Within the SWMF, couplings to other models provide more realistic electric fields and allow for the exploration of the impact of the plasmasphere on the global magnetosphere.  
DGCPM can obtain electric field from the IE module, opening up a greater range of empirical and first-principles-based electric fields \citep{Ridley:2014a, Borovsky:2014a}.
The density of the plasmasphere can be passed to the GM component following the same algorithm used to couple ring current density and pressure \citep{Glocer:2020a}. It is coupled individually to the HEIDI drift physics model, discussed below \citep{Liemohn:2004a}, and used extensively to explore ring current-plasmasphere interactions \citep{Liemohn:2006a,Liemohn:2008a, Ridley:2014a}.
When at least three modules are used (GM, IE, and PS), the effect of plasmasphere drainage plumes on dayside reconnection can be explored in a self-consistent manner.

\subsubsection{Rice Convection Model}
\label{subsubsec:rcm}

The Rice Convection Model (RCM, \cite{Wolf:1974a, Toffoletto:2003a, Sazykin:2002a}) is a guiding center drift model of a set of assumed-isotropic populations, each with a given energy invariant, $\lambda$, such that,

\begin{equation}
    \lambda = W\mathcal{V}^{{2}/{3}}
    \label{rcm:lambda}
\end{equation}

\noindent where $W$ is the kinetic energy of the particles within the population and $\mathcal{V}$ is the magnetic flux tube volume per unit magnetic flux.
The RCM then solves for the evolution of the flux tube density content, $\eta$, for each energy invariant population, by solving the continuity equation,

\begin{equation}
    \frac{\partial \eta}{\partial t} - \overline{\textbf{v}}_\sD \cdot \grad \eta = 0
    \label{rcm:cont}
\end{equation}

\noindent where $\eta$ is a function of time and space and is defined by,

\begin{equation}
    \eta = \int \frac{n ds}{B} = n \mathcal{V}
    \label{rcm:dens}
\end{equation}

\noindent Here, $\overline{\textbf{v}}_\sD$ is the full electromagnetic drift velocity of the population and is given by,

\begin{equation}
    \overline{\textbf{v}}_\sD = \frac{\overline{\bE}\times\overline{\bB}}{B^2}+
        \frac{\lambda\overline{\bB}\times\grad\mathcal{V}^{-{2}/{3}}}{qB^2}
    \label{rcm:velocity}
\end{equation}

\noindent where $q$ is the charge of the individual particles that constitute the energy invariant population.
The electric and magnetic fields must be prescribed via external models or empirical relations.
Equation (\ref{rcm:cont}) is solved via an ionospheric grid where each grid point represents the foot point of a magnetospheric flux tube.

The RCM version integrated into the SWMF is unique in its configuration.
Flux tube volume, $\mathcal{V}$, is obtained from the GM component, as are initial and boundary conditions for $\eta$ for each energy invariant population.
Electric potential is obtained via the IE solution.
The total density and pressure at each RCM grid point is handed to GM where it is treated as a source term to the MHD values along each flux tube, nudging the MHD solution towards the RCM solution.
This is done via,

\begin{equation}
    p_{\sG\sM}^{\prime} = p_{\sG\sM} + \min(1, \frac{dt}{\tau_{couple}})(p_{\sR\sC\sM}-p_{\sG\sM})
    \label{rcm:couple}
\end{equation}

\noindent where $p_{\sG\sM}$ and $p_{\sR\sC\sM}$ refer to the total fluid pressure as calculated by the respective modules and $\tau_{couple}$ is a time constant, typically set to $10s$. If other inner magnetosphere (IM) models are used the $p_{\sR\sC\sM}$ term is replaced by the appropriate pressure obtained from the IM model.
Equation (\ref{rcm:couple}) may also be used to return density values alongside pressure.
The coupled version of RCM has no explicit source terms outside of advection through the outer boundary; a simple decay is added with a 10 hour e-folding rate.
It should be noted that stand-alone RCM versions have evolved different capability sets than the one described here \citep[\eg][]{Yang:2014a, Chen:2019c}.

\subsubsection{The Ring Current Atmospheres Interaction Model (RAM) Family}
\label{subsubsec:ram}

The original  Ring current Atmospheres interaction Model (RAM) is a fully kinetic bounce-averaged drift model of the ring current \citep{Fok:1993a, Jordanova:1994a, Liemohn:1998a}.
It solves the kinetic equation to yield the bounce-averaged distribution function as a function of azimuth, radial distance, energy and pitch angle for several species,

\begin{equation}
\begin{split}
    \Big \langle \frac{d Q_s}{dt} \Big \rangle = & \frac{\partial Q_s}{\partial t}
    + \frac{1}{R^2_\snull} \frac{\partial}{\partial R_\snull}\left(R^2_\snull \Big \langle \frac{d R_0}{dt}Q_s \Big \rangle \right) 
    + \frac{\partial}{\partial \phi}\left( \Big \langle \frac{d \phi}{dt} \Big \rangle Q_s \right) \\
    & + \frac{1}{\sqrt{E}}\frac{\partial}{\partial E} \left( \sqrt(E) \Big \langle \frac{d E}{dt} \Big \rangle Q_s \right) 
    + \frac{1}{h(\mu)\,\mu} \frac{\partial}{\partial \mu}
    \left( h(\mu) \,\mu \Big \langle \frac{d \mu}{dt} \Big \rangle Q_s \right)
    = \Big \langle \frac{d Q_s}{dt} \Big \rangle_{loss}
    \label{ram:dist}
    \end{split}
\end{equation}

\noindent where angle brackets denote average values over the bounce period, $Q_s$ is the distribution function for species $s$, $R_\snull$ is the radial direction in the equatorial plane, $phi$ is azimuth, $E$ is particle's kinetic energy, and $\mu$ is the cosine of the particle's equatorial pitch angle. Finally, $h$ is defined via,

\begin{equation}
    h(\mu) = \frac{1}{2R_\snull}\int^b_a \frac{ds}{\sqrt{1-B(s)/B_m}}
    \label{ram:h}
\end{equation}

\noindent where $a$ and $b$ are mirror points for a particle of a given $\mu$ along magnetic field line $B(s)$ with field strength of $B_m$ at the mirror point.
The relationship between RAM-like models and the RCM (see \sectionname~\ref{subsubsec:rcm}) is described by \cite{Heinemann:2001a}.

RAM has spawned many ``child'' codes, three of which are integrated into the SWMF and each with its own unique capabilities. 
These include The Hot Electron Ion Drift Integrator (HEIDI), the Comprehensive Inner Magnetosphere Model (CIMI), and the RAM with a Self-Consistent Magnetic field (RAM-SCB) codes. 
These models all fall into the same class of bounce-averaged models, but have unique implementations as well as differences in variable choice, grid formulation, and source terms that provide strengths for particular problems. Each is integrated into the SWMF such that fields throughout the ring current domain and plasma conditions about the outer boundary are obtained from GM and IE components; pressure and density values are returned to GM following the approach given by \equationname~(\ref{rcm:couple}).

HEIDI expands upon the original RAM model in several ways. Its usage has mostly focused on large-scale dynamics of the inner magnetospheric pressure and current systems, including tracking all source and loss terms \cite[]{Liemohn:1999a, Liemohn:2002a}. A key development was the inclusion of self-consistent electric field calculations \citep{Liemohn:2004a}, allowing for the analysis of conductance influences on the ring current \citep{Liemohn:2005a}, and the eventual inclusion of self-consistent conductance calculation from electron precipitation \citep{Perlongo:2017a}.  
A broad set of geocoronal models is available within HEIDI, allowing for deep investigations of ring current decay \citep{Ilie:2013a}. Also, HEIDI includes a robust definition of non-dipolar drift suitable for an arbitrary magnetic field description \citep{Ilie:2012a} and can now account for the effects of the inductive electric field as well, giving a more dynamic picture of ring current development \citep{Liu:2021a}. It runs within the SWMF with RIM and work is in progress toward full coupling with BATS-R-US.

The CIMI model represents the combination of two earlier models, the Comprehensive Ring Current Model (CRCM) and the Radiation Belt Environment (RBE) code, to create a complete model of the plasmasphere, ring current, and radiation belt populations of the inner magnetosphere \citep{Fok:2014a}. The earlier RBE and CRCM codes were coupled into the SWMF \citep{Glocer:2009a, Glocer:2013a}, and CIMI uses an improved version of these couplers that allows it to couple to BATS-R-US configured with single fluid MHD, anisotropic MHD, and multi-fluid MHD \citep{Glocer:2018a, Glocer:2020a}.   The CIMI grid is located at the magnetic field ionospheric foot points, which simplifies the calculation of the $E \times B$ drift as the ionospheric potential does not need to be mapped to the equator. In addition to using different spatial coordinates, CIMI solves a version of \equationname~(\ref{ram:dist}) recast in a different set of velocity space coordinates. Namely, the version of CIMI in SWMF uses $\mu_a$ and $K$ coordinates that correspond to the first and second invariants of motion. This approach lets CIMI represent the advection portion of the transport in a conservation form which can be treated with standard finite volume methods. Specifically, the advection portion of the code can be written as \cite[]{Fok:2021a}: 

\begin{equation}
    \frac{\partial F_s}{\partial t} + \frac{\langle \dot{\lambda_i}\rangle F_{s}}{\partial \lambda_{i}}+\frac{\langle \dot{\phi_i}\rangle F_s}{\partial \phi_i}= S_i
    \label{eq:cimi}
\end{equation}

\noindent where $\lambda_i$ and $\phi_i$ are latitude and azimuthal angle of the field line foot point,  $F_s$ is the bounce averaged distribution function times a Jacobian for a particular $\mu_a$ and $K$ (see \cite{Fok:2021a} for details). 
In addition to advection on the left hand side of \equationname~(\ref{eq:cimi}), wave diffusion terms are often included on the right hand side, shown here only as $S_i$ to represent wave-particle interactions with Chorus, hiss, and EMIC waves, which are critical to modeling local acceleration of radiation belt electrons as well as scattering into the loss cone and subsequent precipitation. 
Additional components of $S_i$ include charge exchange loss, loss cone loss, and the effect of Coulomb collisions. 
Complete descriptions of these terms are given in \cite{Fok:2021a} along with new forms of \equationname~(\ref{eq:cimi}) in different coordinates that will eventually be included in the SWMF version. 
Finally, it is interesting to note that when CIMI is coupled to BATSRUS, the magnetic and electric fields are naturally self-consistent. 
\cite{Meng:2013a} demonstrated that the magnetic field calculated in BATS-R-US  using the coupled CIMI pressure is consistent with a force balance. Similarly, the pressure feedback from CIMI drives currents in BATS-R-US that contribute the ionospheric potential and hence the convection in CIMI.  

RAM with Self-Consistent Magnetic field (RAM-SCB) uses an Euler potential representation of the magnetic field to achieve a self consistent magnetic field configuration inside the model's domain \cite[]{Zaharia:2006a, Jordanova:2010a}.
It has an energy range of approximately 100\,$eV$ to 500\,$keV$.  
Loss terms include charge exchange, Coulomb collisions and  atmospheric loss at low altitudes.  
The RAM model was updated to use nondipolar field geometries \cite[]{Jordanova:2006a, Jordanova:2010a}.
This improvement allows for integration of the  3-D force balance magnetic field model (SCB, 
\cite[]{Zaharia:2004a, Zaharia:2008a}).  
This model balances the $\overline{\bj}\times\overline{\bB}$ force with the divergence of the general pressure tensor to calculate the magnetic field configuration within its domain.  
The domain ranges from near the Earth's surface, where the field is assumed dipolar, to the shell created by field lines passing through the equatorial plane at a radial distance of 6.5 $R_{\sE}$.  
Anisotropic pressure both at the outer boundary and inside the code's domain is required and is provided by RAM.  
By relying on anisotropic pressure calculated by RAM, the force balance model creates a more stretched, more realistic field than isotropic MHD models that do not capture the ring current pressure build up and are typically very dipolar within 6.6 $R_{\sE}$.
Initial coupling of these two codes is detailed in \cite{Zaharia:2005a}, \cite{Jordanova:2006a}, and \cite{Zaharia:2006a}; details about the full coupling can be found in \cite{Zaharia:2010a}.  RAM provides anisotropic pressure to the 3D equilibrium code, which in return calculates the field aligned integrals required by RAM to calculate particle drift paths.  
The addition of self consistency creates significant differences in the ring current drift paths \cite[]{Jordanova:2006a} and a depression in the nightside magnetic field \cite[]{Zaharia:2006a}.
RAM-SCB continues to see improvements in its algorithms and implementation related to robustness, efficiency, and performance during extreme driving \cite[]{Engel:2019a}.
A comprehensive discussion of its coupling within the SWMF is given by \cite{Welling:2018a}.

\subsection{Energetic Particle Models}
\label{subsec:ep}

\subsubsection{SEP Models}
\label{subsubsec:sepmodels}
The acceleration and transport processes of energetic particles in interplanetary space is described by the focused transport equation in which the particle's gyrophase is averaged out and the particle's motion is reduced to the guiding center's motion along the magnetic field and diffusion due to magnetic turbulence \citep{Skilling:1971a, Kota:2000a, Kota:2005b, Qin:2006a, Zhang:2009a, Zhao:2016b, Zhao:2017a}

\begin{equation}
\frac{\partial f}{\partial t}+\mu v \frac{\partial f}{\partial s} + (\overline{\bu}\cdot \grad) f +
\frac{dp}{dt}\frac{\partial f}{\partial p}+\frac{d\mu}{dt}\frac{\partial f}{\partial \mu}-\frac{\partial}{\partial \mu}\left(D_{\mu\mu}\frac{\partial f}{\partial \mu}\right)=Q,
\end{equation}

\noindent where $f$ is the particle's distribution function, $\mu$ is the cosine of the particle's pitch angle, $D_{\mu\mu}$ is the pitch angle diffusion coefficient, $\overline{\bu}$ is solar wind velocity, $p$ is the particle's momentum in the solar wind frame, $s$ is the direction along the magnetic field, $v$ is the particle's speed, and $Q$ is the source term. 

In the diffusive limit, where the distribution function is assumed to be isotropic, the focused transport equation reduces to the original \cite{Parker:1965a} equation 

\begin{equation}
    \frac{\partial f}{\partial t} + (\overline{\bu} \cdot \grad) f-\frac{1}{3} (\divg \overline{\bu} \frac{\partial f}{\partial \ln{p}} = \divg(\bar{\bar{\bD}}\cdot\grad f),
\end{equation}

\noindent where $\bar{\bar{\bD}} = D\,\bb\,\bb$ is the diffusion tensor along the magnetic field, with $D$ being the scalar diffusion coefficient. In the Lagrangian coordinates advecting with the background plasma, the above governing equation is reduced to 

\begin{equation}
    \frac{df}{dt} + \frac{1}{3} \frac{d\ln{\rho}}{dt} \frac{\partial f}{\partial \ln{p}} = B \frac{\partial}{\partial s}\left(\frac{D}{B} \frac{\partial f}{\partial s} \right),
\end{equation}

\noindent where $\rho(s,t)$ and $B(s,t)$ are plasma density and total magnetic field magnitude along the magnetic field lines.
The 3-D problem is then reduced to a set of independent 1-D problems on those time-evolving Lagrangian grids \cite[]{Sokolov:2004a}.
In M-FLAMPA the plasma and turbulence parameters along the magnetic field lines are extracted dynamically from the the BATS-R-US models (the SC, IH and OH components).
The diffusion process is treated as pitch angle scattering of the particles by the magnetic \alf waves calculated self consistently within the BATS-R-US simulation.

\subsubsection{GCR Models}
\label{subsubsec:gcrmodels}
Transport of GCRs in the heliosphere is affected by the solar modulation, which results in a dynamical change of the GCR energy spectrum and anisotropy as they propagate in the heliosphere  \citep[e.g.,][]{Vainio:2008a}. 
The theory of modulation is based on solving the \cite{Parker:1965a} equation with cosmic ray drift

\begin{equation}
	\frac{\partial f}{\partial t} + (\overline{\bu}\cdot\grad) f + \left(\overline{\Bv}_\sD \cdot \grad\right) f -\frac{1}{3} (\divg\overline{\bu})\frac{\partial f}{\partial\ln p}= \divg(\bar{\bar{\bD}}\cdot\grad f)Q,
\label{Parker-Equation}
\end{equation}

\noindent where $f (r,p,t)$ is the omnidirectional distribution function of GCRs, $p$ the particle's momentum, $r$ the heliocentric distance, $\overline{\bu}$ the solar wind velocity, $\bar{\bar{\bD}}$ the symmetric part of the diffusion tensor, $\overline{\Bv}_\sD$ the pitch-angle averaged guiding center drift velocity \cite[\cf][]{Gombosi:1998book}, and $Q$  defines the source of the GCRs. 

Equation~(\ref{Parker-Equation}) describes the main transport processes: 1) diffusion of particles due to their scattering off magnetic inhomogeneities, 2) convection in the out-streaming solar wind, 3) two types of drifts: the gradient-curvature drift in the regular heliospheric magnetic field, and drift along the heliospheric current sheet, and 4) adiabatic energy losses in the expanding solar wind. These processes are defined by the geometrical structure, polarity, strength, and the level of turbulence in the IMF and solar wind, which are ultimately driven by variable solar activity, leading to the temporal variability of the modulation on different timescales. 

\subsection{Simulating Virtual Magnetic Observatories}
\label{subsec:groundmf}
There is an increasing number of ground magnetometer stations that provide magnetic field measurements. These observations can be directly compared to simulated observations obtained from SWMF simulations. The large number and high cadence of ground observations necessitates development of a fast and accurate way to generate synthetic magnetometer observations from our simulation results. Here we describe the algorithm that is used in the SWMF suite. 

\subsubsection{Biot-Savart Integral for Currents in the Magnetosphere}
\label{subsubsec:biosavart}
The contribution from the magnetospheric current system to the ground magnetic field is given by the Biot-Savart integral (neglecting the displacement current):

\begin{equation}
\label{eq:i0}
    \overline{\bB}_\sM(\overline{\bx}_\snull) = \frac{\muo}{4\pi} \int\limits_{|\overline{\bx}|>R_\sM} \overline{\bJ}(\overline{\bx}) \times \frac{\overline{\bx}_\snull - \overline{\bx}}{|\overline{\bx}_\snull - \overline{\bx}|^3} \, dV,\qquad \overline{\bJ}(\overline{\bx})=\frac1{\mu_0}\grad_{\overline{\bx}}\times \overline{\bB}(\overline{\bx}),
\end{equation}

\noindent where $\overline{\bx}_\snull$ is the point where we calculate the synthetic magnetic field, $\overline{\bB}_\sM(\overline{\bx}_\snull)$ is the contribution to this field from the magnetosphere currents, $R_\sM$ is radius of the magnetosphere-ionosphere boundary (in our case the inner boundary of the magnetosphere model, $R_\sM\approx2.5 R_\sE$ in most simulations) and $dV=d^3\overline{\bx}$ is the volume element. The current density at a point in the simulation domain is expressed from Amp{\`e}re's law and the gradient operators $\grad_{\overline{\bx}}$ and $\grad_{\overline{\bx}_\snull}$  differentiate over coordinates $\overline{\bx}$ and $\overline{\bx}_\snull$, respectively. The gradient of the inverse distance function is

\begin{equation}
\label{eq:overxx0}
    \grad_{\overline{\bx}} \frac1{|\overline{\bx}_\snull-\overline{\bx}|} = -\grad_{\overline{\bx}_\snull} \frac1{|\overline{\bx}_\snull-\overline{\bx}|} = \frac{\overline{\bx}_\snull - \overline{\bx}}{|\overline{\bx}_\snull-\overline{\bx}|^3}.
\end{equation}

\noindent Now the Biot-Savart integral (\equationname~\ref{eq:i0}) can be written as

\begin{equation}
\label{eq:i1}
    \overline{\bB}_\sM(\overline{\bx}_\snull) = \frac{1}{4\pi} \grad_{\overline{\bx}_\snull} \times \int\limits_{|\overline{\bx}|>R_\sM} \frac{\grad_{\overline{\bx}} \times \overline{\bB}(\overline{\bx})} {|\overline{\bx}_\snull-\overline{\bx}|} dV
\end{equation}

\noindent Next, we expand the double vector product using the $\grad_{\overline{\bx}}\grad_{\overline{\bx}_\snull}|\overline{\bx}_\snull-\overline{\bx}|^{-1}=\grad_{\overline{\bx}_\snull}\grad_{\overline{\bx}}|\overline{\bx}_\snull-\overline{\bx}|^{-1}$, $\grad_{\overline{\bx}}\cdot \overline{\bB}(\overline{\bx})=0$,  and  $\grad^2_{\overline{\bx}}|\overline{\bx}_\snull-\overline{\bx}|^{-1}=-4\pi\delta^3(\overline{\bx}_\snull - \overline{\bx})$ identities:

\begin{align}
\label{eq:i6}
    \overline{\bB}_\sM(\overline{\bx}_\snull) =& \frac{1}{4\pi} \int\limits_{|\overline{\bx}|>R_\sM} \grad_{\overline{\bx}} \left[\overline{\bB}(\overline{\bx}) \cdot \grad_{\overline{\bx}_\snull} \left(\frac{1}{|\overline{\bx}_\snull-\overline{\bx}|}\right) \right]\,dV 
    -\frac{1}{4\pi} \int\limits_{|\overline{\bx}|>R_\sM}\grad_{\overline{\bx}}\cdot \left[\overline{\bB}(\overline{\bx}) \grad_{\overline{\bx}_\snull} \left(\frac{1}{|\overline{\bx}_\snull - \overline{\bx}|}\right)\right]\,dV
\qquad \nonumber \\ &
    -\frac{1}{4\pi}\int\limits_{|\overline{\bx}|>R_\sM}\grad_{\overline{\bx}}\cdot\left[\grad_{\overline{\bx}_\snull} \left(\frac{1}{|\overline{\bx}_\snull-\overline{\bx}|}\right)\overline{\bB}(\overline{\bx})\right] \, dV
    +\int\limits_{|\overline{\bx}|>R_\sM} \overline{\bB}(\overline{\bx}) \, \delta^3(\overline{\bx}_\snull - \overline{\bx}) \,dV
\end{align}

\noindent If the observation point, $\overline{\bx}_\snull$, is outside the simulation domain the last integral is zero. Since we are considering magnetic perturbations on the ground -- that is outside the simulation region -- this last integral can be neglected. Finally, we introduce the unit vector  $\overline{\Bn}_\sR = \overline{\bx}/|\overline{\bx}|$ (note, that positive $\overline{\Bn}_\sR$ points into the domain of integration) and apply Gauss' theorem to \equationname~(\ref{eq:i6}):

\begin{equation}
\label{eq:eqv1}
    \overline{\bB}_\sM(\overline{\bx}_\snull) = \int\limits_{|\overline{\bx}|=R_\sM} \left[\sigma_m(\overline{\bx}) \frac{\overline{\bx}_\snull - \overline{\bx}} {|\overline{\bx}_\snull - \overline{\bx}|^3} + \frac{\muo}{4\pi}\mathbf{i}_m(\overline{\bx}) \times \frac{\overline{\bx}_\snull - \overline{\bx}} {|\overline{\bx}_\snull - \overline{\bx}|^3} \right] \, dS,
\end{equation}

\noindent where $dS$ is an area element on the spherical surface, $|\overline{\bx}|=R_\sM$. Note that \equationname~(\ref{eq:eqv1}) replaces the effect of all currents in the simulated magnetosphere with a surface current, $\mathbf{i}_m(\overline{\bx}) = -\Bn_\sR \times \overline{\bB}(\overline{\bx})/\muo$ and magnetic surface charge, $\sigma_m(\overline{\bx}) = -\Bn_\sR \cdot \overline{\bB}(\overline{\bx})/4\pi$. For the special case when $\overline{\bx}_\snull=0$ \equationname~(\ref{eq:eqv1}) reduces to

\begin{equation}
\label{eq:eqv2}
    \overline{\bB}_\sM(0) = \int\limits_{|\overline{\bx}|=R_\sM}  \frac{\overline{\bB}(\overline{\bx})\,dS}{4\pi R_\sM^2}
\end{equation}

\noindent This result agrees with the well-known property of a potential field at the center of a sphere that equals the average of the field over a spherical surface \cite[see][]{Jackson:1975a} as long as the field given by \equationname~(\ref{eq:eqv1}) is created by currents located outside the sphere. 

\subsubsection{Magnetic Field Perturbations Caused by Field-Aligned Currents}
\label{subsubsec:mfcurrent}
Another source of geomagnetic variations, which is particularly significant at high geomagnetic latitudes, is the magnetic field produced by the currents connecting the magnetosphere-ionosphere boundary, $|\overline{\bx}|=R_\sM$ to the ionosphere and closing there. The currents in the \textit{gap region}, between $|\overline{\bx}|=R_\sM$ and the ionosphere, $R_\sI \le |\overline{\bx}^\prime|\le R_\sM$ are field-aligned, which  means that the current density is parallel or anti-parallel to the terrestrial magnetic field: $\overline{\bJ}(\overline{\bx}^\prime) \,\parallel \, \overline{\bB}_\snull(\overline{\bx}^\prime)$, where $\overline{\bx}^\prime$ represents a point in the gap region. This assumption allows us to derive the magnetic field from the field aligned currents. 

Through the boundary of each surface element at the M-I boundary, $dS$, a flux tube may be traced to the ionosphere boundary. The total current enclosed by this flux tube is

\begin{equation}
\label{eq:mi1}
    dI = \frac{1}{\muo} \Bn_\sR \cdot \left[\grad_{\overline{\bx}} \times \overline{\bB}(\overline{\bx})\right] \,dS
\end{equation}

\noindent The field line (and current line) is described by $d\overline{\bx}^\prime/d\ell = \pm \overline{\bb}_\snull(\overline{\bx}^\prime)$, where $d\ell$ is the path length element, and $\bb_\snull = \overline{\bB}_\snull / |\overline{\bB}_\snull|$. Expressing $d\ell = \pm dR^\prime / [\Bn_{R^\prime} \cdot \bb_\snull (\overline{\bx}^\prime)]$ in terms of the element of radial coordinate in the gap region, $R^\prime=\sqrt{|\overline{\bx}^\prime|^2}$, one can express the equation for the current line in terms of $dR^\prime$:

\begin{equation}
\label{eq:line}
    \frac{d\overline{\bx}^\prime}{dR^\prime} = \frac{\bb_\snull(\overline{\bx}^\prime)} {\Bn_{R^\prime} \cdot \bb_\snull(\overline{\bx}^\prime)}.
\end{equation}

The magnetic field line given by \equationname~(\ref{eq:line}) should be integrated from $R_\sM$ down to $R_\sI$, starting from each point $\overline{\bx}$ at the M-I interface. This way a multitude of field lines, $\overline{\bx}^\prime(\overline{\bx},R^\prime)$ are obtained in the $R_\sI\le R^\prime\le R_\sM$ domain. Equation~(\ref{eq:eqv1}) can now be generalized to account for the Biot-Savart integral from the multitude of field aligned currents, $dI$, in the gap region, 

\begin{equation}
\label{eq:gap}
    \overline{\bB}_\text{Gap}(\overline{\bx}_\snull) = \frac{1}{4\pi} \int\limits_{|\overline{\bx}| = R_\sM} \Bn_\sR \cdot \left[\grad_{\overline{\bx}} \times \overline{\bB}(\overline{\bx})\right] \int\limits_{R_{\rm I}}^{R_{\rm MI}} \frac{\bb_0 (\overline{\bx}^\prime(\overline{\bx},R^\prime))}{\mathbf{n}_{R^\prime}\cdot\bb_0(\overline{\bx}^\prime(\overline{\bx},R^\prime))} \times \frac{\overline{\bx}_\snull - \overline{\bx}^\prime(\overline{\bx},R^\prime)} {|\overline{\bx}_\snull - \overline{\bx}^\prime(\overline{\bx},R^\prime)|^3}dR^\prime dS.
\end{equation}

\noindent Note, that the integral over $dR^\prime$ in \equationname~(\ref{eq:gap}) is a complicated vector function. However, its value only depends on $\overline{\bx}$ and $\overline{\bx}_\snull$, and consequently, for any given computational grid and set of surface points, this function can be calculated only once (at the beginning of the simulation). After this, the contribution from the currents in the gap region, similarly to that from the magnetosphere currents, is given by the surface integral over the M-I interface. We also note that only the derivatives of tangential components of the magnetospheric field at the M-I interface contribute to the radial component of the current density. Therefore, only the magnetic field at the M-I interface contributes to the surface magnetic variation, but not its radial gradient.

Equation~(\ref{eq:gap}) is a very important result for computational efficiency. It says that the Biot-Savart integral along magnetic field lines going through the gap region can be written as the field-aligned current multiplied by a constant that only depends on the field line that is approximated and the location of the point where the magnetic perturbation is calculated. These constants can be precalculated and stored (properly distributed among the processors) so that the integrals become a simple weighted sum, which is much faster to calculate than the integrals.

The electric current is a sum of contributions from separate current bundles (flux tubes), therefore one can use the Biot-Savart integral to obtain the perturbation magnetic field at a surface point, $\overline{\bx}_\snull$, produced by the flux tube current, $I$, for a flux tube described by $\overline{\bx}^\prime(\overline{\bx}_\sM,R^\prime)$:

\begin{equation}
\label{eq:perturb}
    \delta\overline{\bB}(\overline{\bx}_\snull) = \frac{\muo I}{4\pi}
    \int\limits_{R_\sI}^{R_\sM} {\left[ \frac{d\overline{\bx}^\prime(\overline{\bx}_\sM,R^\prime)}{dR^\prime} \times \frac{\overline{\bx}^\prime(\overline{\bx}_\sM,R^\prime) - \overline{\bx}_\snull}{|\overline{\bx}^\prime(\overline{\bx}_\sM,R^\prime) - \overline{\bx}_\snull|^3} \right] dR^\prime},
\end{equation} 

\noindent where the magnetic field line passing through the point $\overline{\bx}_\sM$ at the M-I interface is parameterized in the gap region with the radial coordinate, $R^\prime = |\overline{\bx}^\prime|$ ($R_\sI \le |\overline{\bx}^\prime| \le R_\sM$, and $d\overline{\bx}^\prime = (d\overline{\bx}^\prime / dR^\prime) \,dR^\prime$).

The expression for the magnetic field line in the gap region greatly simplifies if the Earth's  magnetic field is described in the dipole approximation. In this case the differential equation for the magnetic field line can be easily solved by projecting this vector equation on the direction of $\overline{\be}_\sM$ and on the two perpendicular directions. The solution is

\begin{equation}
\label{eq:xprime}
    \overline{\bx}^\prime(\overline{\bx}_\sM,R^\prime)=\left\{
    \overline{\bx}_\sM + \overline{\be}_\sM \left[\sgn(z_\sM) \sqrt{\frac{R_\sM^3}{R^\prime} + z_\sM^2-R_\sM^2} - z_\sM\right]\right\} \left(\frac{R^\prime}{R_\sM}\right)^\frac{3}{2}
\end{equation}

\noindent where we used the notation, $z_\sM = \overline{\bx}_\sM \cdot \overline{\be}_\sM$. This expression can be further simplified if we use one of the standard geocentric coordinate systems with the $z$-axis aligned with the direction of the magnetic dipole moment (the magnetic axis), such as MAG or SM \cite[see][]{Franz:2002a}.

\subsection{Geomagnetic Indexes}
\label{subsec:indexes} 
Geomagnetic indexes, including Dst, Kp, and the AE family of values, are a regular product of the SWMF Geospace.
Dst is approximated via a single Biot-Savart integral of all currents flowing in the GM component (typically, BATS-R-US at Earth, see \sectionname~\ref{subsubsec:biosavart}).
Kp and AE indexes leverage virtual magnetometer stations to more closely reflect the calculation of their real-world counterparts.

Real-world Kp is the average of local-K index values calculated from 13 mid-latitude ground observatories, rounded to the nearest third.
Local-K is a  range index of the maximum minus the minimum disturbance at a single observatory over set three-hour windows (0-3 UT, 3-6 UT, etc), scaled to a 9-value integer via a semi-logarithmic transformation.
The scale factors are station specific; additional adjustments are made for season. 
In the SWMF, the calculation is optimized for simplicity and performance.
For the Kp calculation, twenty-four stations are used, spread equally about local time and all placed at a constant geomagnetic latitude of 60$^{\circ}$, and are scaled such that $K=9$ corresponds to $\ge 600 nT$.
Virtual Kp is then the average of these 24 stations, rounded to the nearest third.
Because the value is written to file during the simulation, set three-hour windows are not possible.
Instead, a rolling three-hour window that ends at the current simulation time is used.
Therefore, SWMF virtual Kp range windows line up exactly with the real-world index when the universal time hour is a multiple of three.
No seasonal adjustments are made for virtual Kp.
The use of the 24-station approach provides a minor improvement in predictive performance compared to using the real-world stations and scalings.

The Auroral Electrojet (AE) family of indexes are high-latitude data products.
The real-world indexes are the product of the geomagnetic north-south perturbations obtained from 13 observatories.
AU is the maximum perturbation of the 13 stations as a function of time (reported minutely); AL is the minimum; AE is the difference of AU and AL, and AO is the average of AU and AL.
The calculation of virtual AE follows this closely except for the location of the contributing stations:
24 stations at a constant magnetic latitude of 70$^{\circ}$ evenly spaced in local time are used.
Further work is required to optimize the location and number of stations used for virtual AE indexes.

\subsection{New Diagnostics}
\label{subsec:diagnostics}
The optically thin solar corona emits in the XUV, visible and IR wavelengths. Currently, there is only one mission (Parker Solar Probe) that takes in-situ, local measurements of the solar corona, all information of the global corona measured by ground-based observatories (visible and IR) or via spacecrafts carrying remote-sensing instrumentation (XUV). With synthetic observations of the solar corona one can evaluate the solar corona model performance, decompose and analyse physical process and the formation the radiation output. Within the SWMF the solar corona can be visualized via synthetic narrow-band imaging (line-of-sight, LOS) or synthetic spectra (SPECTRUM). 

\subsubsection{Line-of-Sight Images}
\label{subsubsec:los}

SWMF has the capability	of generating various synthetic line-of-sight (LOS) plots, such as EUV images. The response	$\mathcal{R}$ of each pixel of the image is treated as a LOS integral of a function $f$ through the plasma:

\begin{equation}
  \mathcal{R} = \int f(\ell)\, d\ell,
 \label{eqn:los}
\end{equation}

\noindent where $\ell$ follows along the LOS.

The LOS algorithm is implemented in the following parallel way: For each LOS ray and each grid block we determine first the segment of the ray that intersects the block. Then for each ray and block the function $f$ is tri-linear interpolated along the LOS and integrated according to equation (\ref{eqn:los}) using a trapezoidal rule. The step size of the integration is proportional to the cell size of the block. Once the integration is finished for all blocks, we add for each LOS ray all integrals over the block segments via MPI reduce.

\subsubsection{SPECTRUM}
\label{subsubsec:spectrum}
The Spectral Calculations for Global Space Plasma Modeling (SPECTRUM) code \cite[]{Szente:2019a} calculates emissions from the optically thin solar corona by combining AWSoM(-R) simulation results with the CHIANTI database \cite[]{Dere:1997a, Dere:2019a}. Doppler-shifted, nonthermal line broadening due to low-frequency Alfv{\'e}n waves and anisotropic proton and isotropic electron temperatures can be individually taken into account during the calculations. The synthetic spectral calculations can then be used for model validation, for interpretation of solar observations, and for forward modeling purposes. SPECTRUM is implemented within the Space Weather Modeling Framework (SWMF) and is publicly available.

SPECTRUM is a post-processing tool within the SWMF: it processes output after a simulation is completed. It is a stand-alone Fortran code that can process output files originating from any global coronal model, assuming that the data set is formatted appropriately. Currently, SPECTRUM can handle either a Cartesian-grid or the BATS-R-US unstructured grid. SPECTRUM uses the same LOS integration technique as described in \sectionname~\ref{subsubsec:los}. 
\begin{figure}
  \begin{floatrow}
    \ffigbox[\FBwidth]
      {\includegraphics[trim={3.5cm .5cm 1cm 4.5cm},clip,width=0.9\linewidth]{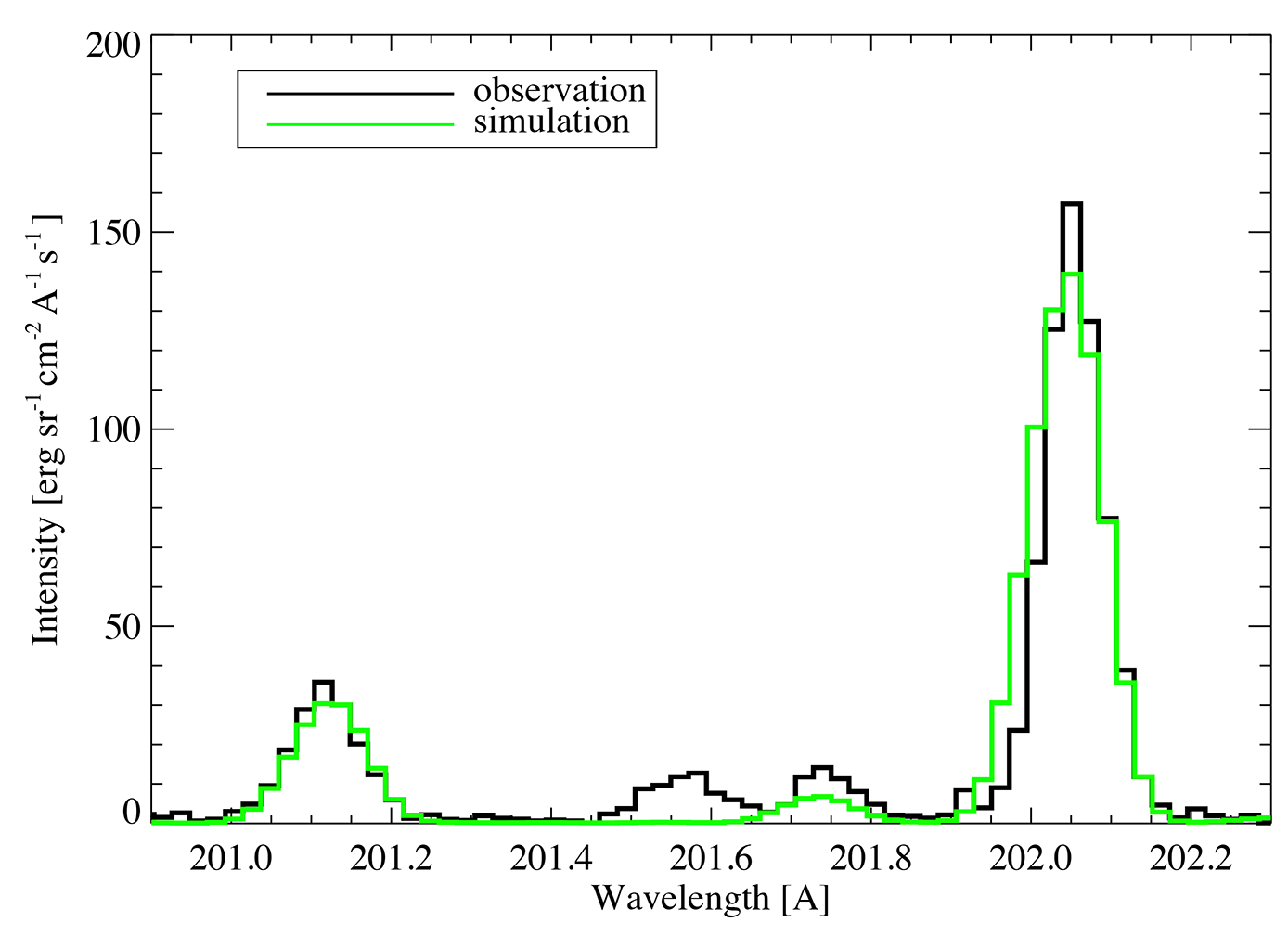}}
      {\caption{Synthetic spectra (green) compared to observation (black) taken by Hinode/EIS at 12 November 2007 12:32:02 UT of the Northern coronal hole during the Carrington Rotation~2063 (from \cite{Szente:2019a}). Line profiles of Fe~XI 201.734~\AA, Fe~XIII 201.121~\AA\ and 202.044~\AA\ are closely predicted in intensity, width and Doppler-shift; while there is no line in the CHIANTI database between 201.5-201.6~\AA, which explains the missing peak in the model's result. }
      \label{fig:spectra:line}}
    \ffigbox[\Xhsize]
      {\includegraphics[trim={4.5cm .5cm 6.cm 5.cm},clip,width=0.9\linewidth]{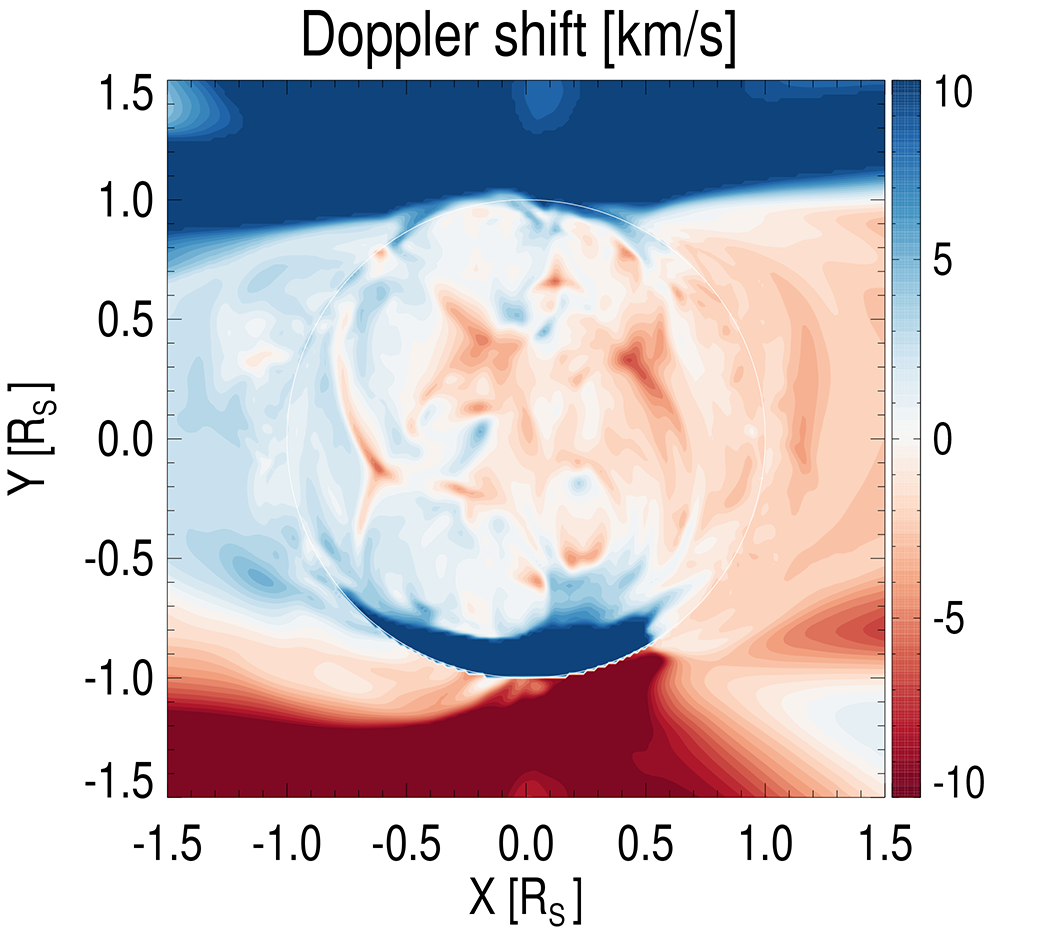}}
      {\caption{Doppler map of line Fe~XIII 202.044~\AA\ is obtained from synthetic spectral image using the magnetic boundary of Carrington Rotation~2082, as could be perceived from Earth after the line-of-sight integration of individual emission lines (from \cite{Szente:2019a}). Blueshifts show regions where the solar wind is moving toward the observer, redshifts are present where plasma is moving away from the observer. The denser the plasma, the more dominant its effect is on the overall integrated Doppler shift of the observed line.}
      \label{fig:spectra:doppler}}
    \end{floatrow}
\end{figure}

The spectral calculation is performed the following way.
It is assumed that the ion emissions coming from a given volume element $dV$ follow a Gaussian profile, centered at wavelength $\lambda_0$ with line width $\Delta\lambda$:

\begin{equation}
\label{eqn:spectrum:gaussian}
\phi \left( \lambda \right) = \frac{1}{\sqrt{2\pi}\Delta\lambda}
  e^{- \left(\frac{\lambda - \lambda_0}{2\Delta\lambda}\right)^2} .
\end{equation}

The total flux at an instrument's detector at distance $d$ is the sum of all the emission along the LOS from each volume element:

\begin{equation}
  \label{eqn:spectrum:flux}
  F = \frac{1}{4\pi d^2} \int_V N\left(X_{j}^{+m} \right) A_{ji}h\nu_{ij}dV ,
\end{equation}

\noindent where $N\left(X_{j}^{+m}\right)$ is the density of the emitting  $X_j^{+m}$ ions, $A_{ji}$ is the Einstein coefficient and $\nu_{ij}$ is the transition frequency from $j$ to $i$. Rewriting the expression into separate density- and temperature dependent terms (details in \cite[]{Szente:2019a}):

\begin{equation}
  \label{eqn:spectrum:fluxwithG}
  F = \frac{1}{4\pi d^2} \int_V G\left(T_e,N_e \right) N_e^2 dV ,
\end{equation}

\noindent $G\left(T_e, N_e\right)$ is the contribution function, slowly varying with density and strongly dependent on temperature. The contribution function for each ion is calculated and saved into a lookup table using tables and procedures from CHIANTI v9 \cite[]{Dere:1997a, Dere:2019a} from SolarSoft. 

SPECTRUM takes two input files, one is the tabulated contribution-function values and the BATS-R-US output saved from an AWSoM simulation. The following calculations are performed on a line-by-line basis, for each cell along the line-of-sight.
First we apply Doppler-shift to the line center:

\begin{equation}
  \label{eqn:spectrum:doppler}
  \lambda_{shifted} = \left(1 -\frac{u_{LOS}}{c_{light}} \right) \lambda_0 ,
\end{equation}
where $u_{LOS}$ is the line-of-sight bulk plasma velocity (positive toward the observer) and $c_{light}$ is the speed of light. 

The line width is the sum of instrumental broadening, and a thermal- and a non-thermal component:

\begin{equation}
  \label{eqn:spectrum:linebroadening}
  \Delta\lambda^2 = \Delta\lambda_{instrument}^2 + \lambda^2\frac{u_{th}^2+u_{nth}^2}{c_{light}^2}.
 \end{equation}    
  
Thermal broadening is calculated considering the contribution along the line-of-sight direction of the anisotropic temperature, the non-thermal component is due to the low-frequency Alfv\'en wave contribution along the line-of-sight.

The SPECTRUM code output is synthetic spectra or synthetic spectral images. An example is shown in Figures~\ref{fig:spectra:line} and \ref{fig:spectra:doppler}.

\section{Algorithms}
\label{sec:algorithms}
Continuous development of the numerical algorithms is a necessity in order to maintain state-of-the-art numerical models. Large interdisciplinary teams provide an ideal environment to learn about and adopt the best algorithms in a wide range of applications. Over the last two decades the algorithms of the SWMF and the models in it have improved tremendously. This section highlights some of the most important developments.

\subsection{Advanced Spatial Discretization Methods}
\label{subsec:5thorder}
The numerical error of the solution depends on several factors. In general the numerical error $\varepsilon$ generated by the spatial discretization (which usually dominates) in a single time step at a given grid cell can be approximately written as

\begin{equation}
    \varepsilon = k (\Delta x)^n
\label{eq:numerror}
\end{equation}

\noindent where $k$ is some coefficient depending on the numerical method, $\Delta x$ is the grid resolution and $n$ is the spatial order of the scheme. There are at least three ways to reduce numerical error: reduce the coefficient $k$, reduce the grid resolution $\Delta x$ or increase the order of the scheme $n$.

The most straightforward approach is to increase grid resolution. Doing this uniformly over the whole computational domain is very expensive. In fact, the computational cost scales roughly with $(\Delta x)^{-4}$ for a three-dimensional simulation, because the number of grid cells is $\propto(\Delta x)^{-3}$ and the time step $\Delta t$ has to be kept proportional to $\Delta x$. A much better approach is to increase the grid resolution only where it is necessary. Thanks to its CFD heritage, BATS-R-US was born with block-adaptive mesh refinement. This algorithm allows refining the grid where necessary, and coarsen it where possible. Using block-based adaptation instead of cell-based adaptation (or fully unstructured grids) has distinct advantages for high performance massively parallel codes. In the past 20 years, the original block-adaptive grid implementation has been improved, extended, and in fact, completely rewritten into the Block Adaptive Tree Library (BATL) \cite[]{Toth:2012a}). BATL can use an arbitrary number of ghost cells, works in 1, 2 or 3 dimensions, allows for non-Cartesian grids, allows grid adaptation based on geometric and physics-based criteria, and it has very efficient algorithms that scale well to a large number of CPU cores. 

Another way to change grid resolution is to use non-Cartesian grids. For example, a spherical grid naturally has smaller cells in the longitude and latitude directions near the surface of the central body (Sun, planet, moon) than further away, which is advantageous for typical applications. A further refinement is to use a non-uniform grid spacing by applying some non-linear stretching. A typical example is to make the grid points linear in the logarithm of the radius instead of the radius itself. Using $\ln(r)$ as a generalized coordinate will increase the radial resolution near the central body, which is usually beneficial. One can in fact use custom designed coordinate mapping to resolve specific regions, for example AWSoM uses a special stretching to concentrate cells around the transition region of the Sun. Combining generalized coordinates and adaptive mesh refinement provides great flexibility in using the optimal grid for a given problem. 

The original version of BATS-R-US used a second order total variation diminishing (TVD) scheme, which was state-of-the-art in the 1990s \cite[]{Powell:1999a}. But computational fluid dynamics has evolved since then. Inspired by other codes, such as LFM \cite[]{Lyon:2004a}, we decided to extend BATS-R-US to use higher order schemes. The space physics applications require the solution of complicated partial differential equations typically in 3 spatial dimensions. The solutions often contain discontinuities, such as shocks or current sheets. TVD schemes excel in maintaining monotonic profiles across shock waves, but at discontinuities the TVD scheme falls back to the first order upwind scheme, which means that the accuracy is only linearly improving with the reduction of the grid cell size. 

In search of a suitable high-order scheme, we had the following requirements: 
\begin{outline}[enumerate]
	\setlist[enumerate,1]{itemsep=-1ex, topsep=0ex, labelwidth = 1ex, labelsep= 1ex,leftmargin=5ex,listparindent=2ex}
    \1 
    Minimal oscillations near discontinuities.
    \1 
    Conservative scheme that gives correct jump conditions.
    \1 
    High order at grid resolution changes and high order for non-Cartesian grids.
    \1 
    Small stencil to allow for small grid blocks.
    \1 
    Only moderately more expensive than the second order TVD scheme.
    \1 
    General method that is high order for various system of equations including non-linear terms.
\end{outline}

To meet these requirements is very challenging. We looked at existing codes and explored the options published in the literature and presented at meetings. It is important to note that higher than second order accurate finite volume schemes require a high order accurate integral (quadrature) of the fluxes over the cell faces and a high order accurate quadrature of the source terms in the cell volume, which makes them rather complicated and expensive in multi-dimensional simulations. The LFM \cite[]{Lyon:2004a} and GAMERA \cite[]{Zhang:2019b} codes, for example, are only higher than second order accurate in the finite difference sense for linear systems of equations. In addition, the use of a second order accurate update of the induction equation renders the overall scheme to be second order accurate only when the magnetic field plays an important role. Nevertheless, for the linearly high order donor cell algorithm the coefficient $k$ is small in \equationname~(\ref{eq:numerror}), which makes the LFM/GAMERA scheme exceptionally accurate, although still second order only ($n=2$). 
After considerable experimentation, we have opted for a conservative finite difference scheme based on the fifth order accurate monotonicity preserving (MP5) limiter. 

\begin{figure}[htb]
    \floatbox[{\capbeside
        \thisfloatsetup{capbesideposition={left,top},
        capbesidewidth=0.35\textwidth}}]{figure}[\FBwidth]
    {\caption{Comparison of synthesized EUV images of the model with observational \textit{STEREO A}/EUVI images. The columns are from left to right for 171\AA, 195\AA, and 284\AA. Top panels: synthesized EUV images for 2nd order scheme. Middle panels: synthesized EUV images for 5th order scheme. Bottom panels: observational \textit{STEREO A}/EUVI images. The observation time is 7 March 2011 20:00 UT.}
    \label{fig:EUV5vs2}}
    {\includegraphics[width=0.6\textwidth]{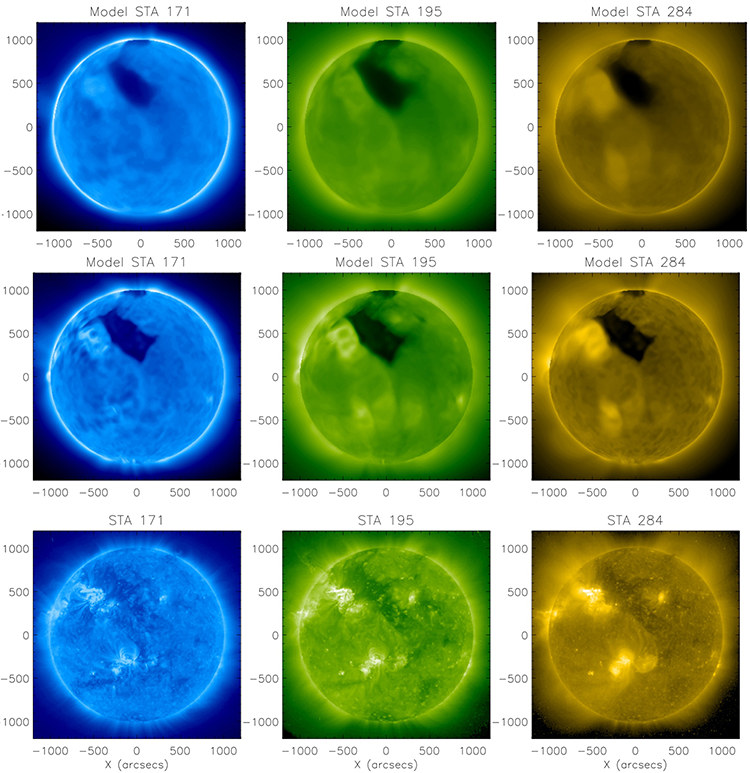}}
\end{figure}

We have developed a new 5th order scheme \cite[]{Chen:2016a} that satisfies all the requirements listed above. It is 5th order accurate for all terms in the MHD equations, it works for Cartesian and non-Cartesian grids alike, and it remains globally 5th order accurate with adaptive mesh refinement included. The stencil is quite compact, so only 3 ghost cells are needed, which means that the grid blocks can be as small as $6\times6\times6$ cells, which allows flexible adaptation. Using a third order Runge-Kutta scheme, the 5th order scheme is only about 3 times more expensive than the two-stage 2nd order TVD method. We can further reduce the computational cost by restricting the 5th order scheme to a part of the computational domain. In \figurename~\ref{fig:EUV5vs2}, we computed the {\it STEREO} images for the three Fe emission lines 171\AA, 195\AA, and 284\AA. The top row is for the AWSoM solar wind model using a 2nd order scheme, while the middle row is for using the 5th order scheme, which gives more detail and compares favorably with the observations (bottom panels)

\subsection{Advanced Time Integration Methods}
\label{subsec:timestepping}

Most numerical models employ an explicit time stepping scheme, where the values at the next time step in a given grid cell are calculated from the current values in the vicinity of this cell, the stencil. Explicit schemes are simple and fast, but the time step $\Delta t$ is limited by the Courant-Friedrich-Lewy (CFL) condition: 
\begin{equation}
    \Delta t < C \frac{\Delta x}{c_{\max}}
\label{eq:cfl}
\end{equation}
\ie\ it cannot exceed the distance of the neighboring cells $\Delta x$ divided by the fastest characteristic wave speed $c_{\max}$ of the system of equations. The proportionality factor $C$ depends on the numerical scheme, but it is typically less than unity. The CFL condition is a simple but fundamental consequence of causality. When the solution changes due to the fastest waves, there is no magical bullet, the explicit method is optimal. In many cases, however, the solution changes at a much lower rate, because the fastest modes are not present in the solution. For example, the magnetosphere of the Earth typically changes due to changes in the solar wind and not due to propagating fast magnetosonic waves. 

The BATS-R-US code has immediately benefited from the aerospace 
CFD expertise: a simple but incredibly efficient way to accelerate convergence to a steady state solution is local time stepping. Each grid cell takes the largest time step allowed by the Courant condition. While propagating time at different rates in different grid cells is not physical, the final steady state will be still correct, as it finds a balance of the divergence of the fluxes and the source terms, where the time step is a simple multiplier that makes no difference. Local time stepping has been used routinely in the aerospace community, but was virtually unknown in the space physics community. In combination with adaptive mesh refinement, BATS-R-US routinely obtains exact or approximate steady state solutions 10 to 1000 times faster than the simple explicit method on a static grid \cite[]{Toth:2012a}.

The Courant condition due to a fast wave mode is a specific example of stiffness. Stiffness means that the partial differential equations contain potentially large terms that happen to cancel each other out. Simple explicit time integration will blow up if the time step exceeds some restrictive limit. Implicit time integration offers a way to speed up the calculation: the fluxes and source terms are calculated from the values based on the next time step. Obviously, the values at the next time step are not yet known, hence the name, implicit. Typically, an implicit time integration scheme requires to solve a system of equations. The simplest example is a stiff source term, for example collisional terms among multiple species. Since these source terms do not involve spatial derivatives, one can solve the equation for state variables (for example densities) at each grid point independently, which is why this is called the point-implicit method. When the stiff terms involve spatial derivatives, for example heat conduction, the system of equations involve all the grid cells together. Typically, we employ an iterative scheme to solve a linearized system. Since the rest of the equations are solved explicitly, this method is called semi-implicit. Finally, one may solve the full system of equations implicitly with an iterative scheme, which alleviates all the stability restrictions. Solving a large system of equations is, of course, computationally expensive. It only makes sense if the time step can be increased sufficiently to beat the efficiency of the explicit method. The time step of an implicit scheme is always limited by accuracy considerations. The various implicit schemes in BATS-R-US originate from an interdisciplinary project of applied mathematicians, computer scientists and plasma physicists in the Netherlands in the 1990s \cite[]{Toth:1998a}.
\begin{figure}[htb]
    \floatbox[{\capbeside
        \thisfloatsetup{capbesideposition={left,top},
        capbesidewidth=0.4\textwidth}}]{figure}[\FBwidth]
    {\caption{The layered software structure of BATS-R-US. The arrows point from the module that is using data or methods from the other module. There are multiple versions of the equation and user modules. The various time stepping schemes are independent of the details of the equations being solved. (from \protect\cite{Toth:2012a})}
    \label{fig:bats-structure}}
    {\includegraphics[width=0.5\textwidth]{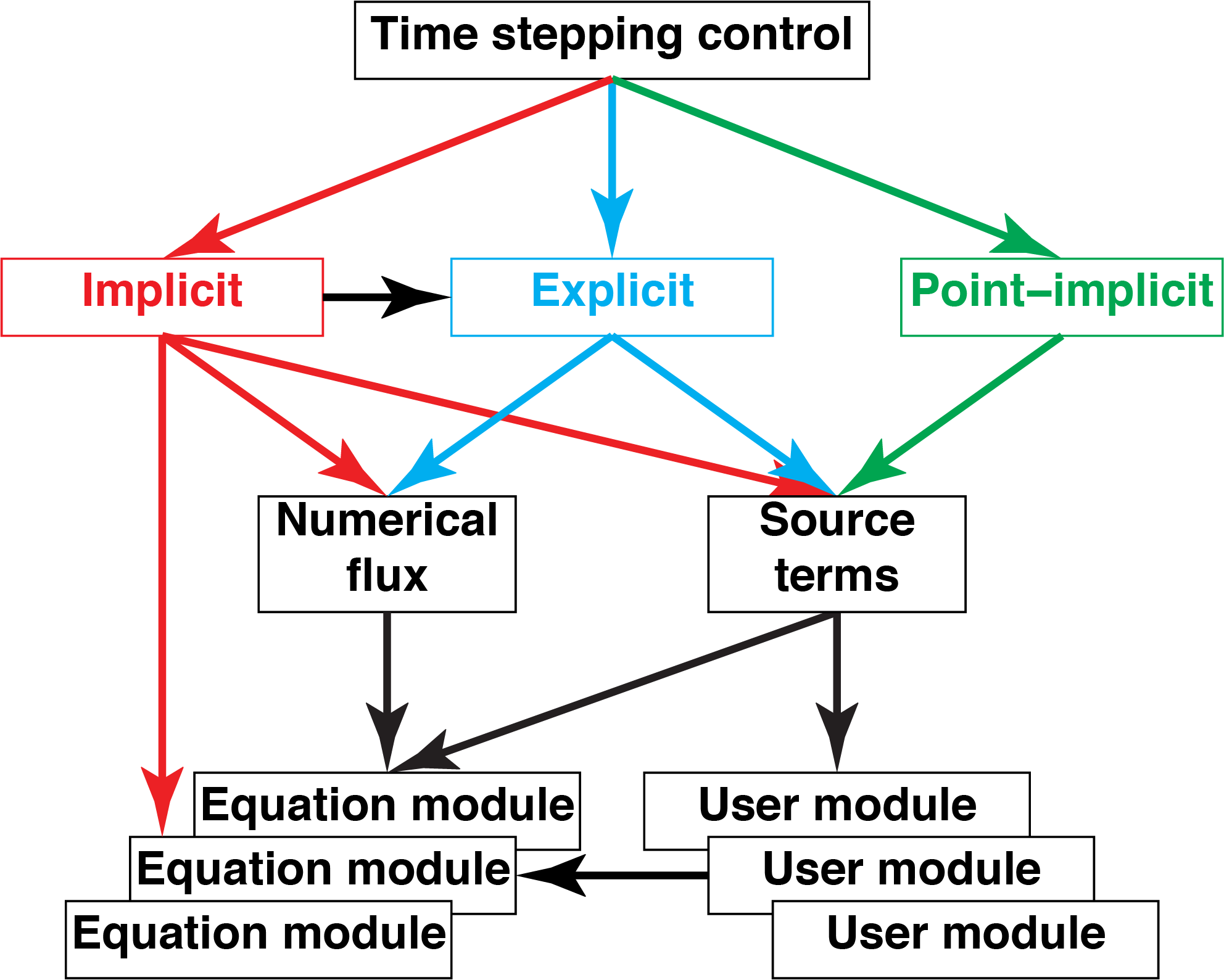}}
\end{figure}

One does not have to choose a certain time integration scheme for the whole computational domain. In fact, in some applications the point-implicit method is used only where the stiff source terms are present (for example the collisional terms are only important in the ionosphere of Mars), the semi-implicit scheme may be limited to the region where the Hall term is important (for example the magnetotail), and the implicit scheme for the full set of equations may also be combined with the explicit method in an adaptive manner based on the stability constraint for a given time step \cite[]{Toth:2006a}. In practice, we are using all of these schemes in various combinations. The optimal choice depends on the application and it can be orders of magnitude faster than the simple explicit time stepping. Figure~\ref{fig:bats-structure} shows how BATS-R-US implements the various time integration schemes in a hierarchical manner. This allows using the different schemes independently or combined for the various equation sets and applications.

A particularly interesting application of advanced time stepping algorithms is for particle-in-cell (PIC) codes. Explicit PIC models are limited by the Courant condition for light waves. In addition, the grid resolution is also limited: the model has to resolve the Debye length, which can be exceedingly small compared to the scales of the full system. The usual remedy is to artificially increase the electron mass and reduce the speed of light, but explicit PIC remains extremely expensive even with these tricks. Using an implicit algorithm removes both the spatial and temporal limitations: the semi-implicit particle-in-cell algorithm \cite[]{Brackbill:1982a} can use arbitrarily large grid cells and the time step is limited by the Courant condition for the thermal velocities of the particles rather than the speed of light.

This advance was a game changer for modeling large systems with a kinetic model. A further major improvement was the energy conserving semi-implicit method (ECSIM) developed by  \cite{Lapenta:2017a}. Energy conservation is crucial to maintain the long-term stability of PIC models. Before ECSIM, the original remedy was applying some smoothing on the electric field that required experimentation with the amount of smoothing and resulted in somewhat diffused solutions. ECSIM, however, did not take care of enforcing Gauss' law that the divergence of the electric field equals the net charge density, which limited its applicability. This problem was solved by \cite{Chen:2018a} who developed the Gauss' Law satisfying ECSIM algorithm. GL-ECSIM is now the workhorse PIC algorithm used in the various kinetic models (iPIC3D, AMPS and FLEKS) in the SWMF.

\subsection{Hybrid Schemes}
\label{subsec:hybrid}
The SWMF allows applying different models in different physics domains. This flexibility is crucial to model a complex multi-scale and multi-physics system. A similar approach is used in BATS-R-US to apply different numerical methods (for example high order vs. second order scheme) or even different physics (for example Hall MHD vs. ideal MHD) in different regions to achieve optimal performance. The regions can be selected using geometry- and/or physics-based criteria. We have developed a general library that can define regions in the computational domain using addition and subtraction of simple geometrical objects (boxes, cylinders, spheres, cones, paraboloids, shells, rings, etc.) This allows the user to define regions of complicated shape from the input parameter file. When the region is dynamic, it can be defined by some local physical quantities based on the numerical solution. For example, one can use some threshold for the current density to apply adaptive mesh refinement. The two approaches can also be used in combination, for example the mesh refinement based on current density can be restricted to a certain part of the magnetotail. These capabilities are now available for a variety of options: adaptive mesh refinement, high vs. low order scheme, explicit vs. implicit schemes, Hall term, resistivity, heat conduction, viscosity, conservative vs. non-conservative energy equation, semi-relativistic correction, etc. To minimize numerical artifacts at the interfaces, we allow for a linear tampering in the critical parameter when applicable, for example the coefficient of the Hall term is 0 outside the Hall region, 1 inside, and varies linearly from 0 to 1 at the interface of a finite width. 

Using different schemes with different computational costs in the computational domain poses new requirements for the load balancing algorithm. Our approach is to assign a type for each grid block based on the combination of numerical schemes used. Blocks of the same type use the same combination of schemes, so their computational cost is similar. Then we load balance the various types of grid blocks independently. As long as there are enough grid blocks to fill the CPU cores, this approach works well.

\subsection{Achieving and Maintaining High Performance}
\label{subsec:highperform}
Both BATS-R-US and the SWMF were designed to achieve high performance on massively parallel super computers. BATS-R-US uses a block-adaptive grid for multiple reasons: it makes load balancing simple, it provides fixed loop sizes over the grid cells that can be easily optimized by the compiler, and the amount of data associated with each grid block can fit into the cache memory. The original design of BATS-R-US was based on the Message Passing Interface (MPI) library, which is still the most used parallelization tool on current supercomputers. All these design features improve performance and parallel scaling. In fact, in 1997 BATS-R-US achieved 13 Gflops on 512 cores of a Cray T3D computer. At that time this was among the largest supercomputers and $>10$ Gflop performance with excellent parallel scaling was a heroic achievement. The current version demonstrated nearly perfect weak scaling up to 250,000 cores of a Cray supercomputer, as shown in \figurename~\ref{fig:bats-scale}. While this is very respectable and more than sufficient for current supercomputers, we have to prepare the code for future architectures with even more cores and less memory per core. One reason we cannot run the code on even more cores is that the data structure describing the block-adaptive grid keeps growing with the problem size. To mitigate this issue, we have implemented a hybrid MPI+OpenMP parallelization \cite[]{Zhou:2020a}: MPI is used to parallelize over the CPU nodes, while OpenMP is used to use multi-threading over the cores on a single node. This means that large data structures can be shared by multiple threads, which reduces the memory use substantially. Using this hybrid parallelizaton, BATS-R-US could run up to 500,000 cores of the Blue Waters super computer while still maintaining excellent performance. 

The next frontier is porting large simulation codes to GPUs. Using support from NSF, we have started to work on porting BATS-R-US to GPUs. Our current approach is using the OpenACC library to parallelize loops over grid cells and run them on separate GPU threads. This work is in a preliminary phase now, but we already have some simple tests running on one GPU. 

\begin{figure}[ht]
    \floatbox[{\capbeside
        \thisfloatsetup{capbesideposition={left,top},
        capbesidewidth=0.4\textwidth}}]{figure}[\FBwidth]
    {\caption{The cell and particle update rates as a function of number of cores for the SWMF running the two-way coupled BAT-S-R-US and iPIC3D models. The problem size scales in proportion to the number of parallel processes. The dotted lines represent linear scaling.}
    \label{fig:epic-scale}}
    {\includegraphics[width=0.5\textwidth]{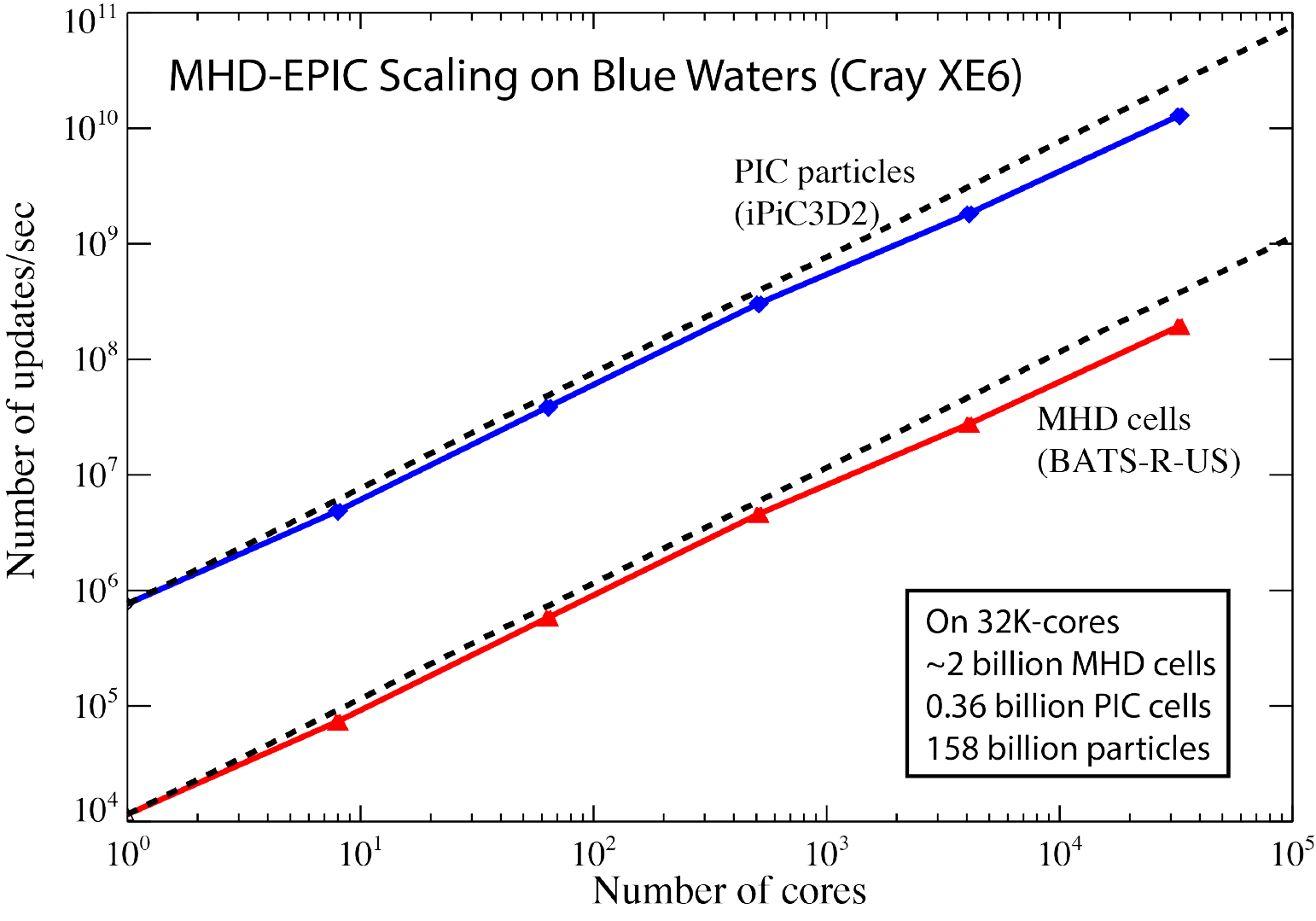}}
\end{figure}

The SWMF was designed to be as light weight as possible. The models can run serially or concurrently and synchronization is only performed when necessary \cite[]{Toth:2006b}. The current SWMF also supports models running with OpenMP: each model can use a different number of threads. Typically one thread per core is used, but hyper-threading is also supported. Coupling between the models also needs to be efficient, especially when a large amount of information is exchanged frequently. We have developed efficient and flexible coupling libraries that allow direct parallel coupling between two massively parallel models. As \figurename~\ref{fig:epic-scale} shows, the coupled BATS-R-US and iPIC3D models scale to 32,000 cores with minimal loss of efficiency. The two models exchange the MHD quantities calculated from the PIC distribution function for every MHD grid cell inside the PIC region every time step.

\section{MHD-EPIC and MHD-AEPIC}
\label{sec:epic}
The magnetohydrodynamics with embedded particle-in-cell (MHD-EPIC) algorithm allows global MHD simulations performed with the kinetic physics properly handled by a PIC model in regions of interest \cite[]{Daldorff:2014a}. The PIC domain can cover the regions where kinetic effects are most important, such as reconnection sites. In the newly developed MHD with adaptively embedded PIC (MHD-AEPIC) algorithm, the PIC domain consists of small blocks that can be adaptively activated and deactivated to cover the dynamically changing regions of interest. While keeping the expensive PIC model restricted to a small region, the BATS-R-US code can efficiently handle the rest of the computational domain where the MHD or Hall MHD description is sufficient. Since the PIC model is able to describe the electron behavior self-consistently, our coupled MHD-EPIC and MHD-AEPIC models are well suited for investigating the nature and role of magnetic reconnection in space weather phenomena.

\begin{figure}[hbt]
  \begin{floatrow}
    \ffigbox[\FBwidth]
      {\includegraphics[width=1\linewidth]{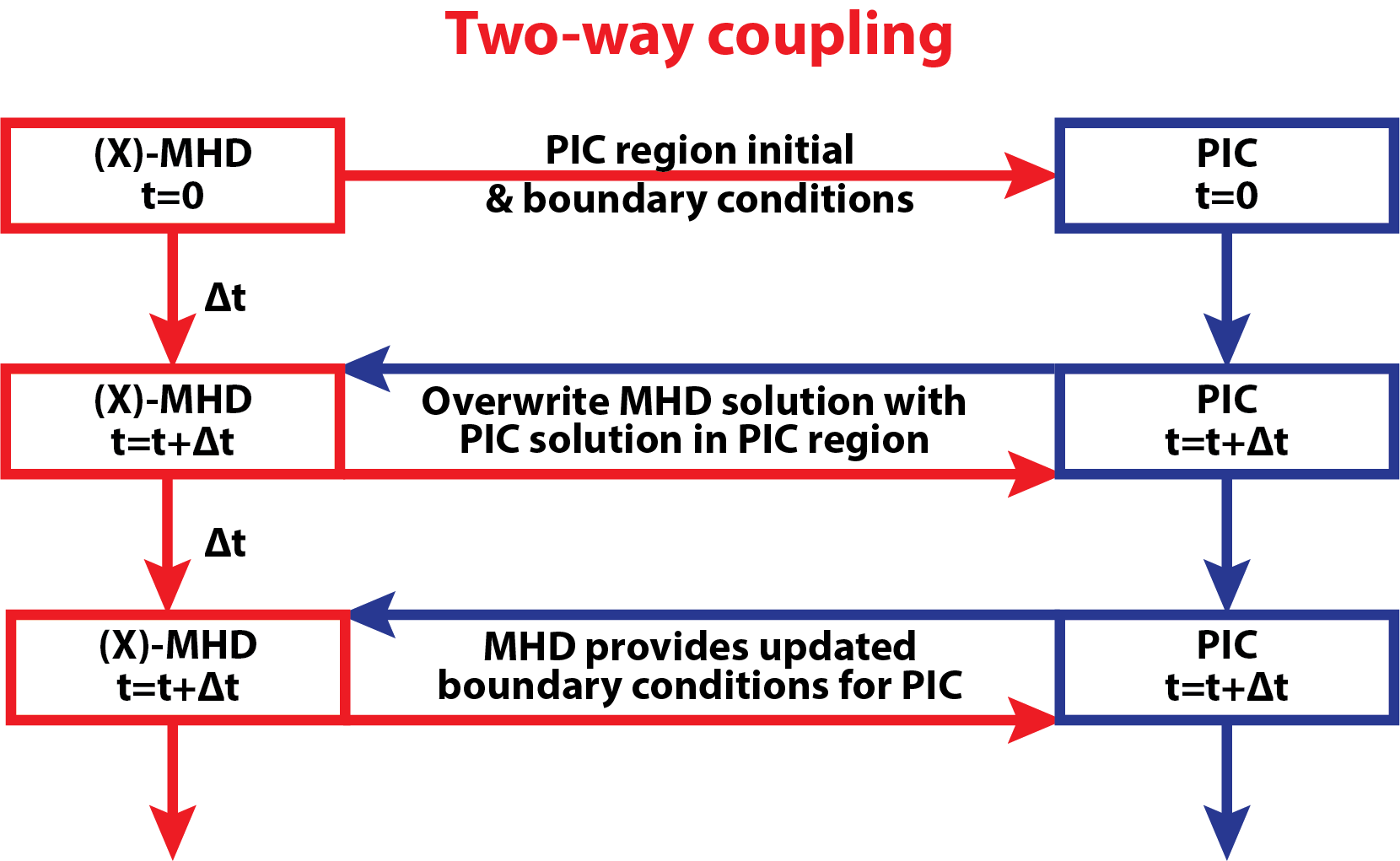}}
      {\caption{The overall flow of MHD-PIC coupling. At $t=0$ the MHD code sends the MHD state inside and around the PIC region to the PIC code. Both the MHD and PIC codes then advance by one or more time steps until both models reach the next coupling time. Information is exchanged both ways, but this time the PIC code only uses the MHD solution as a boundary condition, while the MHD code overwrites its solution in the PIC region with the PIC solution. This process continues until they reach the final simulation time or until the PIC region is removed. \cite[after][]{Daldorff:2014a}}
      \label{fig:epic1}}
    \ffigbox[\Xhsize]
      {\includegraphics[width=0.9\linewidth]{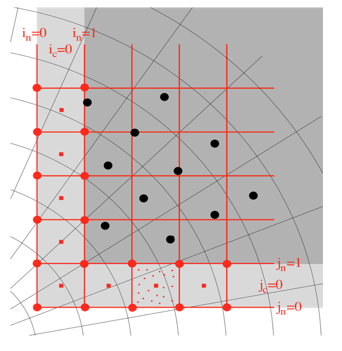}}
      {\caption{Spatial discretization of the MHD-EPIC coupling. The Cartesian grid of the PIC region is indicated with the darker gray area. The lighter gray area shows the ghost cell/node region of the PIC grid. The large red dots are node values obtained from the MHD variables. The small red dots illustrate particles created in the ghost cells of the PIC grid. The small red squares are the ghost cell centers of the PIC grid where the magnetic field is set from the MHD solution. The black dots indicate the MHD cell centers where the solution is obtained from the PIC code. The MHD grid can be either Cartesian or spherical. \cite[after][]{Chen:2017a}}
      \label{fig:epic2}}
  \end{floatrow}
\end{figure}

\figurename~\ref{fig:epic1} illustrates the overall flow of the coupling algorithm between BATS-R-US and iPIC3D \cite[]{Daldorff:2014a}, while \figurename~\ref{fig:epic2} shows the spatial discretization of the coupling. It is important to point out that the BATS-R-US -- iPIC3D coupling via SWMF is genuinely 2-way: all physical quantities are self-consistently advanced by both codes and the relevant information is fully exchanged in every time-step.

MHD-EPIC \cite[]{Daldorff:2014a, Chen:2017a, Toth:2017a, Zhou:2020a} uses the node centered number densities, velocities, pressures and magnetic field (large red dots in \figurename~\ref{fig:epic2}) to create the macro-particles inside the ghost cells of the PIC grid, as illustrated by the small red dots in the light gray area in \figurename~\ref{fig:epic2}. The particles that leave the PIC region (dark gray area in the figure) are discarded at the end of the PIC time step. New macro-particles are generated for each species in each ghost cell of the PIC domain with the appropriate (bi-)Maxwellian distribution functions using the MHD solution. The locations of the new particles are random with a uniform distribution over the ghost cell. For each ghost cell the corresponding number density, velocity and pressure are linearly interpolated from the surrounding MHD values (large red dots in the figure) to the given location. In this two-way coupled method the MHD values in the cell centers covered by the nodes of the PIC grid (black dots in \figurename~\ref{fig:epic2}) are fully overwritten by the PIC solution. The magnetic field can simply be interpolated from the PIC field. For the other MHD variables MHD-EPIC \cite[]{Daldorff:2014a, Chen:2017a, Toth:2017a, Zhou:2020a} takes various moments of the distribution function represented by the macro-particles.

The SWMF and the BATS-R-US codes require a Fortran compiler and the MPI library. The iPiC3D code requires a C++ compiler and the MPI library and optionally the parallel HDF5 library. Very good scaling up to 32k MPI processes on 1024 XE nodes of the Blue Waters supercomputer were achieved, as shown in \figurename~\ref{fig:bats-scale} and \figurename~\ref{fig:epic-scale}.

Currently, SWMF, BATS-R-US and iPiC3D mainly use pure MPI parallelism that works fine up to 1024 XE nodes and 32 thousand MPI processes (see \figurename~\ref{fig:bats-scale} and \figurename~\ref{fig:epic-scale}). While this is more than sufficient for most applications, the codes encounter some limitations when running with 65k and more MPI processes. Using OpenMP parallelism on the nodes can reduce memory usage and the number of messages sent between the MPI processes. The OpenMP+MPI approach does not improve the performance relative to the pure MPI parallelization, but the hybrid approach allows running larger problems on a larger number of nodes \cite[]{Zhou:2020a}.

\end{document}